\documentclass[a4paper,11pt]{article}
\usepackage{pos}
\usepackage{bbold}
\usepackage{graphicx}
\usepackage{physics}
\usepackage{subcaption}
\usepackage{cleveref}

\newcommand{\beq}{\begin{equation}}
\newcommand{\eeq}{\end{equation}}
\newcommand{\beqs}{\begin{eqnarray}}
\newcommand{\eeqs}{\end{eqnarray}}
\newcommand{\nn}{\nonumber}

\newcommand\Fig[1]{Fig.~\ref{fig:#1}}
\newcommand\Eq[1]{Eq.~(\ref{eq:#1})}
\newcommand{\gsim}{\mathrel{\raisebox{-
.6ex}{$\stackrel{\textstyle>}{\sim}$}}}
\newcommand{\lsim}{\mathrel{\raisebox{-
.6ex}{$\stackrel{\textstyle<}{\sim}$}}}

\title{\boldmath Progress in $Sp(2N)$ lattice gauge theories}
\ShortTitle{$Sp(2N)$ lattice gauge theories}

\author[a]{E.~Bennett}
\author*[b,i,1]{J.~Holligan}
\author[c]{D.~K.~Hong}
\author*[d,2]{H.~Hsiao}
\author*[c,e,3]{J.-W.~Lee}
\author[d,f]{C.-J.~David~Lin}
\author*[g,a,4]{B.~Lucini}
\author[a]{M.~Mesiti}
\author[b]{M.~Piai}
\author*[h,5]{D.~Vadacchino.}

\note{Presented {\em ``$Sp(4)$ and $Sp(6)$ quenched meson spectrum,''}combined in this work.}
\note{Presented {\em ``Mesonic spectrum of Sp(4) gauge theory with $N_f=3$ dynamical 
antisymmetry fermions,''} combined in this work.}
\note{Presented {\em ``$Sp(4)$ lattice gauge theory with fermions in multiple representations,''} combined in this work.}
\note{Presented {\em ``$Sp(2N)$ gauge theories on the lattice: status and perspectives,''} combined in this work.}
\note{Presented {\em ``Topology and scale setting for $Sp(2N)$ pure gauge theories,''} combined in this work.}

\affiliation[a]{Swansea Academy of Advanced Computing, Swansea University (Bay Campus),\\Fabian Way, SA1 8EN Swansea, Wales, United Kingdom}
\affiliation[b]{Department of Physics, Faculty of Science and Engineering, 
Swansea University (Park Campus),\\Singleton Park, SA2 8PP Swansea, Wales, United Kingdom}
\affiliation[c]{Department of Physics, Pusan National University, Busan 46241, Korea}
\affiliation[d]{Institute of Physics, National Yang Ming Chiao Tung University, 1001 Ta-Hsueh Road, Hsinchu 30010, Taiwan}
\affiliation[e]{Extreme Physics Institute, Pusan National University, Busan 46241, Korea}
\affiliation[f]{Center for High Energy Physics, Chung-Yuan Christian University, Chung-Li 32023, Taiwan}
\affiliation[g]{Department of Mathematics, Faculty of Science and Engineering, Swansea University (Bay Campus), Fabian Way, SA1 8EN Swansea, Wales, United Kingdom}
\affiliation[h]{School of Mathematics and Hamilton Mathematics Institute, Trinity College,\\D02 PN40 Dublin 2, Ireland}
\affiliation[i]{Physical Sciences Complex, University of Maryland, 
College Park, Maryland, USA, 20742}

\emailAdd{e.j.bennett@swansea.ac.uk}
\emailAdd{holligan@umd.edu} 
\emailAdd{dkhong@pusan.ac.kr}
\emailAdd{thepaulxiao@gmail.com}
\emailAdd{jwlee823@pusan.ac.kr}
\emailAdd{dlin@nycu.edu.tw}
\emailAdd{b.lucini@swansea.ac.uk}
\emailAdd{michele.mesiti@swansea.ac.uk}
\emailAdd{m.piai@swansea.ac.uk}
\emailAdd{vadacchd@tcd.ie}

\abstract{
Lattice studies of gauge theories with symplectic gauge groups
provide valuable information
about gauge dynamics, and complement the results of lattice investigations focused on 
unitary gauge groups.
These theories play a central role in
phenomenological contexts such as composite Higgs
and strongly interacting  dark matter models.
We report on recent progress of our lattice 
research programme, starting from 
the glueball mass spectrum and the
topology of the pure gauge theory. We present our 
results on the mass spectrum of mesons 
in the quenched approximation,
by varying the number of colours in the symplectic group.
For the $Sp(4)$ theory, we focus on
results obtained with dynamical fermion matter content
comprising both fundamental and 2-index 
antisymmetric representations of the gauge group, as
dictated by a well known model of composite
Higgs with partial top compositeness.}

\FullConference{%
 The 38th International Symposium on Lattice Field Theory, LATTICE2021
  26th-30th July, 2021
  Zoom/Gather@Massachusetts Institute of Technology
}

\begin{document}
\maketitle

\section{Introduction}
\label{sec:intro}

Our collaboration has been developing an extensive 
programme of lattice explorations of $Sp(N_c=2N)$ gauge 
theories~\cite{Bennett:2017kga,Lee:2018ztv,Bennett:2019jzz,
Bennett:2019cxd,Bennett:2020hqd,Bennett:2020qtj}, reaching 
far beyond pre-existing  explorative and pioneering work 
on the subject~\cite{Holland:2003kg}. In the long run, we 
want to achieve a level of understanding and control over 
$Sp(2N)$ gauge theories with varying fermionic field content 
that is comparable to the current state of the art
for $SU(N_c)$ theories, and use the results for phenomenological purposes.
This contribution reports non-trivial, preliminary results 
collected over the past year
within this research programme, and highlights the next 
steps of development we envision
for  the near future.

The main motivation for interest in $Sp(2N)$ gauge theories
comes from Composite Higgs 
Models (CHMs)~\cite{Kaplan:1983fs,
Georgi:1984af,Dugan:1984hq}---see 
the reviews in Refs.~\cite{Panico:2015jxa,
Witzel:2019jbe,Cacciapaglia:2020kgq}, 
and the summary tables in
Refs.~\cite{Ferretti:2013kya,Ferretti:2016upr,
Cacciapaglia:2019bqz}.
In particular, Ref.~\cite{Barnard:2013zea} highlights 
how $Sp(2N)$ theories with $N>1$, 
and field content consisting of an admixture of Dirac fermions transforming on 
the fundamental and 2-index antisymmetric representation of the gauge group,
provide a compelling 
origin form CHMs with $SU(4)/Sp(4)$ coset, and naturally incorporate also 
an enhancement mechanism for the mass of the top 
through partial
compositeness. Importantly, these CHMs can be studied on the lattice
with existing technology.

Thanks to the peculiarities 
of the $Sp(2N)$ groups 
and their pseudo-real representations, our lattice studies 
provide an essential contribution to the 
growing literature on non-perturbative lattice investigations of candidate CHMs. 
We refer the reader to the literature on lattice theories
based on the gauge group 
$SU(2)$~\cite{Hietanen:2014xca,Detmold:2014kba,
Arthur:2016dir,Arthur:2016ozw,Pica:2016zst,Lee:2017uvl,
Drach:2017btk,Drach:2020wux,Drach:2021uhl}
(that yield CHMs with $SU(4)/Sp(4)$ coset, but no top compositeness), 
and $SU(4)$~\cite{Ayyar:2017qdf,Ayyar:2018zuk,Ayyar:2018ppa,Ayyar:2018glg,
Cossu:2019hse,Shamir:2021frg} (which are closely related to
CHMs with $SU(5)/SO(5)$ coset), 
as well as to the recent idea~\cite{Appelquist:2020bqj} 
 (see also Refs.~\cite{Vecchi:2015fma,
Ma:2015gra,BuarqueFranzosi:2018eaj}) 
of using lattice results on the $SU(3)$ theory with $N_f=8$
fundamental fermions~\cite{Aoki:2014oha,
Appelquist:2016viq,Aoki:2016wnc,
Gasbarro:2017fmi,Appelquist:2018yqe}
to build CHMs with $SU(8)\times SU(8)/SU(8)$ coset. 

The beautiful, enhanced symmetry structure of $Sp(2N)$ gauge theories has other
applications, prominently in the context of 
models of dark matter arising from strongly 
coupled dynamics~\cite{Hochberg:2014dra, Hochberg:2014kqa,
Hochberg:2015vrg}---see 
for example Refs.~\cite{Bernal:2017mqb, Berlin:2018tvf, 
Bernal:2019uqr, Tsai:2020vpi, Maas:2021gbf, Zierler:2021cfa}.
More generally, underlying theoretical questions 
about the nature of gauge dynamics, the
mechanism of confinement, the physics of non-trivial 
bound states such as glueballs, mesons,
baryons and more exotic composite objects (such as the chimera baryons 
described in the body of this paper), and the approach to the large-$N_c$ limit,
with its relation to other non-perturbative approaches 
arising for instance in string theory,
are all topics that are amenable to numerical studies within the context of $Sp(2N)$
lattice gauge theories---an incomplete list of relevant references includes
Refs.~\cite{Lucini:2001ej,
Lucini:2012gg,Lucini:2004my,Lucini:2010nv,
Athenodorou:2015nba,Lau:2017aom,Hong:2017suj,
Yamanaka:2021xqh,Athenodorou:2021qvs,Hernandez:2020tbc}.

This document is organised as follows. 
We start with a critical summary of earlier results,
which includes new, unpublished material, in Sect.~\ref{sec:review}. We
draw the attention of the reader 
to the comparison of our results with 
the most up-to-date literature on $SU(N_c)$ theories.
We present a summary plot 
of the spectrum  of the quenched $Sp(4)$ 
gauge theory, 
both for glueballs and  mesons made of fundamental as well as 
antisymmetric fermionic matter. 
After summarising some important technical aspects of 
our lattice study in Sect.~\ref{sec:WF_scale_setting},
we then present our new, preliminary results.
In Sect.~\ref{sec:topology} we discuss the topology 
of pure Yang-Mills theories
with $Sp(2N)$ group for $N=1, 2, 3, 4$, together 
with a preliminary extrapolation to the large-$N$ limit,
and a comparison with $SU(N_c)$ theories.
In Sect.~\ref{sec:quenchedmeson} we show the spectra of 
mesons in the quenched approximations for
$Sp(2N)$ theories with $N=2, 3, 4$, and with fermions
in three different representations: the fundamental, 
the 2-index antisymmetric, and the 2-index symmetric representation.

The theory proposed in 
Ref.~\cite{Barnard:2013zea} as candidate CHM
with top compositeness returns to take central stage  in Sect.~\ref{sec:dynamical_sp4}.
We focus on the  $Sp(4)$  gauge theory with 
two Dirac fermions in the fundamental representation 
and three in the antisymmetric representation.
We set the stage for
the calculation of meson spectra and chimera baryons 
by first analysing the partially quenched theory
(in which fundamental representation fermions are non dynamical), 
and show our results in Sect.~\ref{sec:as}.
Sect.~\ref{sec:multirep} contains the first steps of the 
investigation of the fully dynamical theory,
in particular we present a scan of the lattice 
parameter space relevant for the study of the theory
 with mixed fermionic content.
We conclude with a short outlook in Sect.~\ref{sec:conclusion}.

\begin{figure}[tb]
\centering
\includegraphics[scale=0.6]{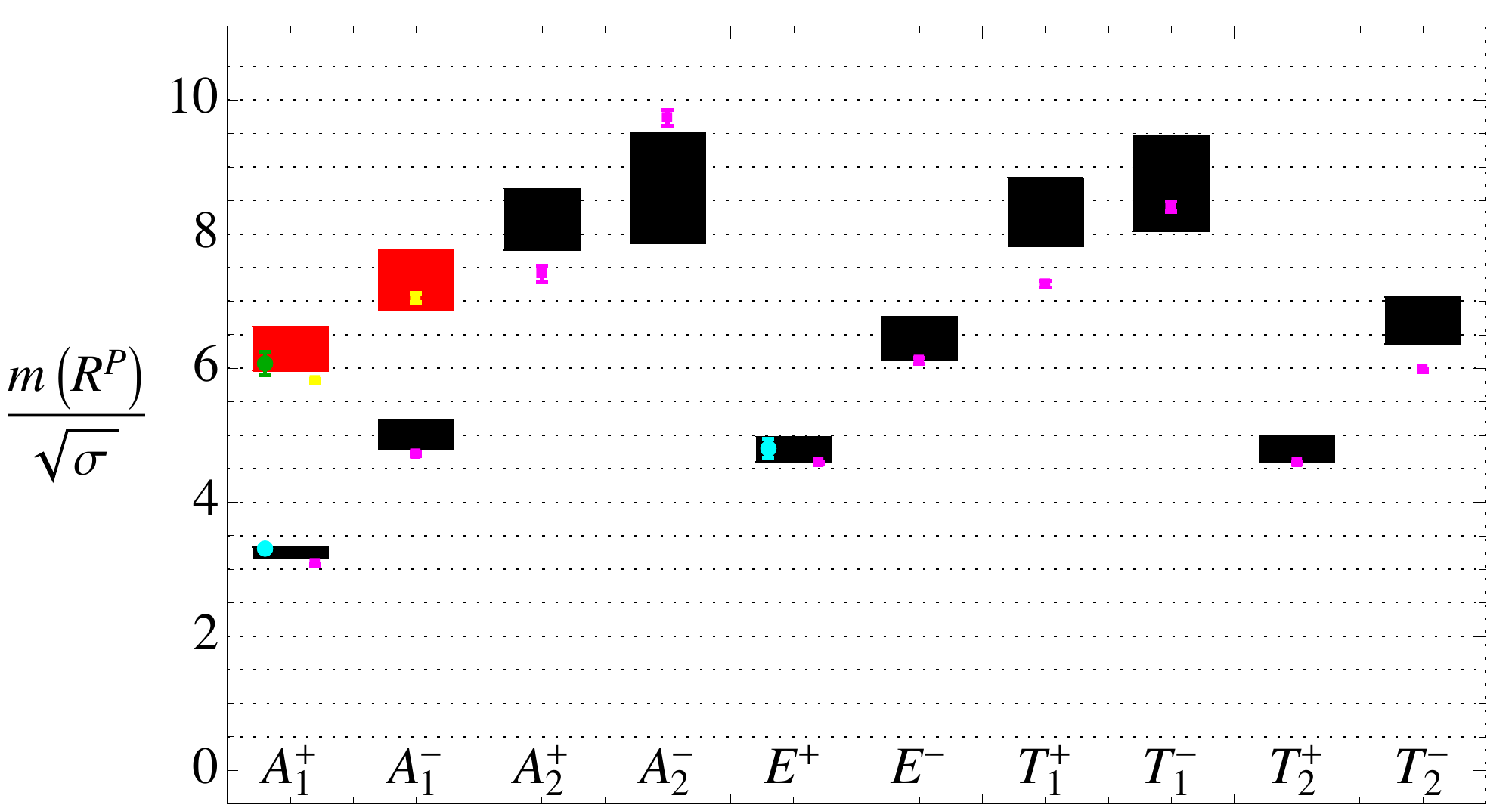}
\caption{Large-$N_c$ extrapolation of the glueball mass spectrum,
in units of the square root of the string tension,
for states transforming under irreducible representations of 
the octahedral group---as indicated along the abscissa. 
The boxes are our results for $Sp(N_c)$~\cite{Bennett:2020qtj}, 
with the ground states in each channel in black and the measured 
excitations in red. The points circles are taken from 
the literature on $SU(N_c)$. 
In particular, cyan and green points are from Ref.~\cite{Lucini:2004my}, 
while magenta and yellow 
points are from Ref.~\cite{Athenodorou:2021qvs}.\label{fig:spectrum_m_sigma}}
\end{figure}

\begin{figure}
\centering
\includegraphics[scale=0.45]{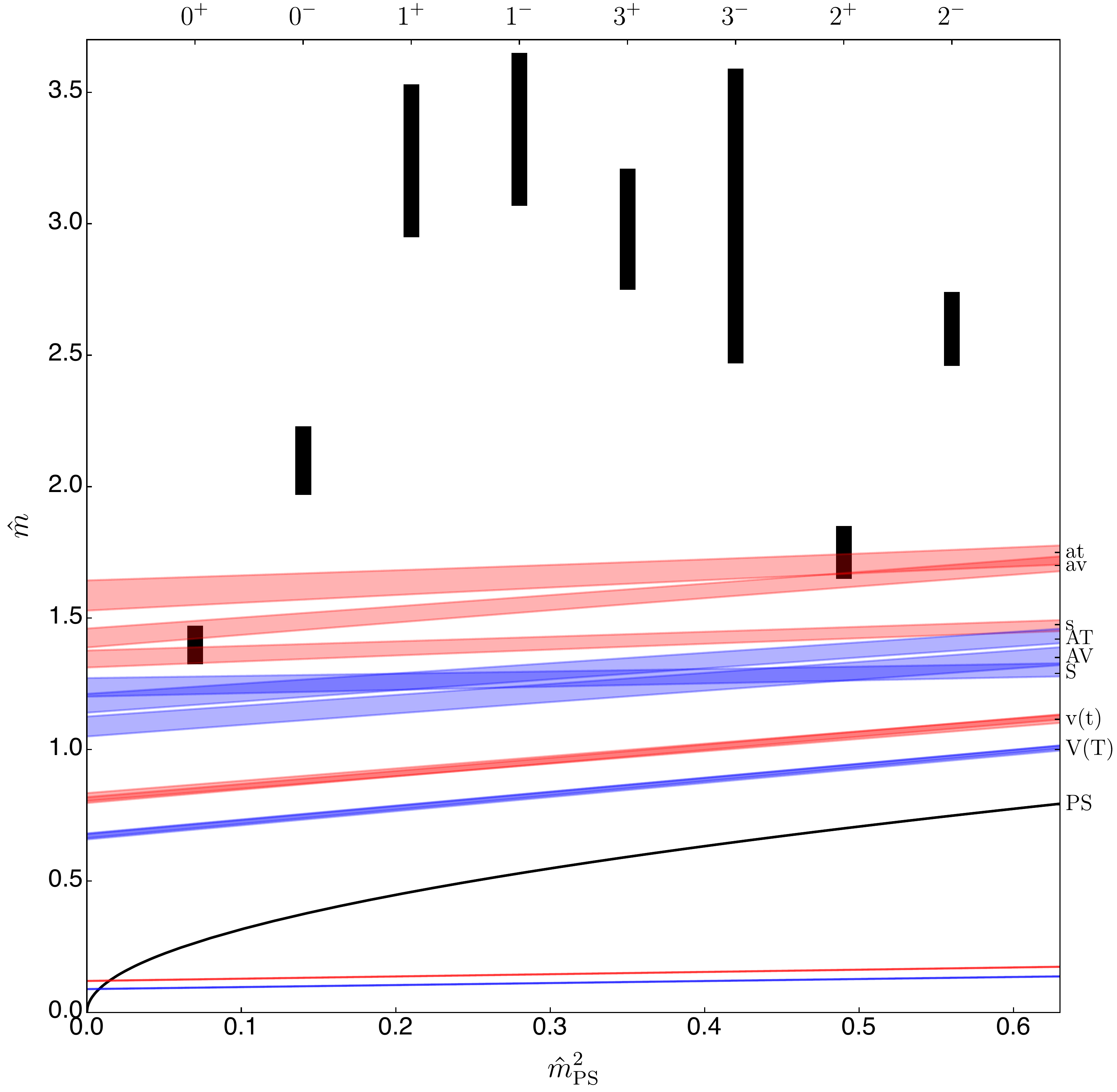}
\caption{Masses and decay constants of mesons in the quenched approximation
for fermions in the fundamental (blue bands) and antisymmetric (red bands)
representation of $Sp(4)$,
 as a function of the pseudoscalar mass squared.
 Yang-Mills glueball spectra (black boxes) are independent of
 the pseudoscalar mass, because of the quenched 
 approximation.~\label{fig:specturm_fundasymglue}
 The pseudoscalar mass is indicated in the plot with a narrow (black) line.
 Individual mesons are labelled on the right vertical axes. 
 Continuum glueball quantum numbers $J^P$ are indicated 
 on the top horizontal axis.~\label{fig:specturm_fundasymglue}
 }
\end{figure}

\begin{figure}
\centering
\includegraphics[scale=0.8]{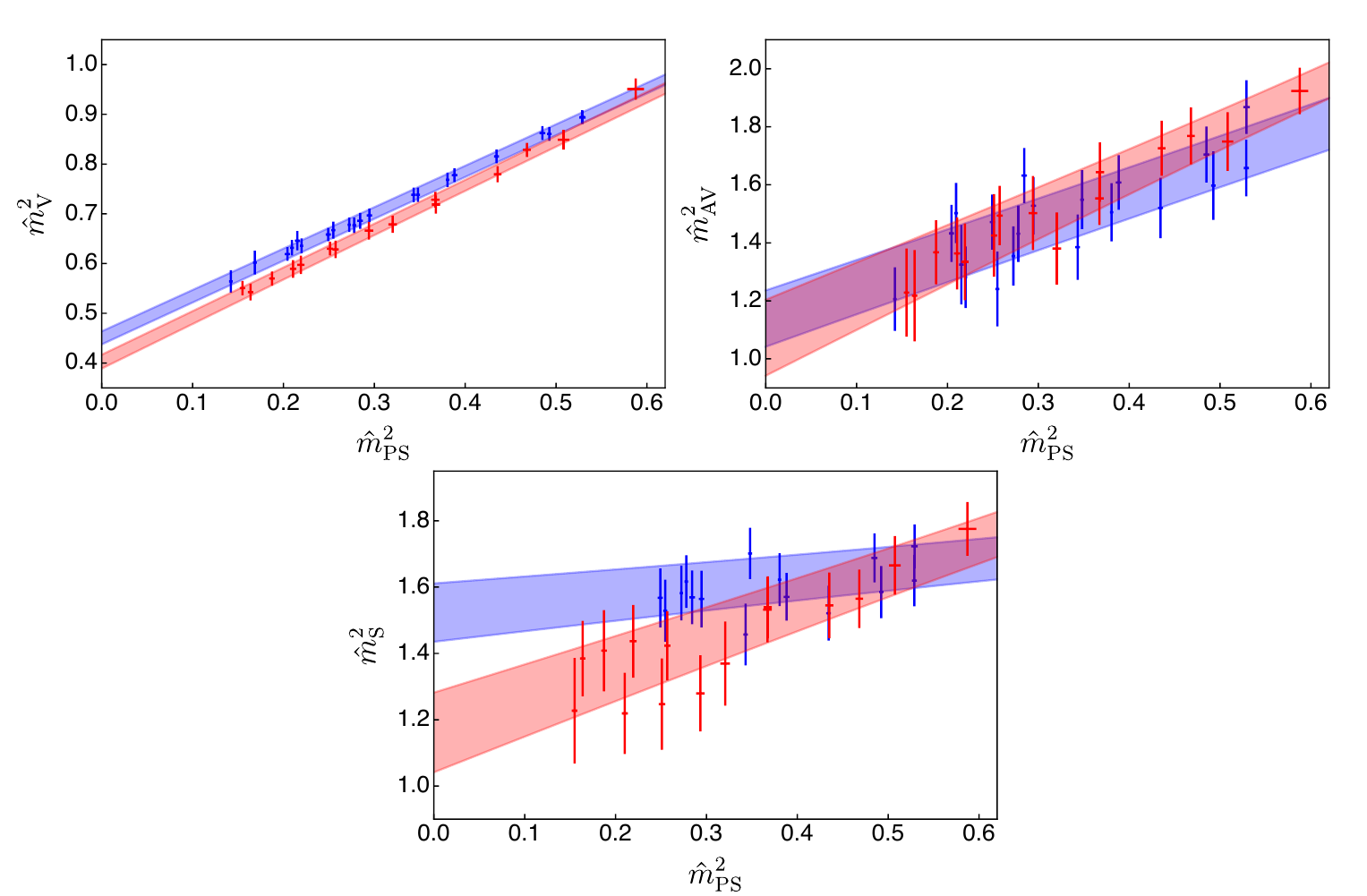}
\caption{Meson mass squared in the continuum limit as a 
function of the pseudoscalar mass squared for the vector meson (top left panel), 
the axial-vector meson (top right) and the flavoured scalar meson (bottom). 
In blue we represent quenched and in red dynamical data. 
The bands are extrapolations to the chiral limit.\label{fig:compare_mesons}}
\end{figure}

\begin{figure}
\centering
\includegraphics[scale=0.8]{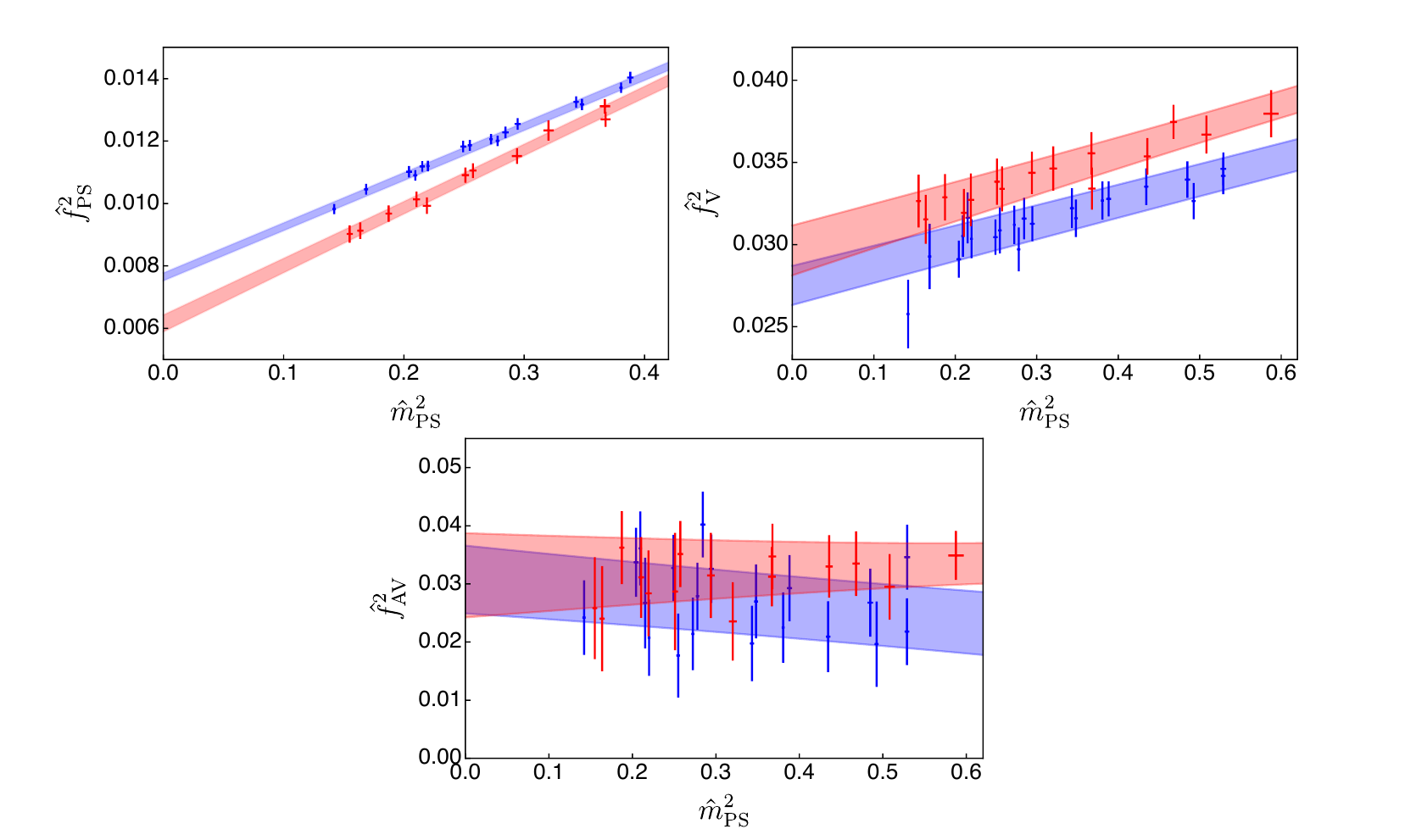}
\caption{Decay constant squared of the pseudoscalar meson (top left panel), 
 the vector meson (top right) and  the axial-vector meson (bottom) 
in the continuum limit, as a function of the pseudoscalar mass squared. 
In blue we represent quenched and in red dynamical data. The bands are extrapolations  to the chiral limit.\label{fig:compare_decayconst}}
\end{figure}

\section{Critical review of previous work}
\label{sec:review}

Prior to  the start of our systematic  programme of investigation 
of $Sp(2N)$ lattice theories~\cite{Bennett:2017kga}, 
very little was known numerically about them, available lattice calculations
being limited to the Yang-Mills 
system for $N=2,3$~\cite{Holland:2003kg}.\footnote{
The conformal window of $Sp(2N)$ theories had been investigated
in Refs.~\cite{Sannino:2009aw,Ryttov:2017dhd}--- 
see also the recent Refs.~\cite{Kim:2020yvr,Lee:2020ihn}---and references therein. 
The EFT description of
$Sp(2N)$ gauge theories had been explored for instance in Ref.~\cite{Appelquist:1999dq}.
}
Although motivated by sophisticated CHMs with  dynamical
matter fields in multiple, distinct 
representations of the gauge group, we started by first studying simpler theories,
that are theoretically better understood and cheaper to simulate, 
and we gradually built upon these as useful reference 
points for our subsequent more demanding investigations.
In this section, we review the body of work we
have produced to date in this programme, 
highlighting its potential implications, both for phenomenological applications and
for  theoretical purposes.
 The order of presentation is chosen to provide a
 pedagogical way to introduce progressively
   relevant concepts and observables with increasing 
   level of complexity. Due to space limits, 
   we provide only a discussion of the main results, referring to the original
    literature for details.

We start from the Yang-Mills theory. We have determined 
the continuum limit of the glueball masses
 in units of the string tension for $N=2,3,4$ and extrapolated those masses 
  to $N \to \infty$. Our results (taken from 
  Ref.~\cite{Bennett:2020qtj}) are reported in Fig.~\ref{fig:spectrum_m_sigma}. We 
  label the states by the irreducible representations of the octahedral group, 
  to which the continuous rotation symmetry reduces on a
  cubical spatial lattice---see
   for instance Ref.~\cite{Lucini:2010nv}. Since $Sp(2N)$ groups are pseudoreal, 
   charge conjugation is always positive.
The $0^{+}$ glueball is expected to be 
   the lightest state in the $A_1^+$ irreducible representation, 
   while the $2^{+}$ appears in both 
   the $E$ and the $T_2$ representations of the octahedral group. 

Theories based upon $Sp(N_c)$ and $SU(N_c)$ group sequences are expected to
yield observables that agree (in the common physical sector) in
the large-$N_c$ limit. We can check this explicitly by
comparing our calculations to the extrapolations to $SU(\infty)$ 
done in Refs.~\cite{Lucini:2001ej,Lucini:2004my} (also discussed in~\cite{Lucini:2012gg}). These earlier works report
only the lowest-lying states in the spectrum 
(the ground state $0^{++}$ and  $2^{++}$, as well as the excited $0^{++*}$ state),
and Fig.~\ref{fig:spectrum_m_sigma} shows good agreement with our
results. 
For the other states in the spectrum,  our 
calculation yield the first determination of the masses in the large-$N$ limit.
A large-scale calculation of the $SU(\infty)$ glueball 
spectrum has appeared recently~\cite{Athenodorou:2021qvs}, and we 
display its results in Fig.~\ref{fig:spectrum_m_sigma}. 
While the errors are visibly 
smaller than ours, the two sets of results are in good agreement, 
all central values being compatible within two standard deviations.

To get a first sense of the behaviour of the spectrum in the 
presence of fermions, we have computed masses and decay constants 
of mesons, in the continuum limit and as a function of the
pseudoscalar meson mass, but
in the quenched approximation~\cite{Bennett:2019cxd}.
Fig.~\ref{fig:specturm_fundasymglue} displays those quantities 
together with Yang-Mills glueball masses. 
We set the scale by the quantity $w_0$, 
extracted from the gradient flow as described in Ref.~\cite{Bennett:2017kga}. 
While the general features of the meson spectrum do not show 
any striking qualitative representation-dependent features,
yet, at fixed pseudoscalar mass, states in the antisymmetric 
representation are heavier than states in the fundamental representation. 
Glueballs are generally heavier than mesons in both representations, 
except that the lightest glueballs have masses comparable to 
those of the heaviest meson channels we considered.

Dynamical effects due to fundamental representation fermions in masses 
and decay constants of mesons have been studied in Ref.~\cite{Bennett:2019jzz}. 
We display some representative meson masses
in Fig.~\ref{fig:compare_mesons} and decay constants in
Fig.~\ref{fig:compare_decayconst}, again in units of the 
gradient flow scale $w_0$. In the  range of mass of 
the pseudoscalar meson currently investigated, 
unquenching effects are generally small, 
the only exceptions being the mass of the flavoured
scalar and the decay constant of the pseudoscalar meson.

In the rest of this proceedings contribution, 
we will focus on new, unpublished results,
for a large set of observable quantities, and in a broader set of
$Sp(2N)$ gauge theories.

\section{Topology and Scale setting in $Sp(N_c)$ pure gauge theories}
\label{sec:topology}

The topological susceptibility of gauge theories is an
interesting quantity for several reasons. 
It enters the Witten-Veneziano formula~\cite{Witten:1979vv,Veneziano:1979ec}, 
that explains the large-$N_c$ behavior of the mass of the $\eta'$ meson. 
It is the coefficient of the $O(\vartheta^2)$ term in a power 
expansion of the free energy around $\vartheta=0$, and as such might 
have implications for the strong-CP problem. 
It gives a quantitative account of the importance of instanton configurations 
in the study of chiral symmetry breaking. 
As $Sp(N_c)$ gauge groups might be relevant to the
UV-completion of composite Higgs 
and composite dark matter models, 
the study of their topological susceptibility might find further applications,
and furthermore it offers an additional way to compare $SU(N_c)$ 
and $Sp(N_c)$ theories at large $N_c$.

The configuration space of gauge fields
on a four dimensional torus can be partitioned into 
topological sectors, each characterized by the value of the topological charge $Q =\int \mathrm{d}^4 x q(x)$, where
\begin{equation}
    q(x) \equiv \frac{1}{32 \pi^2} \epsilon^{\mu\nu\rho\sigma} \mathrm{Tr}~ F_{\mu\nu} F_{\rho\sigma}~.
\end{equation}
$Q$ takes values in the third homotopy group of the gauge group, and, since 
$\pi_3(Sp(2N_c))=\mathbb{Z}$, we expect $Q$ to be integer 
valued---as for $SU(N_c)$. The topological susceptibility is defined as
\begin{equation}
    \chi = \int \mathrm{d}^4 x \langle q(x) q(0) \rangle~.
\end{equation}
The evaluation of $\chi$ on the lattice is known 
to be non-trivial for a variety of reasons. First of all,
the configuration space of lattice fields is simply 
connected and its topology is thus trivial; the partitioning
of said space in topological sectors arises as a dynamical 
effect in the vicinity of the continuum limit~\cite{Luscher:1981zq}.
Second, as the continuum limit is approached, the lattice 
topological susceptibility is dominated by quantum UV
fluctuations~\cite{Campostrini:1989dh}. Lastly, the 
integrated auto-correlation time of the topological charge is known
to diverge as $a\to 0$, and is observed to do so progressively
faster as $N_c$ is increased~\cite{DelDebbio:2002xa}. 

One way to overcome most of the aforementioned technical
difficulties makes use of the gradient flow~\cite{Luscher:2010iy} 
and its lattice incarnation, the Wilson flow. At finite flow time, 
the Wilson flow is a smoothened version 
of the lattice field, which removes
UV fluctuations as $a\to 0$, improving the convergence to the continuum limit
for the topological susceptibility and other physical quantities. 
Pure gauge quantities are renormalized at finite flow-time~\cite{Luscher:2013vga};
the value of energy-dependent observables along the
flow can also be related to the renormalized coupling calculated 
in the gradient flow renormalization scheme.
The Wilson Flow is thus a natural  quantity to use to set the 
scale of lattice calculations. 

The exception is the last of the problems listed above, 
known as topological freezing, 
which afflicts the calculations performed near the continuum limit. It is
related to the intrinsic difficulty of evolving a global property
as the topological charge by a finite sequence of local updates. 
Dedicated algorithms have been suggested (see, e.g.,~\cite{Bonanno:2020hht,Borsanyi:2021gqg,Cossu:2021bgn}) to address this problem. However, these algorithms are computationally expensive. In our preliminary work, we quantify and monitor the effects of topological freezing through the measurement of the integrated correlation time of the topological charge, which we call $\tau_Q$. 

In this section, we report on our preliminary analysis of
the properties of the Wilson flow and on the calculation 
of the topological susceptibility in $Sp(N_c)$ pure gauge theories 
at finite $N_c=2N$ and in the $N_c\to\infty$ limit.

\subsection{The lattice theory}

The $Sp(N_c)$ lattice gauge theory is defined on a four
dimensional Euclidean lattice of spacing $a$ by the Wilson action:
\begin{equation}
S_\mathrm{W} \equiv \beta \sum_x \sum_{\mu<\nu} \left( 1-
\frac{1}{N_c}\Re \mathrm{Tr}\, \mathcal{P}_{\mu\nu} \right)~,
\label{eq:gauge_action}
\end{equation}
where $x$ labels the sites, $\mu$ and $\nu$ the directions on 
the lattice and $\beta=\tfrac{2 N_c }{g_0^2}$ is the inverse coupling.
The quantity $\mathcal{P}_{\mu\nu}(x)$ is called the elementary plaquette:
\begin{equation}\label{eq:plaquette}
\mathcal{P}_{\mu\nu}(x) \equiv U_\mu(x) U_\nu(x+\hat{\mu})
U^\dag_\mu(x+\hat{\nu}) U^\dag_\nu(x)\,,
\end{equation}
written in term of the link variables $U_\mu(x)$.
We consider lattices that have the same linear extent $L$ 
and periodic boundary conditions in all space-like directions. 

\subsection{Wilson flow and scale setting for $Sp(N_c)$ gauge theories}
\label{sec:WF_scale_setting}

The Wilson Flow $V_\mu(t,\,x)$ is defined as the solution of the equation
\begin{equation}
\frac{ \partial V_\mu( t,\,x) }
{\partial t} = -g_0^2 \left\{
\partial_{x,\,\mu} S_\mathrm{W}\left[V_\mu\right]\right\}
V_\mu(t,\, x)\,,
\label{eq:WF}
\end{equation}
with initial condition $V_\mu(0,\,x)=U_\mu(x)$. It is demonstrated 
to exist and to be unique on every finite lattice. 
Since $\partial_t S_\mathrm{W}\leq 0$, as the flow-time increases
the field tends to configurations that make the action 
stationary---the classical configurations. Moreover, 
integrating Eq.~(\ref{eq:WF}) is equivalent, at leading order
in $g_0$, to smoothening the field over a sphere 
of mean-square radius $\sqrt{8t}$. In practice, this 
can be used to remove short-distance singularities,
and the need to introduce counterterms for composite operators.
Lastly,  it was shown in Ref.~\cite{Luscher:2010iy} that,
at leading order in $g_0$,
\beq
\alpha(\mu) = k_\alpha t^2 \langle E(t) \rangle = k_\alpha \mathcal{E}(t)\,,
\label{eq:GF_scheme}
\eeq
where $\alpha(\mu)$ is the renormalized coupling at energy scale $\mu=\frac{1}{\sqrt{8t}}$, $k_{\alpha}$ is a calculable constant,
and $E(t)$ is defined as follows:
\beq
E(t) = -\frac{1}{2} \Tr \left(G_{\mu\nu} G_{\mu\nu}\right)\,,
\eeq
with $G_{\mu\nu}$  any lattice operator that tends to the field-strength 
tensor in the continuum limit. 
Alternative definitions yield differences proportional 
to  powers of the lattice spacing, and are used to estimate
the magnitude of discretization errors. 
In this work, we consider the plaquette operator in Eq.~(\ref{eq:plaquette}) 
and compare to the four-plaquette clover operator, to this purpose.

The scales $t_0$ and $w_0$ are defined by fixing reference 
values for $\mathcal{E}_0$ and $\mathcal{W}_0$ and solving
\begin{equation}\label{eq:scales}
\mathcal{E}(t_0) = \mathcal{E}_0~,\qquad
\mathcal{W}(t=w_0^2) = \mathcal{W}_0\,,
\end{equation}
where $\mathcal{W}(t) = t \frac{d}{dt} \mathcal{E}(t)$. 

But in our study we compute $t_0$ and $w_0$ for  $Sp(N_c)$ theories
with several values of $N_c$, and we are interested 
to the ('t Hooft) limit $N_c\to\infty$, 
reached by holding  $\lambda \equiv N_c \alpha$ fixed.
Because at leading order in $\alpha$ one expects the perturbative result
\begin{equation}\label{eq:pert_scaling}
    t^2 \langle E \rangle = \frac{4 C_2(F)}{32 \pi^2} \lambda ( 1 + O(\alpha))  
\end{equation}
to hold, with $C_2(F) = \frac{N_c + 1}{4}$  the quadratic Casimir 
of the fundamental representation of $Sp(N_c)$,  we 
decided to scale $\mathcal{E}_0$ and $\mathcal{W}_0$ according
to the following relations:
\begin{equation}\label{eq:scaling_e0w0}
\mathcal{E}_0 = c_e \frac{ N_c +1}{4},\qquad
\mathcal{W}_0 = c_w \frac{ N_c +1}{4}~,
\end{equation}
where $c_e,\,c_w$ are constants. 
Whether this scaling law, suggested by perturbative theory, 
is reproduced is general 
is one of the question we start to address in the following.

\begin{table}[]
\centering
\caption{Numerical choices of lattice parameters ($N_c$, $L/a$, $N_{conf}$,
$\beta$), of the reference values of $\mathcal{E}_0$
and $\mathcal{W}_0$ (rescaled with $N_c$ as explained in the main text)
and of our measurements of $t_0/a^2$, $w_0/a$, $\chi_L t_0^2$, and  
$\chi_L w_0^4$. }
\label{tab:setup_topology}
\begin{tabular}{cccccccccc}
\hline
$N_c$& $L/a$& $N_{conf}$ & 
$\beta$ & $\mathcal{E}_0$ & $t_0/a^2$ & $\chi_L t_0^2$ & $\mathcal{W}_0$ & $w_0/a$ & $\chi_L w_0^4$ \\
\hline
$2$&$12$&$2000$ &$2.4 $ &$0.375 $ &$ 3.455(11) $ & $ 3.93(12) $ & $ 0.525 $ & $ 1.9780(47) $ & $ 3.88(13) $ \\
$2$&$16$&$2000$ &$2.475$ & $0.375 $ &$ 5.609(16) $ & $ 5.10(17) $ & $ 0.525 $ & $ 2.4816(50) $ & $ 5.01(16) $ \\
$2$&$20$&$4000$ &$2.55$ & $0.375 $ &$ 9.095(20) $ & $ 4.78(11) $ & $ 0.525 $ & $ 3.1379(47) $ & $ 4.78(11) $ \\
$2$&$24$&$2000$ &$2.6$ & $0.375 $ &$ 12.490(40) $ & $ 5.10(19) $ & $ 0.525 $ & $ 3.6621(77) $ & $ 5.10(18) $ \\
$4$&$20$&$975$  &$7.7$ &$ 0.625 $ &$ 4.3620(61) $ & $ 23.2(1.3) $ & $ 0.875 $ & $ 2.2645(21) $ & $ 23.2(1.1) $ \\
$4$&$20$&$2002$ &$7.72$ & $ 0.625 $ &$ 4.6043(44) $ & $ 19.63(66) $ & $ 0.875 $ & $ 2.3281(17) $ & $ 19.58(63) $ \\
$4$&$20$&$2391$ &$7.76$ & $ 0.625 $ &$ 5.1101(51) $ & $ 15.47(44) $ & $ 0.875 $ & $ 2.4532(17) $ & $ 15.46(43) $ \\
$4$&$20$&$2002$ &$7.78$ & $ 0.625 $ &$ 5.3689(63) $ & $ 14.09(48) $ & $ 0.875 $ & $ 2.5146(21) $ & $ 14.04(54) $ \\
$4$&$20$&$2002$ &$7.8$ & $ 0.625 $ &$ 5.6386(75) $ & $ 11.79(39) $ & $ 0.875 $ & $ 2.5766(26) $ & $ 11.79(40) $ \\
$4$&$20$&$2202$ &$7.85$ & $ 0.625 $ &$ 6.3559(88) $ & $ 9.45(39) $ & $ 0.875 $ & $ 2.7381(27) $ & $ 9.39(37) $ \\
$4$&$24$&$697$  &$8.2$ & $ 0.625 $ &$ 13.771(48) $ & $ 4.56(28) $ & $ 0.875 $ & $ 4.047(11) $ & $ 4.59(24) $ \\
$6$&$18$&$952$  &$15.75$ &$ 0.875 $ &$ 4.1301(47) $ & $ 15.16(85) $ & $ 1.225 $ & $ 2.2560(18) $ & $ 15.39(81) $ \\
$6$&$16$&$499$  &$15.9$ & $ 0.875 $ &$ 5.014(15) $ & $ 6.23(43) $ & $ 1.225 $ & $ 2.4926(54) $ & $ 6.16(40) $ \\
$6$&$16$&$751$  &$16.1$ & $ 0.875 $ &$ 6.395(20) $ & $ 3.70(26) $ & $ 1.225 $ & $ 2.8235(64) $ & $ 3.70(25) $ \\
$6$&$20$&$763$  &$16.3$ & $ 0.875 $  &$ 8.003(18) $ & $ 4.92(27) $ & $ 1.225 $ & $ 3.1618(52) $ & $ 4.90(29) $ \\
$8$&$16$&$1299$ &$26.7$ &$ 1.125 $ &$ 4.1931(49) $ & $ 8.71(37) $ & $ 1.575 $ & $ 2.3129(20) $ & $ 8.78(36) $ \\
$8$&$16$&$800$  &$27.0$ & $ 1.125 $ &$ 5.2555(92) $ & $ 5.60(31) $ & $ 1.575 $ & $ 2.5949(35) $ & $ 5.60(29) $ \\
$8$&$16$&$300$  &$27.2$ & $ 1.125 $ &$ 6.089(22) $ & $ 3.58(42) $ & $ 1.575 $ & $ 2.7964(77) $ & $ 3.54(41) $ \\
\hline
\end{tabular}
\end{table}

\subsection{Topological charge and susceptibility}

The topological charge density at flow-time $t$ is defined as
\begin{equation}
\label{eq:lat_top_charge}
q_L(t,\,x) \equiv \frac{1}{32\pi^2} \epsilon^{\mu\nu\rho\sigma}
\mathrm{Tr} ~V_{\mu\nu}(t,\,x) V_{\rho\sigma}(t,\,x)~,
\end{equation}
where $V_{\mu\nu}(t,\,x)$ is the plaquette operator at flow-time $t$,
built from the solutions of Eq.~(\ref{eq:WF}).
The topological charge is hence $Q_L(t)=\sum_x q_L(t,\,x)$. The topological susceptibility is defined as 
\begin{equation}\label{eq:topo_suscept}
a^4 ~\chi_L (t) \equiv \left\langle \sum_x q_L(t,\,x) q_L(t,\,0) \right\rangle_\beta = \frac{\langle Q_L(t)^2 \rangle}{L^4}~.
\end{equation}
and is computed from the values of $Q_L$ as measured on the lattice. 

General arguments suggest that the sequence of $SU(N_c)$ and $Sp(N_c)$ theories 
should yield observables (in their common sectors)
that agree in the large-$N_c$ limit. 
One such observable is expected to be the topological susceptibility, and 
we will next proceed to report our preliminary 
results for $Sp(2N)$ theories with $N=1,\,2,\,3,\,4$, and 
a preliminary extrapolation towards large-$N_c$.

\subsection{Numerical results}
\label{sec:num_scale_setting}

Ensembles of thermalized configurations of the lattice $Sp(N_c)$ Yang-Mills
theories, at different values of $\beta$ and $L/a$ for $N_c=2,\,4,\,6,\,8$, 
were generated using a heat-bath  update algorithm supplemented by 
over-relaxation,
and stored for later analysis. The choices of  
lattice parameters are reported in the first four
columns of Table~\ref{tab:setup_topology}. 
The value of the number of full lattice sweeps between 
successive measurements, denoted $N_\mathrm{sw}$, was chosen on the basis 
of preliminary technical runs that were used to determine the
auto-correlation time
$\tau_Q$; we tuned the choice of $N_\mathrm{sw}$ in the runs
so that $\tau_Q/N_\mathrm{sw} \lesssim 2$ for every value of $\beta$ and $N_c$. 
These choices ensured that  $N_{conf} \sim O(1000)$ topologically uncorrelated 
configurations be available for each ensemble.

\begin{figure}[t]
\centering
\includegraphics[scale=0.8]{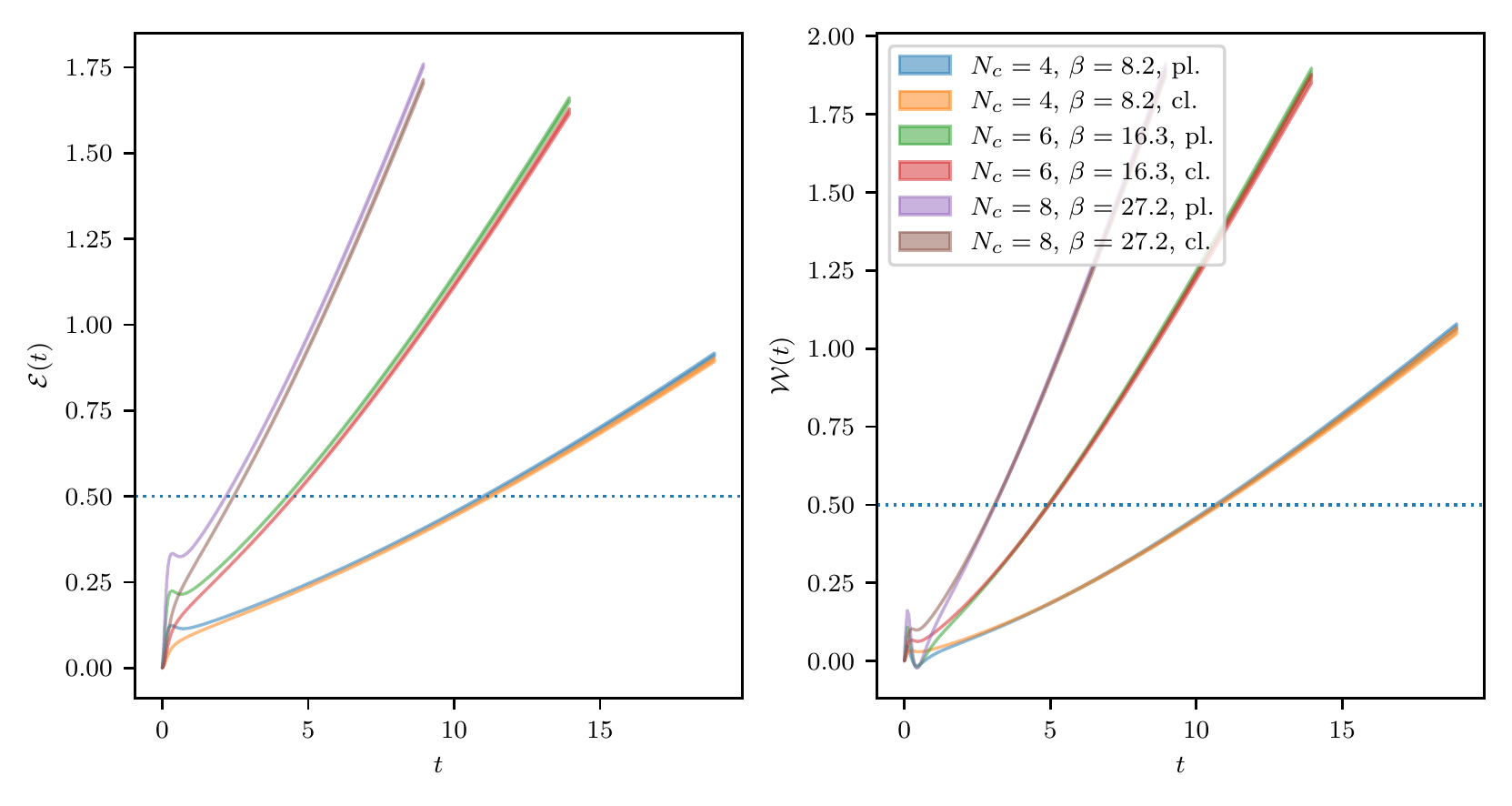}
\caption{The quantities $\mathcal{E}(t)$---defined in Eq.(\ref{eq:GF_scheme})---and
$\mathcal{W}(t) = t \frac{d}{dt} \mathcal{E}(t)$, computed from the 
plaquette and clover expressions,  
and obtained from the numerical integration of the Wilson flow equation. 
Different  colours denote different choices of
lattice parameters, as in Table~\ref{tab:setup_topology}). 
The dotted horizontal line denotes an illustrative choice 
reference value for $\mathcal{E}_0=\mathcal{W}_0$---not the one 
used for the measurements.\label{fig:NONscaled_flows}
}
\end{figure}

Each configuration is used as the initial condition of the Wilson-flow equation. 
Numerical integrations are performed with a 
third-order Runge-Kutta integrator~\cite{Luscher:2010iy}. 
The range of integration $[0,t_{\text{max}}]$ is chosen to avoid 
finite size effects, with $\sqrt{8 t_\text{max}}<L/(2)$, to obtain
 the field $V_{\mu}(t,\,x)$  for $t<t_\text{max}$.
$\mathcal{E}(t)$ and $\mathcal{W}(t)$ are 
computed for the plaquette and four-plaquettes clover expressions.
The results are displayed in Figure~\ref{fig:NONscaled_flows}---$E(t)$ 
in the left panel, $W(t)$ in the right one---for
three different ensembles, corresponding 
to the largest available value of $\beta$ for each of $N_c=4,\,6,\,8$. 
The curves naturally regroup in pairs, corresponds to either the plaquette 
(pl.) or four-plaquette clover (cl.) choice for $G_{\mu\nu}$. 
The difference between the two curves in a pair provides an estimate
of discretization effects. 
At  sufficiently large value of $t$, both for $\mathcal{E}(t)$ and
$\mathcal{W}(t)$, the difference between the two curves in each pair
is roughly a constant. For $\mathcal{W}(t)$,
the two curves appear to approach each other, 
confirming that $\mathcal{W}(t)$ is less affected by discretization 
effects than $\mathcal{E}(t)$.

\begin{figure}[t]
\centering
\includegraphics[scale=0.8]{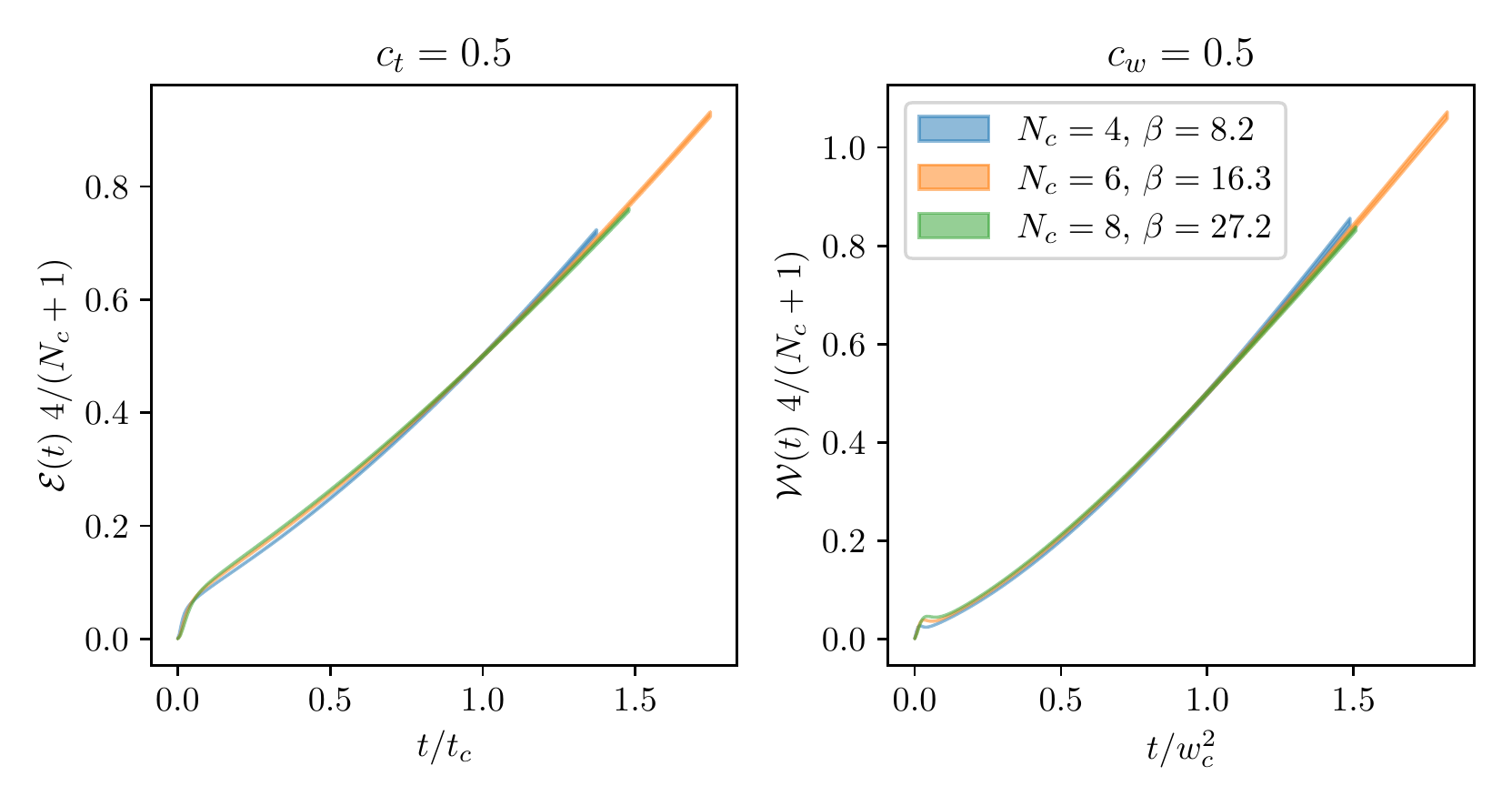}
\caption{
The quantities $\mathcal{E}(t)$---defined in Eq.(\ref{eq:GF_scheme})---and
$\mathcal{W}(t) = t \frac{d}{dt} \mathcal{E}(t)$, 
obtained from 
numerical integration of the Wilson flow with clover 
expressions for $E(t)$. 
Different  colours denote different choices of
lattice parameters, as in Table~\ref{tab:setup_topology}). 
With respect to Figure~\ref{fig:NONscaled_flows}, the axes have been rescaled 
in order to exhibit the scaling of the curves. 
The vertical axis has been divided by $\tfrac{N_c+1}{4}$, 
while the horizontal axis has been rescaled with $t_c$ 
and $w_c^2$, measured at the reference values 
of $\mathcal{E}_0$ and $\mathcal{W}_0$ obtained 
with 
 Eq.~(\ref{eq:scaling_e0w0}), for $c_w=c_e=0.5$.
\label{fig:rescaled_flows}
}
\end{figure}

To test the validity of the scaling in Eq.~(\ref{eq:scaling_e0w0}),
we set $c_e=c_w=0.5$ and compute the corresponding scales $t_c$ and $w_c$, 
defined in Eq.~(\ref{eq:scales}). 
We then divide $\mathcal{E}(t)$ and $\mathcal{W}(t)$  by
$C_2(F)$, scale the flow time with $t_c$ and $w^2_c$, 
and vary $N_c$. Figure~\ref{fig:rescaled_flows},
obtained with  the largest available values of $\beta$,
shows that the resulting curves, for $N_c=4,\,6,\,8$,
agree over a large interval of values of $t$.

In summary, we have studied
the behaviour of the Wilson flow for
$Sp(N_c)$ Yang-Mills theories, 
paying attention both to
$\mathcal{E}(t)$ and $\mathcal{W}(t)$. We
have identified an
interesting, approximate scaling law, suggesting that
the values of $t_0$ and $w_0^2$ 
obtained from Eq.~(\ref{eq:scales}) only 
depend on the 't Hooft coupling $\lambda(\mu)$ 
at scale $\mu=1/\sqrt{8t}$ and not on $N_c$.
We expect finite-$a$ and large-$N_c$ corrections to alter these (perturbative)
scaling properties at smaller $\beta$ values.

\begin{figure}[t]
\centering
\includegraphics[scale=0.9]{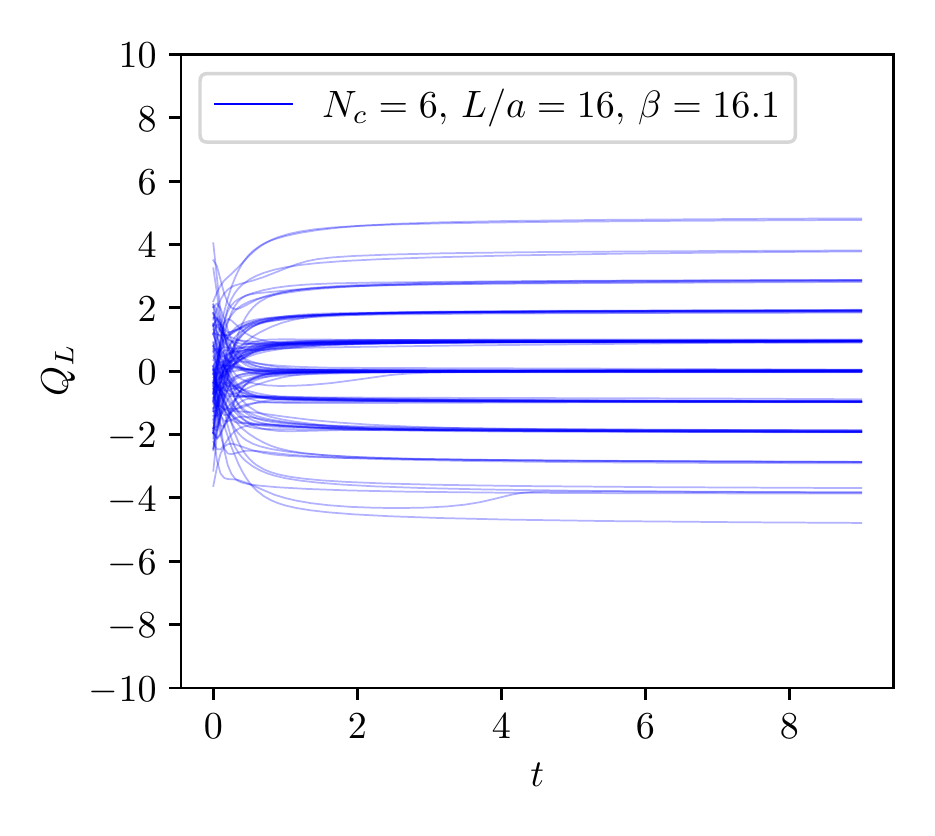}
\caption{The topological charge defined in
Eq.~(\ref{eq:lat_top_charge_TOT}) as a function of the flow time $t$.
Each curve corresponds to a different starting configuration in the 
ensemble with $N_c=6$, $\beta=16.1$, and
$L/a=16$. After an initial transient, the values of $Q_L(t)$
approach integer values. Other ensembles
yield comparable results.\label{fig:qL_vs_t}
}
\end{figure}

Following Ref.~\cite{DelDebbio:2002xa}, we redefine
\begin{equation}\label{eq:lat_top_charge_TOT}
    Q_L(t) \equiv \alpha \sum_x q_L(t,\,x)\,,
\eeq
where $\alpha$ is determined by numerical minimization of the quantity
\begin{equation}
\label{eq:alpha}
\Delta(\alpha)=\langle\left[ \alpha \tilde{Q}_L - \langle \alpha \tilde{Q}_L
\rangle\right]^2\rangle~.
\eeq
In practice, we find the values of $\alpha$ to be close to unity.
The behaviour of $Q_L(t)$ as a function of $t$ for $N_c=6$ at $\beta=16.1$ 
is reported in Figure~\ref{fig:qL_vs_t}. 
Other ensembles yield qualitatively similar results:
after an initial transient, the values of $Q_L(t)$ settle near
integer values. This is a residual effect of the fact that we are simulating 
a discretised system. 

In the following, we will use the topological charge obtained 
from Eq.~(\ref{eq:lat_top_charge}) at fixed flow-time $t=t_0$---defined 
in Eq.~(\ref{eq:scales})---and  corrected according to Eq.~(\ref{eq:alpha}). 
For different values of $N_c$, the reference values $\mathcal{E}_0$ 
and $\mathcal{W}_0$ were chosen in agreement with Eq.~(\ref{eq:pert_scaling}), 
with $c_e=c_w=0.5$. We will drop the dependence of $Q_L$ on the flow-time 
from now onwards.

In order to estimate the uncertainty,
and monitor the magnitude of topological freezing effects, 
we compute the integrated auto-correlation time $\tau_Q$ of the topological charge. 
As anticipated, the evaluation of $\tau_Q$ allows us 
to tune $N_\mathrm{sw}$ so that $\tau_Q/N_\mathrm{sw}\lesssim 2$. 
The calculation  can  be done in two 
independent ways: by binning the data and by
using Madras-Sokal windowing method~\cite{Madras:1988ei}. 
We found agreement in $\tau_Q$ obtained with the two different methods,
for all the available ensembles. The behaviour of $\tau_Q$ in
units of the number of full lattice sweeps is represented 
as a function of $\sqrt{8t_0}/a$ in Figure~\ref{fig:tauQ_vs_t0}, 
for different choices of $N_c$ and $\beta$. 
The behaviour of $\ln{\tau_Q}$ is approximately linear,
as in the case of $SU(N_c)$ gauge group~\cite{DelDebbio:2002xa, 
Athenodorou:2021qvs}. Moreover, the slope in $\ln{\tau_Q}$ grows 
as $N_c$ is increased. As $N_c$ is increased, we hence expect topological 
freezing to set in at smaller values of $\beta$,
making the continuum extrapolations progressively more computationally
expensive, as $N_c$ grows. 

\begin{figure}
\centering
\includegraphics[scale=1.1]{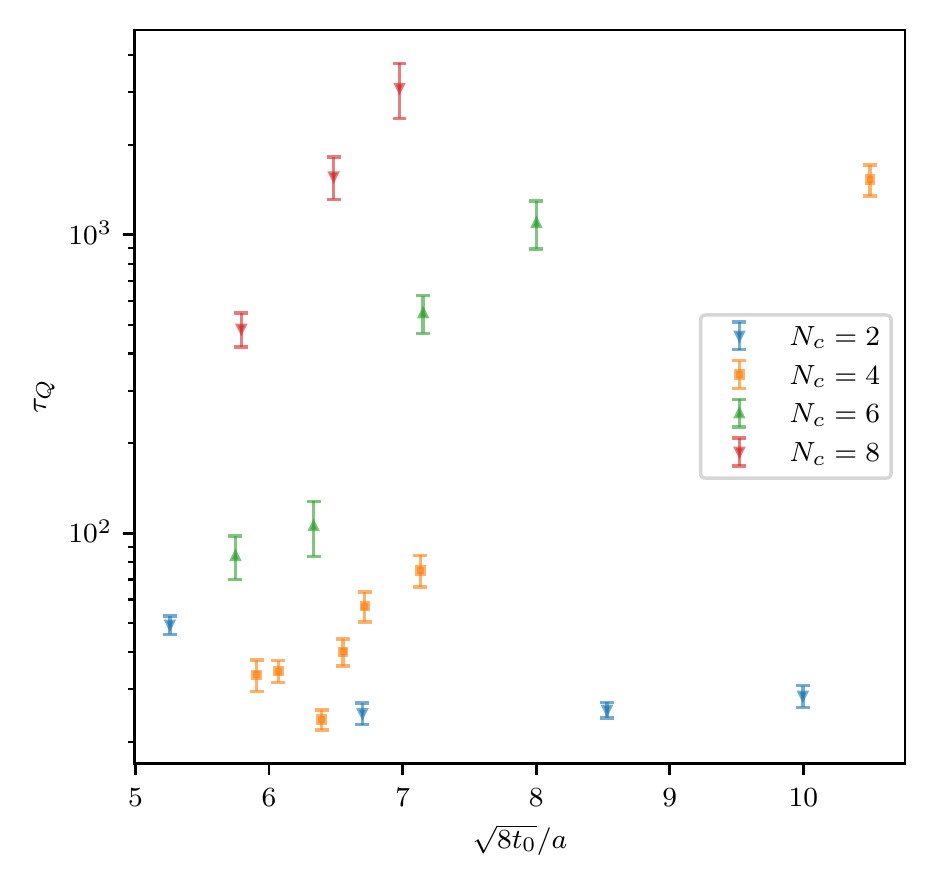}
\caption{Measurements of $\tau_Q$, obtained with the Sokal-Madras 
windowing algorithm, as a function of the inverse lattice spacing
$\sqrt{8t_0}/a$. Note the log-scale on the vertical axis. 
As expected from experience with $SU(N)$, $\tau_Q$ grows exponentially with
$\sqrt{8t_0}/a$. \label{fig:tauQ_vs_t0}
}
\end{figure}

Having tuned our choices of $N_\mathrm{sw}$,
we expect the trajectory of the simulations to sample ergodically 
the space of configurations, and hence
the distribution of $Q_L$ should be symmetric and centered about $Q_L=0$. 
This is confirmed but
the frequency histogram for $Q_L$  and the simulation history of 
the topological charge presented in Figure~\ref{fig:hist_top_charge}, for
the ensemble with $N_c=8$ and  $\beta=27.0$. 
Other ensembles yield similar behaviors.

\begin{figure}
\centering
\includegraphics[scale=0.8]{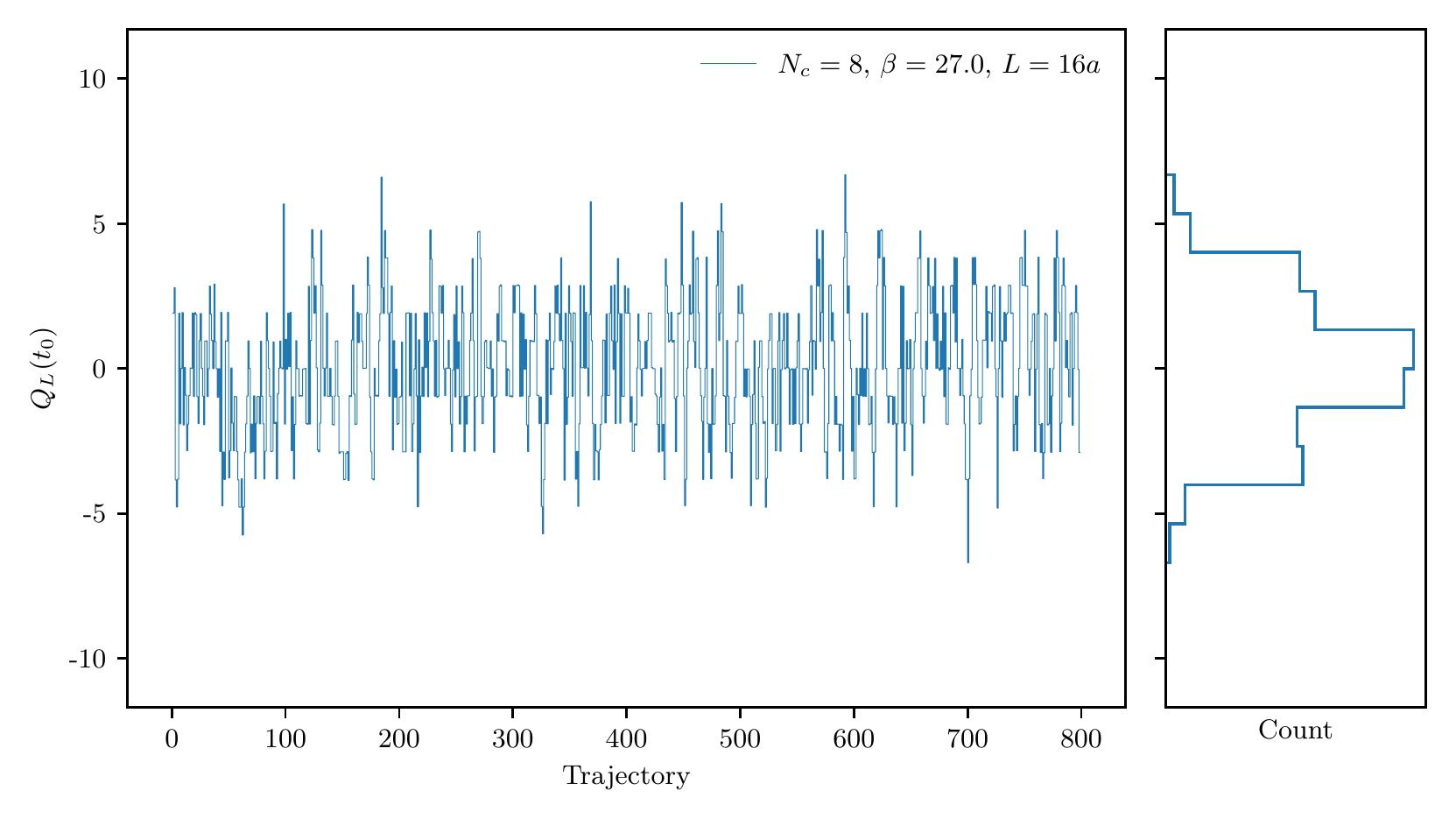}
\caption{Left panel: the simulation-time history 
of the topological charge $Q_L(t_0)$ computed from 
Eq.~(\ref{eq:lat_top_charge_TOT}). For these parameters, 
$\tau_Q=1.3(2)$. Right panel: the frequency histogram 
of $Q_L(t0)$. The latter appears to be symmetric and 
centered around $0$, as expected.\label{fig:hist_top_charge}
}
\end{figure}

The topological susceptibility is computed using 
Eq.~(\ref{eq:topo_suscept}) for $t=t_0$. 
It is represented in units of the $t_0$ scale, as a function
of $a^2/t_0$, in Figure~\ref{fig:top_contlim}. In the same plot we 
also show the same quantities computed 
for $SU(N_c)$ gauge theories with open boundary conditions, taken
from Ref.~\cite{Ce:2016awn}. 
The latter should be less affected by the effects of freezing, 
and as a result the statistical errors are expected to be much
smaller at a similar computational cost~\cite{Luscher:2011kk}. 
For each value of $N_c$, we obtain the 
continuum extrapolation of $\chi_L t_0^2$ by 
performing a linear fit with
\beq
\label{eq:top_contlim}
\chi_L  t_0^2 = t_0\chi_L(N_c,\, a=0) + b_t a^2/t_0\,,
\eeq
the rests of which are summarised in
Table~\ref{tab:topo_contlim},
and the large-$N_c$ extrapolation with
\beq
\label{eq:top_largeN}
\chi_L t_0^2 = t_0^2\chi_L(N_c = \infty,\, a=0 ) + b_N/N_c^2~.
\eeq
This simplified, preliminary analysis yields two 
different values, depending on whether we
include or not of the measurements for $N_c=2$ in the fitting range---see Table~\ref{tab:topo_largeN}.

\begin{table}[]
    \centering
    \caption{Result of the continuum extrapolation 
    at fixed $N_c$ for $N_c=2,\,4,\,6,\,8$. 
    The extrapolations are performed according to Eq.~(\ref{eq:top_contlim}). 
    The fit results are displayed in Figure~\ref{fig:top_contlim}.}
    \begin{tabular}{ccc}
        \hline
        $N_c$ & $\chi_L t_0^2$ & $\chi^2/N_\mathrm{d.o.f.}$ \\
        \hline
        $2$ & $0.002553(81)$ & $0.77$ \\
        $4$ & $0.00227(21)$ & $1.84$ \\
        $6$ & $0.00150(25)$ & $0.72$ \\
        $8$ & $0.00192(59)$ & $1.19$ \\
        \hline
    \end{tabular}
    \label{tab:topo_contlim}
\end{table}

\begin{figure}
\centering
\includegraphics[scale=1.1]{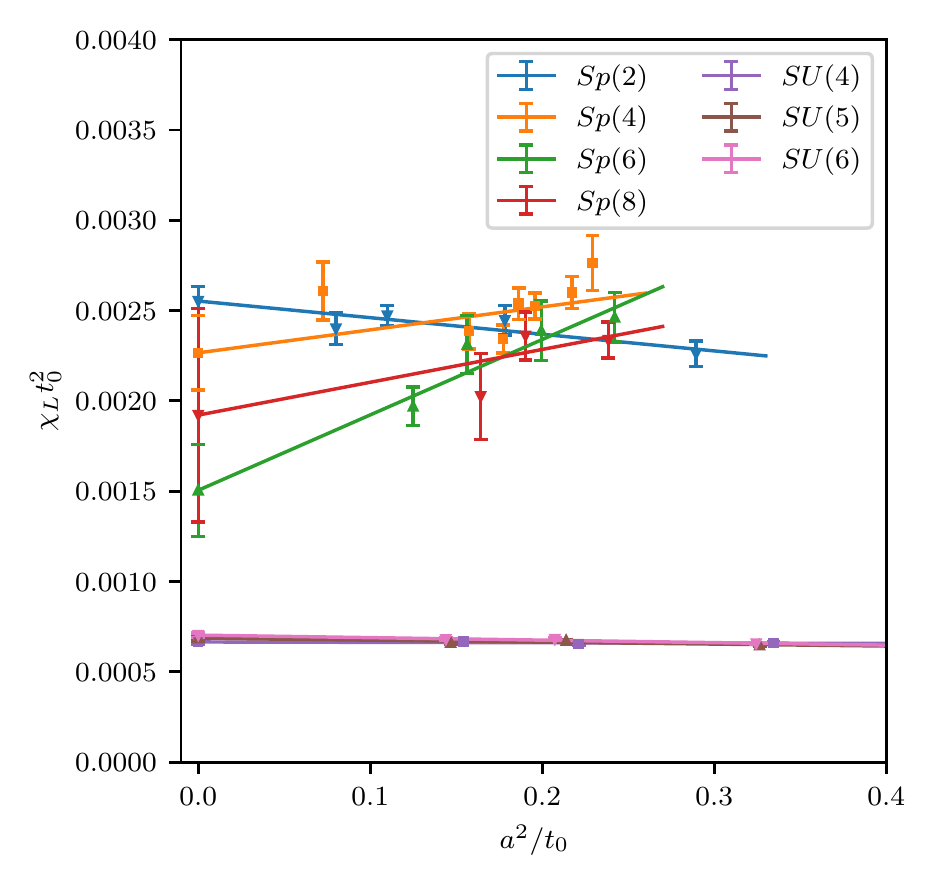}
\caption{The topological susceptibility $\chi_L t_0^2$, as a function of $a^2/t_0$, 
for $Sp(N_c)$ and $SU(N_c)$ gauge theories. 
The results for $SU(N_c)$ are taken from
Ref.~\cite{Ce:2016awn}.\label{fig:top_contlim}}
\end{figure}

\begin{figure}
\centering
\includegraphics[scale=1.1]{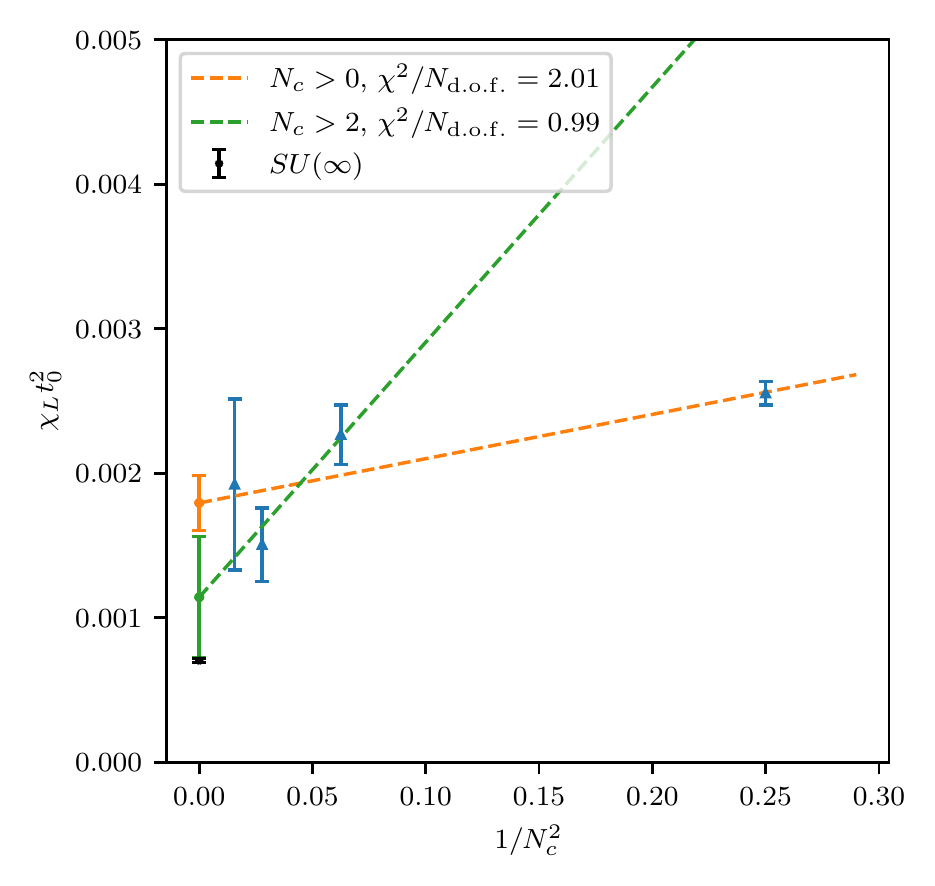}
\caption{Preliminary results for the continuum extrapolation
of the topological susceptibility $\chi_L t_0^2$ for $Sp(N_c)$ gauge theories
 as a function of $1/N_c^2$. 
 The results are extrapolated to $N_c\to\infty$ 
 using the ansatz $\chi_L t_0^2 = a+b/N_c^2$ for $N_c>2$ (green)
 or $N_c>0$ (orange). 
 The result of the extrapolation to $N_c=\infty$ for $SU(N_c)$ gauge theories, taken 
 from Ref.~\cite{Ce:2016awn}, 
 is represented as a black bullet.\label{fig:top_largeN}}
\end{figure}

The physics results are somewhat inconclusive, but 
in the near future we will further refine thee measurements,
the extrapolations to continuum and large-$N_c$ limit, and the
comparison between $Sp(N_c)$ and $SU(N_c)$ theories,
while the results presented in these pages 
can be considered as a proof of principle
of our processes.
In upcoming publication we will also compare with $SU(N_c)$ results taken
from the 
literature, expressed in units of the square-root of the string tension 
$\sqrt{\sigma}$, rather than in units of $t_0$.

\begin{table}[]
    \centering
    \caption{Results of the large-$N_c$ extrapolations for $Sp(N_c)$ theories
    over different fitting ranges, according to
    Eq.~(\ref{eq:top_largeN}). The results of the fits are displayed 
    in Figure~\ref{fig:top_largeN}.}
    \label{tab:my_label}
    \begin{tabular}{ccc}
    \hline
                 & $\chi_L t_0^2$ & $\chi^2/N_\mathrm{d.o.f.}$  \\
    \hline
    $N_c > 0$ & $0.00179(19)$ & $2.01$ \\
    $N_c > 2$ & $0.00114(42)$ & $0.99$ \\
    \hline
    \end{tabular}
    \label{tab:topo_largeN}
\end{table}


\section{The quenched meson spectrum}
\label{sec:quenchedmeson}
As a first step towards a systematic exploration of
phenomenologically interesting models,
which may require different choices of $Sp(2N)$  gauge group, 
we investigate how the mass spectrum of mesons
in the quenched approximation changes as we vary $N$.
We adopt the same methodology as in 
Ref.~\cite{Bennett:2019cxd}, 
except that
fermions transform in the fundamental (F), or 2-index
antisymmetric (AS) or symmetric (S) representations.
the dynamics is controlled
by the same action in \Eq{gauge_action}, 
and we define the massive Wilson Dirac 
operators for (quenched) fermions $\psi$ in the representation $R$
to be 
\beqs
D^{R} \psi(x) &\equiv&
\left(\frac{4}{a}+m_0^{\rm R}\right) \psi(x)
-\frac{1}{2a}\sum_\mu
\left\{\frac{}{}(1-\gamma_\mu)U^{R}_\mu(x)
\psi(x+\hat{\mu})+
\nonumber\right.\\
&&\left.+
(1+\gamma_\mu)U^{R\,\dagger}_\mu(x-\hat{\mu})
\psi(x-\hat{\mu})\frac{}{}\right\},
\label{eq:DiracF}
\eeqs
where $m_0^{R}$ is the bare fermion mass, $a$ 
the lattice spacing, and $\hat{\mu}$ a unit 
vector in direction $\mu$ on the lattice.
For the F representation we denote the  gauge link as
$U^{F}_\mu = U_\mu \in Sp(2N)$. 
For the 2-index representations 
we construct the link variables as 
follows~\cite{DelDebbio:2008zf}:
\beq
\left(U^{R}_\mu\right)_{(ij),\,(kl)}(x)=
\Tr\left[
(e_{R}^{(ij)})^\dagger U_\mu(x) e_{R}^{(kl)} U^T_\mu(x)\right],~~~{\rm with}~i<j,~k<l.
\label{eq:U_R}
\eeq
For $R=S$ and $R=AS$ the orthonormal basis $e_R$ 
is given by symmetric and antisymmetric 
$2N\times 2N$ matrices, respectively. 
For symplectic  groups, $e_{AS}$ 
further satisfies $\Tr\left[ \Omega\, e_{AS}\right] =0$,
with $\Omega$ the symplectic matrix.

We list in  Table~\ref{tab:MesonOps} the 
interpolating  operators used  to compute
the spectrum of mesons
with spin $J$ and parity $P$. 
They are built by combining  spinors $\psi$ and 
$\overline{\psi}\equiv\psi^{\dagger}\gamma^0$ 
with different Dirac structures.
The labels PS, S, V, AV, T, and AT stand for pseudoscalar, 
scalar, vector, axial-vector, tensor, and 
axial-tensor, respectively.
We display the corresponding QCD mesons to facilitate comparison.
\begin{table}[]
    \centering
    \caption{Interpolating operators $\mathcal{O}_S$ used to construct meson states, with their labels and 
    quantum assignments. The indices $i\neq j$ denote flavours, while colour and spin indexes are implicit. 
    The corresponding mesons in QCD are included for comparison.}
    \label{tab:MesonOps}
    \begin{tabular}{|c|c|c|c|}
    \hline
    Label & $\mathcal{O}_M$ & $J^P$ & QCD meson\\
    \hline\hline
    PS & $\overline{\psi^i}\gamma^5\psi^j$ & $0^-$ & $\pi$\\
    \hline
    S & $\overline{\psi^i}\psi^j$ & $0^+$ & $a_0$\\
    \hline
    V & $\overline{\psi^i}\gamma^{\mu}\psi^j$ & $1^-$ & $\rho$\\
    \hline
    AV & $\overline{\psi^i}\gamma^5\gamma^{\mu}\psi^j$ & $1^+$ & $a_1$\\
    \hline
    T & $\overline{\psi^i}\gamma^0\gamma^{\mu}\psi^j$ & $1^-$ & $\rho$ \\
    \hline
    AT & $\overline{\psi^i}\gamma^5\gamma^0\gamma^{\mu}\psi^j$ & $1^+$ & $b_1$\\
    \hline
    \end{tabular}
\end{table}

The physical observables---masses and decay constants---are extracted by examining the appropriate two-point correlation functions
between two distinct coordinates on the lattice. Writing the generic operator as 
$\mathcal{O}_M(x)=\overline{\psi^i}(x)\Gamma^M\psi^j(x)$, the correlators are defined as
\begin{equation}
C_{MM'}(x-y) = 
\langle\mathcal{O}_M(x)O^\dagger_{M'}(y)\rangle=
-\Tr[\Gamma^M D^{-1}_{R,i}(x-y)\Gamma^{M'} D^{-1}_{R,j}(y-x)],
\end{equation}
where $D^{-1}_{R,i}(x-y)$ is the inverse of the Dirac operator
from \Eq{DiracF}, for
fermion representation $R$ and flavor index $i$.   
The fermions transform in the same 
representation of the gauge group, to form a gauge singlet, but
we restrict our attention to flavoured mesons, with $i\neq j$.

We call $C_M(t)$  the spatially-averaged correlator  with $M=M'$, at asymptotically large time separation $t\to\infty$, 
which behaves as
\begin{equation}
    C_M(t)\to \frac{|\mel{0}{\mathcal{O}_M}{M}|^2}{2m_M}\left[e^{-m_Mt}+e^{-m_M(T-t)}\right]\,,
    \label{eq:corr_t}
\end{equation}
where $m_M$ is the mass of the meson in the ground state $\ket{M}$ and $T$ is the the temporal extent of the lattice. 
To extract the decay constants of the PS, V and AV states, 
we use the following parametrisations of the matrix elements for (partially) conserved currents:
\begin{eqnarray}
\mel{0}{\overline{\psi^i}\gamma^5\gamma^{\mu}\psi^j}{PS}&\equiv&\sqrt{2}f_{PS}p^{\mu},
\label{eq:av_ps}\\
\mel{0}{\overline{\psi^i}\gamma^{\mu}\psi^j}{V}&\equiv&\sqrt{2}f_{V}m_V\epsilon^{\mu},\\
\mel{0}{\overline{\psi^i}\gamma^5\gamma^{\mu}\psi^j}{AV}&\equiv&\sqrt{2}f_{AV}m_{AV}\epsilon^{\mu},
\end{eqnarray}
where the decay constant $f_{PS}$ is normalised so that it would yield $f_{\pi}\simeq 93\, \textrm{MeV}$ in QCD. 
$p^{\mu}$ and $\epsilon^{\mu}$ are the momentum and polarisation four-vectors,
 respectively. They obey $\epsilon^{\mu}p_{\mu}=0$ and $\epsilon^{\mu}\epsilon_{\mu}=1$. 
 These definitions, together with \Eq{corr_t}, suffice to extract all measurables from lattice results, except for 
$f_{PS}$. We further need one more correlation function:
\beq
C_{AV,PS}(t)=\sum_{\vec{x}} \langle 0 | \mathcal{O}_{AV}(\vec{x},t)\mathcal{O}^\dagger_{PS}(\vec{0},0)|0\rangle
\overset{t\to\infty}{\longrightarrow}
\frac{\sqrt{2}f_{PS}\mel{0}{\mathcal{O}_{PS}}{PS}^*}{2}\left[e^{-m_{PS}t}-e^{-m_{PS}(T-t)}\right].
\label{eq:c_av_ps}
\eeq
We extract $f_{PS}$ from the simultaneous fits to Eqs.~(\ref{eq:corr_t}) and~(\ref{eq:c_av_ps}).

The effective mass $am_{\rm eff}(t)$ is a function of Euclidean time $t$. 
We seek a late-time plateau in the effective mass and restrict our fits to a region in which 
$am_{\rm eff}(t)$ is approximately constant.
An example is shown in \Fig{meffplot} for the effective mass of vector meson V
 in the quenched $Sp(4)$ on a $48\times 24^3$ lattice, where the lattice parameters are $\beta=7.62$ and $am_0=-0.78$. The plateau starts  at $t=11$. In the figure we also show the result obtained from 
 curve-fitting, denoted by solid lines.
\begin{figure}
    \centering
    \includegraphics[width=90mm]{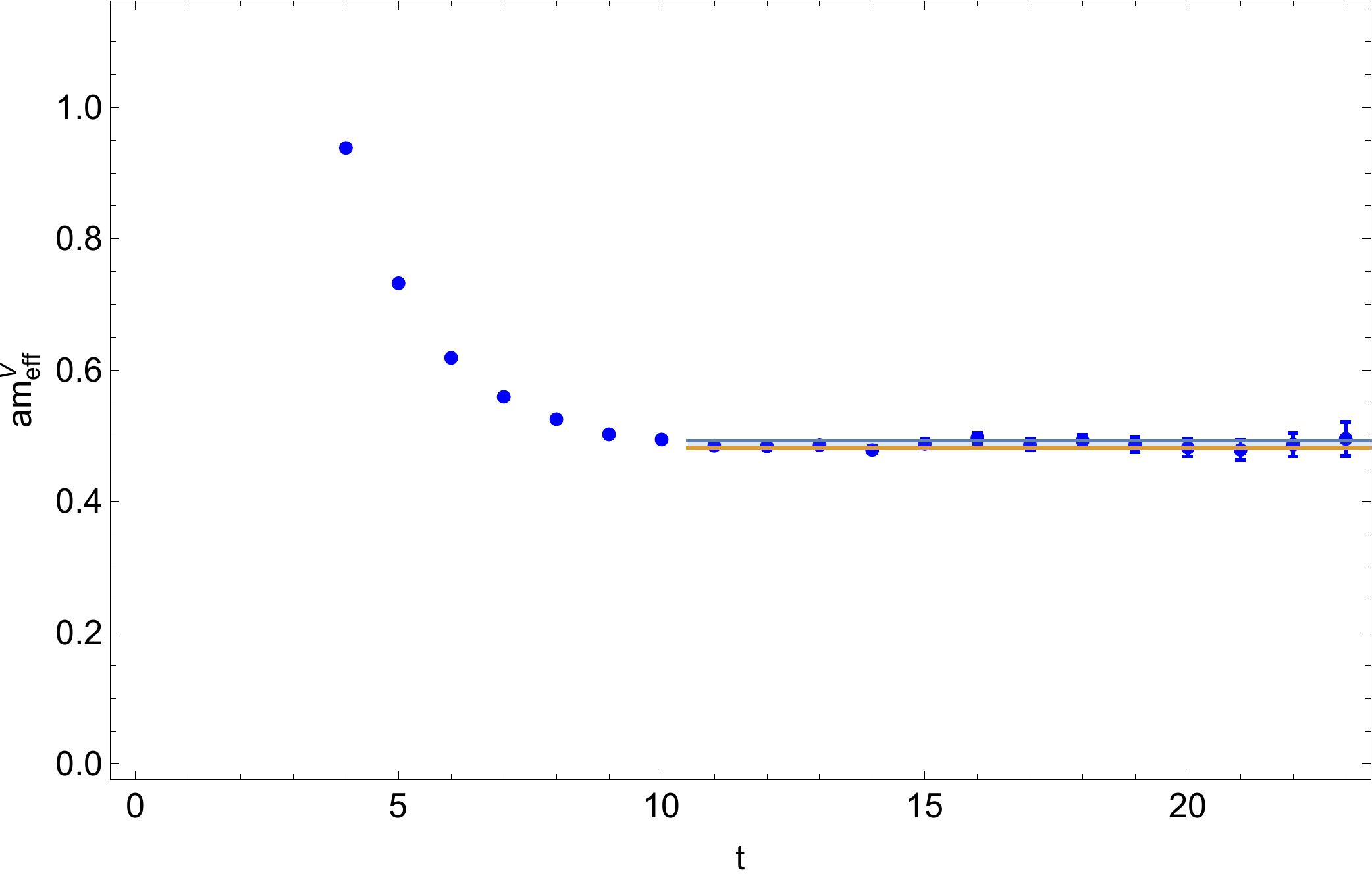}
    \caption{An example of effective mass plotted against Euclidean time for $Sp(4)$. The width of orange and blue lines are the selected time-interval and the vertical distance between them corresponds to the statistical error in the measurement.
This measurement uses lattice size $48\times 24^3$, coupling $\beta=7.62$, and bare Wilson mass $a m=-0.78$. 
the fit that has $\chi^2/N_{d.o.f}=0.84$.}
    \label{fig:meffplot}
\end{figure}

The decay-constants require renormalisation, even in the quenched approximation in which we neglect 
fermion loops. The process introduces multiplicative factors in each meson channel:
\begin{eqnarray}
f^{\rm ren}_{PS}=Z_Af^{\rm bare}_{PS},~
f^{\rm ren}_V=Z_Vf^{\rm bare}_V,~
f^{\rm ren}_{AV}=Z_Af^{\rm bare}_{AV}.
\end{eqnarray}
For $Sp(N_c)$, the two $Z$ quantities are calculated in lattice perturbation theory for Wilson 
fermions at the one-loop level~\cite{Martinelli:1982mw}, along with tadpole improvement~\cite{Lepage:1992xa}, and
 given by
\begin{eqnarray}
Z_V&=&1+C_2(R)(\Delta_{\Sigma_1}+\Delta_{\gamma_{\mu}})\frac{N_c}{8\pi^2\beta\langle P\rangle},\\
Z_A&=&1+C_2(R)(\Delta_{\Sigma_1}+\Delta_{\gamma_5\gamma_{\mu}})\frac{N_c}{8\pi^2\beta\langle P\rangle}.
\end{eqnarray}
The term $C_2(R)$ is the quadratic Casimir operator of representation $R$,
 $\beta$ is the inverse coupling used on the lattice,
  and $\langle P\rangle$ is the average plaquette. The $\Delta$ values are shown in table~\ref{tab:RenormalisationFactors}.
\begin{table}[h]
    \centering
    \begin{tabular}{|c|c|c|}
    \hline
    $\Delta_{\Sigma_1}$ & $\Delta_{\gamma_{\mu}}$ & $\Delta_{\gamma_5\gamma_{\mu}}$\\
    \hline
    -12.82 & -7.75 & -3.0\\
    \hline
    \end{tabular}
    \caption{The $\Delta$ values used in the renormalisation of decay constants.}
    \label{tab:RenormalisationFactors}
\end{table}

\begin{figure}
    \centering
    \includegraphics[width=100mm]{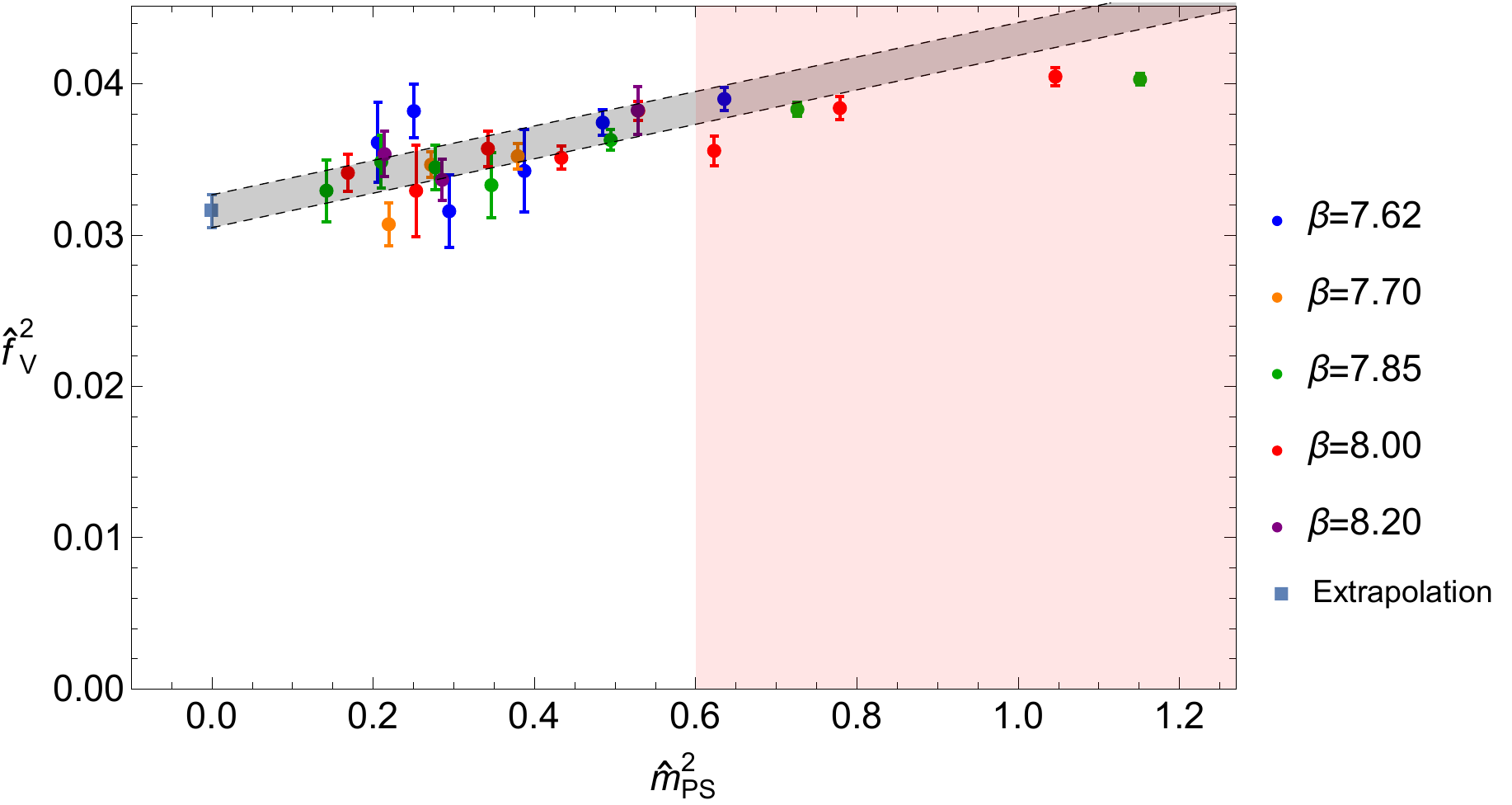}
    \caption{An example of  extrapolation to the chiral limit for the vector decay constant for fermions in the fundamental representation of $Sp(4)$. The reduced chi-squared is $\chi^2/N_{\rm dof}=1.20$. Data points in the pink shaded region are not included in the curve-fitting procedure. The grey band represents the continuum and massless extrapolation
    and the vertical width corresponds to the statistical error. }
    \label{fig:chiralExtpn}
\end{figure}

In order to extrapolate towards the continuum limit,
 we generate ensembles at several values of the lattice coupling $\beta$, 
and express all physically interesting quantities in units of the gradient flow scale $w_0$.
For each ensemble we consider several bare masses of the valence fermion, restricted to satisfy the following conditions:
 $am_M \lesssim 1$, $m_{PS} L \gtrsim 7.5$, and $f_{PS} L \gtrsim 1$.  Finite volume effects
 are negligible if the second such condition holds.
When the  first and third conditions are satisfied, an EFT  description exists in which, 
at the next-to-leading order (NLO), one finds
\begin{eqnarray}
\hat{m}^{2\rm, NLO}_{M}&=&\hat{m}^{2,\chi}_M(1+L^0_{m,M}\hat{m}^2_{PS})+W^0_{m,M}\hat{a},\\
\hat{f}^{2\rm, NLO}_{M}&=&\hat{f}^{2,\chi}_M(1+L^0_{f,M}\hat{m}^2_{PS})+W^0_{f,M}\hat{a}.
\end{eqnarray}
These NLO results display the dependence on the pseudoscalar mass $\hat{m}_{PS}$, which vanishes in the chiral limit. 
All quantities measured in terms of the gradient-flow are 
denoted by the caret ($\hat{\hspace{0.5em}}$) symbol, 
e.g. $\hat{f}=f w_0$. The observables in the chiral limit 
are denoted by the superscript $\chi$. 
$L^0$ and $W^0$ are free parameters to be determined by curve-fitting. 
An example  is shown in \Fig{chiralExtpn}.

\begin{figure}
    \centering
    \includegraphics[width=110mm]{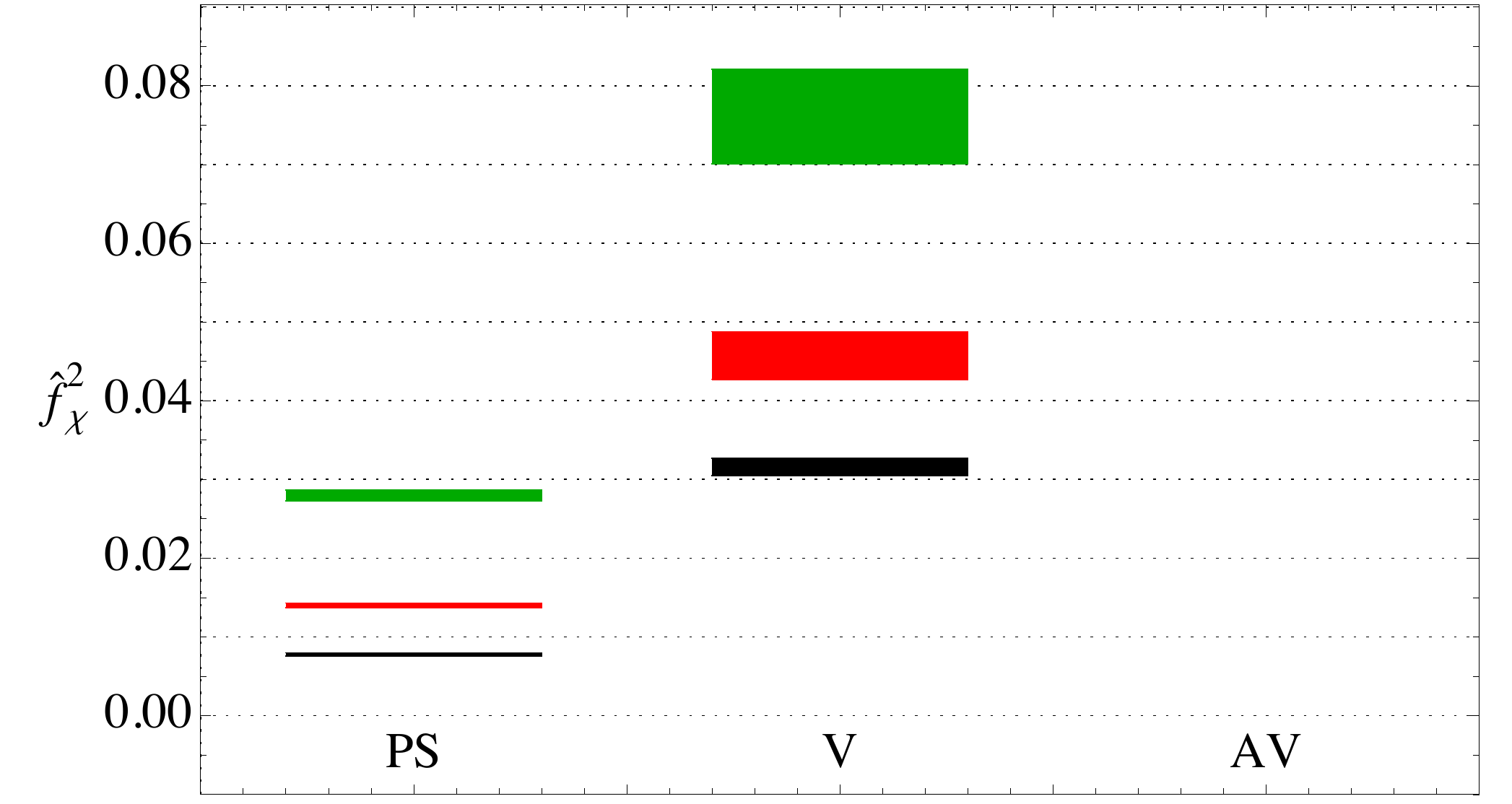}
    \caption{Chiral and continuum extrapolation of the decay constant of mesons with 
     fundamental (black), symmetric (green) and antisymmetric (red) fermions, expressed in units of gradient flow, in $Sp(4)$.}
    \label{fig:Sp4ChiralDecay}
\end{figure}
\begin{figure}
    \centering
    \includegraphics[width=110mm]{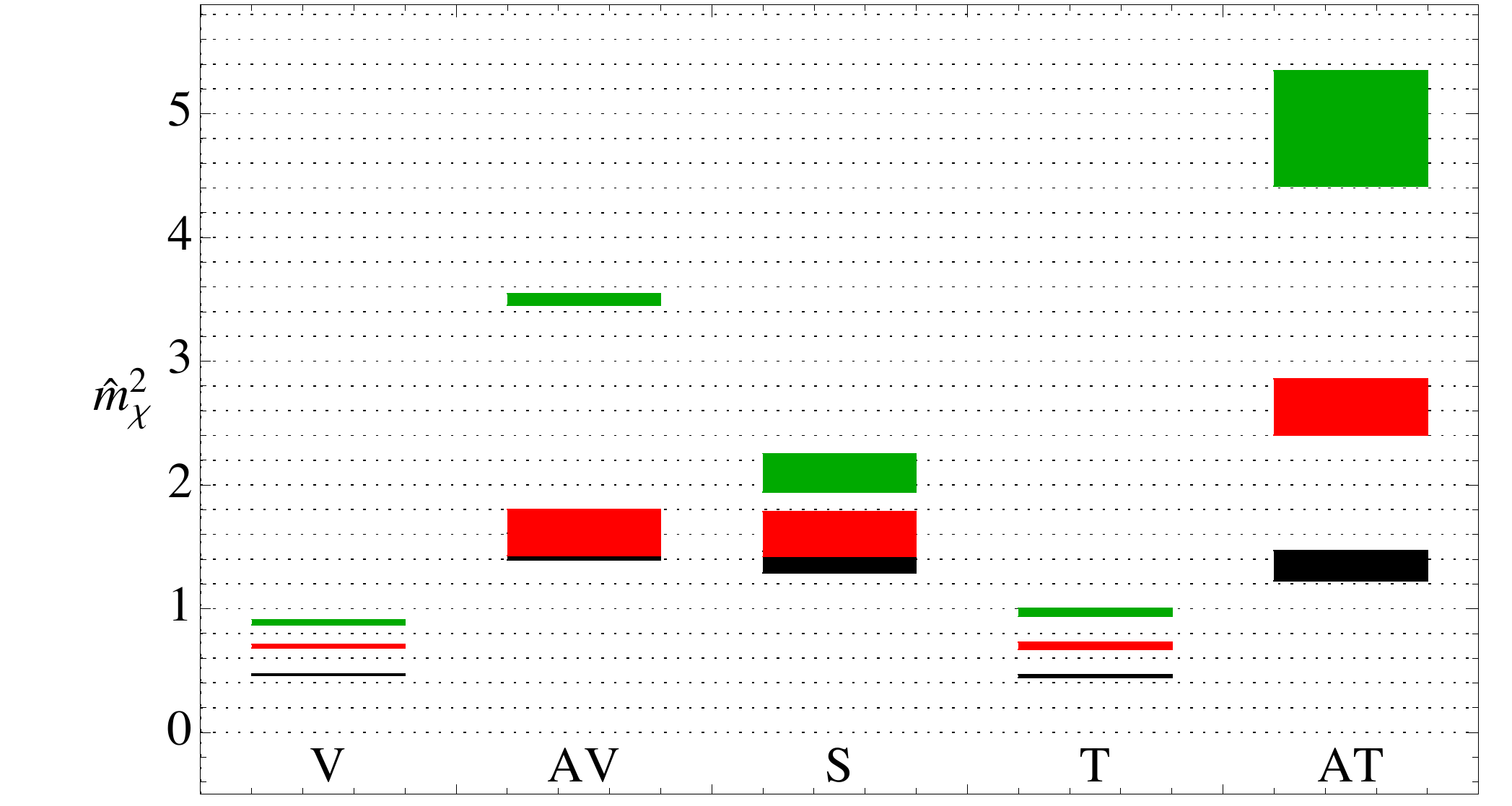}
    \caption{Chiral and continuum extrapolation of the mass squared of mesons with 
     fundamental (black), symmetric (green) and antisymmetric (red) fermions, expressed in units of gradient flow, in $Sp(4)$.}
    \label{fig:Sp4ChiralMasses}
\end{figure}
\begin{figure}
    \centering
    \includegraphics[width=110mm]{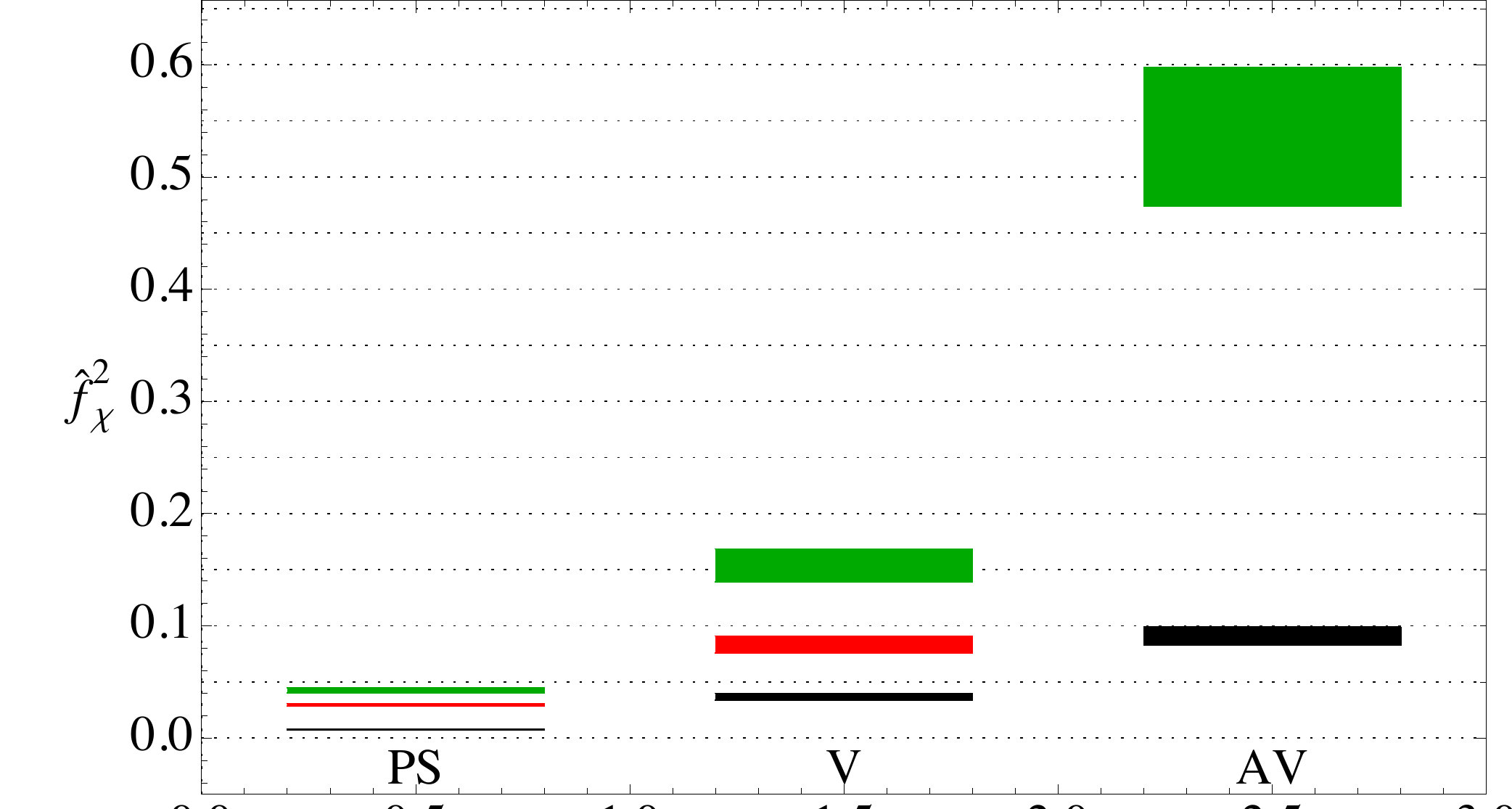}
    \caption{Chiral and continuum extrapolation of the decay constant of mesons with 
     fundamental (black), symmetric (green) and antisymmetric (red) fermions, expressed in units of gradient flow, in $Sp(6)$.}
    \label{fig:Sp6ChiralDecay}
\end{figure}
\begin{figure}
    \centering
    \includegraphics[width=110mm]{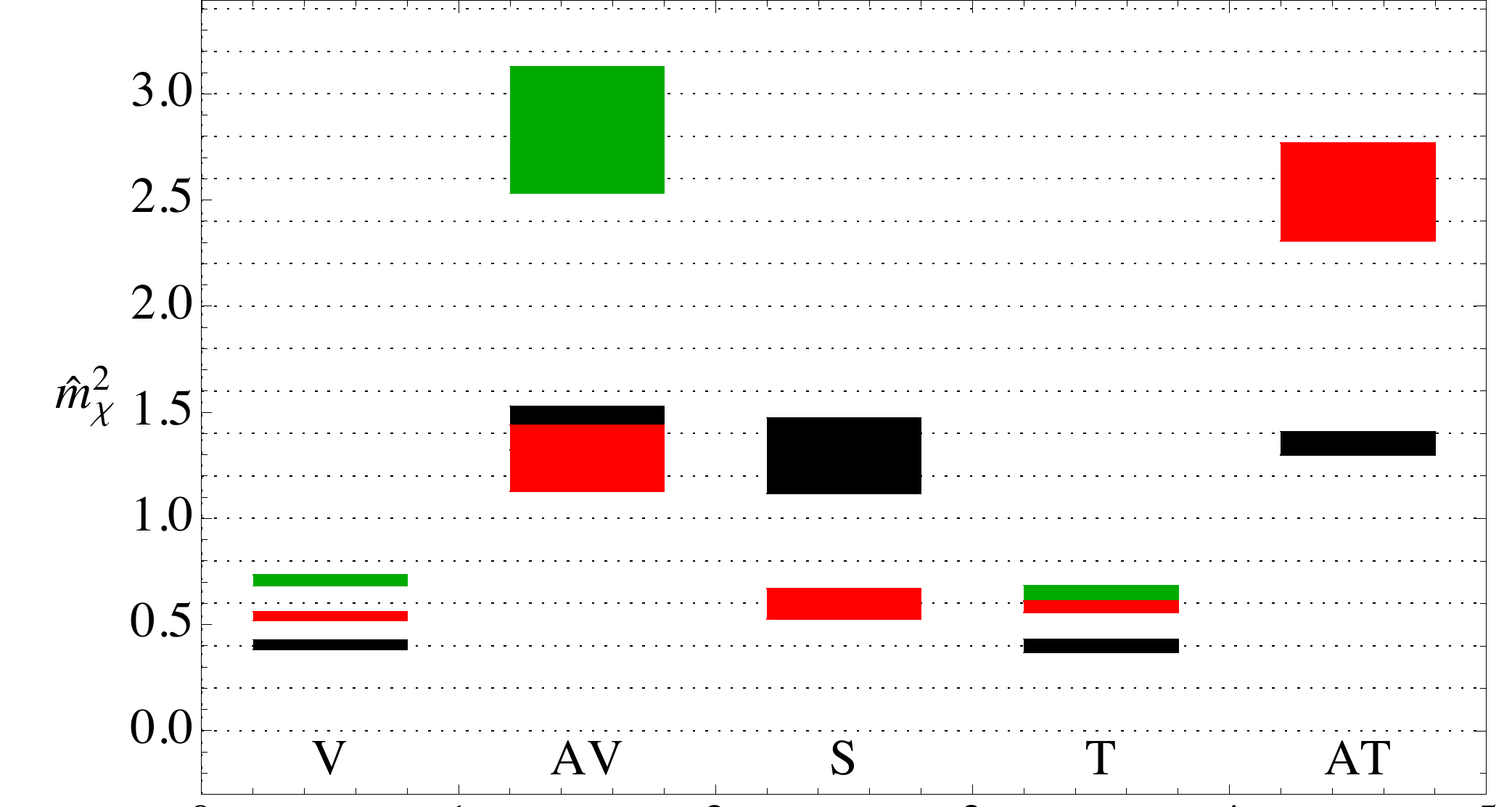}
    \caption{Chiral and continuum extrapolation of the mass squared of mesons with 
     fundamental (black), symmetric (green) and antisymmetric (red) fermions, expressed in units of gradient flow, in $Sp(6)$.}
    \label{fig:Sp6ChiralMasses}
\end{figure}

The results for meson masses and decay constants in the continuum and massless limit for $Sp(4)$ 
and $Sp(6)$ are shown
 in~\Cref{fig:Sp4ChiralDecay,fig:Sp4ChiralMasses,fig:Sp6ChiralDecay,fig:Sp6ChiralMasses}---the 
 width of the boxes represents the statistical uncertainties.
 We observe the emergence of a clear hierarchy between the three representations: masses and decay 
 constants tend to increase from fundamental to antisymmetric to symmetric representation. 
 In the cases of $Sp(4)$ with fundamental and antisymmetric valence fermions, the results are 
 comparable with our published results in Ref.~\cite{Bennett:2019cxd}. 
 Some of the measurements have been omitted from the plots;
  this is due to a poor signal in the extrapolation to the chiral limit, signaling an unreliable result. 
  We will provide these results in future publications, which will also
 report  the extrapolation to the large-$N$ limit.

\section{Fermions in mixed representations of $Sp(4)$}\label{sec:Dynamical}
\label{sec:dynamical_sp4}

In this section we return to the CHM candidate proposed in Ref.~\cite{Barnard:2013zea}.
It is a $Sp(2N)$ gauge theory with  $N_f=2$ Dirac fermions transforming according to the fundamental representation, 
and  $n_f=3$ on the 2-index antisymmetric representation. 
For $N>1$, mesons made of fundamental representation fermions 
yield the PNGBs needed for Higgs compositeness, while (partial) top compositeness arises because of the existence of 
exotic fermionic matter (chimera baryons) composed of two fundamental and one antisymmetric fermion.
From now on,  we restrict attention to the $Sp(4)$ gauge theory.
We discuss some progress we made on the lattice numerical treatment of the theory, and we report
 preliminary results
 for the mesons with dynamical antisymmetric fermions, for chimera baryons in the 
partially quenched theory (in which only the antisymmetric fermions are dynamical) and
for the bulk transitions in the fully dynamical theory with mixed representations.

\subsection{The lattice theory}

The lattice action we adopt contains both fundamental (F) and antisymmetric (AS) fermions:  
\beq
S\equiv S_W+
a^4\sum_{i=1}^{2}\sum_x \overline{Q}_i(x) D^{F} Q_i(x)+a^4\sum_{i=1}^3\sum_x \overline{\Psi}_i(x) D^{AS} \Psi_i(x),
\label{eq:action}
\eeq
where $S_W$ is the Wilson gauge action in \Eq{gauge_action} 
and $D^{R}$ is the massive Wilson Dirac operator for fermions in the representation $R$ defined in \Eq{DiracF}.
We impose periodic boundary conditions in all directions 
for the gauge fields. For the Dirac fields we 
consider periodic and antiperiodic boundary 
conditions for the spatial and temporal extents, respectively.  Simulations were carried out using the HiRep
code~\cite{DelDebbio:2008zf} 
with the standard rational hybrid Monte Carlo (RHMC) 
algorithms, where several advanced techniques such as even-odd preconditioning can also be accessed. 
The implementation of $Sp(2N)$ gauge 
theories with fermions in the fundamental representation is described in Ref.~\cite{Bennett:2017kga}.
Details about the antisymmetric representation 
and for the mixed representation will be discussed 
in upcoming publications

\subsection{Towards composite Higgs and partial top
compositeness}

With field content comprising $N_f=2$ fundamental and 
$n_f=3$ 2-index antisymmetric fermions,
the mass terms break the enhanced global symmetry $SU(4)\times SU(6)$ to the subgroups $Sp(4)\times SO(6)$, 
which is also the subgroup preserved in the vacuum, in the presence of fermion bilinear condensates.
The unbroken generators leave invariant the symplectic matrix $\Omega$ and the symmetric matrix $\omega$, 
which we write as follows:
\beq
\Omega=\Omega_{jk}\,=\,\Omega^{jk}\,\equiv\,
\left(\begin{array}{cccc}
0 & 0 & 1 & 0\cr
0 & 0 & 0 & 1\cr
-1 & 0 & 0 & 0\cr
0 & -1 & 0 & 0\cr
\end{array}\right)\,,
\label{Eq:symplectic}
~~~~
\omega\,=\,\omega_{jk}\,=\,\omega^{jk}\,\equiv\,
\left(\begin{array}{cccccc}
0 & 0 & 0 & 1 & 0 & 0 \cr
0 & 0 & 0 & 0 & 1 & 0 \cr
0 & 0 & 0 & 0 & 0 & 1  \cr
1 & 0 & 0 & 0 & 0 & 0 \cr
0 & 1 & 0 & 0 & 0 & 0 \cr
0 & 0 & 1 & 0 & 0 & 0 \cr
\end{array}\right)\,.
\eeq

We decompose the antisymmetric 
 representation of $Sp(4)$ as $5=4+1$ of $SO(4)$, and we identify the gauged $SU(2)_L$
 of the SM, and its approximate $SU(2)_R$, as $SU(2)_L\times SU(2)_R \sim SO(4) \subset Sp(4)$. 
By doing so, the complex scalar doublet of $SU(2)_L$ is identified with the 4 of $SO(4)$. The 
corresponding operators are flavoured mesons (and diquarks) built of fundamental fermions. 
 
 The chimera baryons that have the same quantum 
 numbers of the top quark 
can mix with it, providing an origin for its mass 
(partial top compositeness). 
 Such quantum numbers can be obtained by combining the $4$ of $SO(4)$ with the antisymmetric fermions $\Psi^k$.
 Since the $SU(3)\times SU(3)$ subgroup of $SU(6)$ breaks into $SU(3)_V$, it is natural to identify
 the QCD gauge group as $SU(3)_c \sim SU(3)_V \subset SO(6)$, and write the following 
fermion operators: 
\beqs
{\mathcal{O}}_{{\rm CB}, 1}\nonumber
 &=&
 \left(\overline{Q^{1\,a}} \gamma^5 Q^{2\,b}+\overline{Q^{2\,a}} \gamma^5 Q^{1\,b}\right) \Omega_{bc} \Psi^{k\,ca}\,,\\
{\mathcal{O}}_{{\rm CB}, 2}&=&
\left(-i\overline{Q^{1\,a}} \gamma^5 Q^{2\,b}+i\,\overline{Q^{2\,a}} \gamma^5 Q^{1\,b}\right)\Omega_{bc} \Psi^{k\,ca}\,,\\
 {\mathcal{O}}_{{\rm CB}, 4}&=&\nonumber
 -i\,\left(\overline{Q^{1\,a}}  Q^{2\,b}_{\,C}+\overline{Q_C^{2\,a}}  Q^{1\,b}\right)\Omega_{bc}\Psi^{k\,ca}\,,\\
 {\mathcal{O}}_{{\rm CB}, 5}&=&\nonumber
 i\,\left(-i\,\overline{Q^{1\,a}}  Q^{2\,b}_C+i\overline{Q^{2\,a}_C} Q^{1\,b}\right)\Omega_{bc}\Psi^{k\,ca}\,.
\eeqs
The $U(1)_A$ partners can be obtained by replacing $\mathbb{1}_4 \rightarrow i\gamma^5$ in the fundamental bilinears. For completeness, the $SO(4)$ singlet that combines with the $4$ into  the $5$ of $SO(5)$ is
\beqs
{\mathcal{O}}_{{\rm CB}, 3}\nonumber
 &=&
 \left(\overline{Q^{1\,a}} \gamma^5 Q^{1\,b}-\overline{Q^{2\,a}} \gamma^5 Q^{2\,b}\right) \Omega_{bc} \Psi^{k\,ca}\,,
 \eeqs
and plays no role in top compositeness.

To construct the interpolating operators on the lattice, we start from  the most generic 
 gauge-invariant fermion composite of two fundamental fermions and one antisymmetric fermion:
\beq
\mathcal{O}_{\rm CB}^{\alpha}(x) = D^{\alpha\beta\gamma\delta} \Omega_{ac} \Omega_{bd} Q^{i\,a}_\beta(x) Q^{j\,b}_\gamma(x) \Psi^{k\, cd}_\delta(x).
\label{chimera_ops_generic}
\eeq
We use Greek letters to denote for the spin indices, to distinguish them from
the colour and flavour indices. 
The simplest operator interpolating the spin-$1/2$ ground  state would be analogous to the 
$\Lambda$-baryon in QCD, which is  
$D^{\alpha\beta\gamma\delta}=-i(\gamma^5 C)^{\beta\gamma}\delta^{\alpha\delta}$ 
with $C=i\gamma^2\gamma^0$ being the charge conjugation matrix.\footnote{
The other common choice of the Dirac structure would be $D^{\alpha\beta\gamma\delta}=-i(C)^{\beta\gamma}(\gamma^5)^{\alpha\delta}$.
} 
The $4$ of $SO(4)$ and $3$ of $SU(3)_V$ are identified by appropriate choices of $i,\,j=1,\,2$ with $i \neq j$, 
and $k=1,\,2,\,3$. We perform the parity projection in the nonrelativistic limit,
\beq
\mathcal{O}_{\rm CB}^{\pm}(x) \equiv \mathcal{P}_{\pm} \mathcal{O}_{\rm CB}(x),~~\textrm{with} ~\mathcal{P}_{\pm}=\frac{1}{2}\left(
1\pm \gamma_0
\right).
\label{eq:chimera_ops}
\eeq

Using the interpolating operator in \Eq{chimera_ops} and its conjugate, we construct the propagator for the Chimera baryon at positive Euclidean time $t$ and vanishing momentum $\vec{p}$
\beqs
\langle \mathcal{O}_{\rm CB}^{\pm}(t) \overline{\mathcal{O}_{\rm CB}^{\pm}(0)} \rangle
&=& \sum_{\vec{x}}
\Omega_{da} \Omega_{bc} \Omega^{c'b'} \Omega^{a'd'}
S_{\Psi}(t,\vec{x})^{ca,c'a'}_{\alpha,\alpha'}(t,\vec{x})\,
 \mathcal{P}_{\pm}^{\alpha',\alpha}\nn \\
&&\times\,
S_{Q}^j(t,\vec{x})^{d,b'}_{\gamma,\gamma'} (t,\vec{x}) (C\gamma^5)_{\gamma\beta}
S_Q^i(t,\vec{x})^{b,d'}_{\beta,\beta'} (t,\vec{x}) (C\gamma^5)_{\gamma'\beta'},
\label{Eq:cb_dq}
\eeqs
where the fermion propagators are
\beq
S_Q(t,\vec{x})^{a,b}_{\alpha,\beta} = \langle Q(t,\vec{x})^{a}_{\alpha} \overline{Q(0)^{b}}_{\beta} \rangle
~{\textrm{and}}~S_\Psi(t,\vec{x})^{ab,cd}_{\alpha,\beta} = \langle \Psi(t,\vec{x})^{ab}_{\alpha} \overline{\Psi(0)^{cd}}_{\beta} \rangle.
\eeq

\subsection{Mesons with $n_f=3$ dynamical antisymmetric Dirac fermions}
\label{sec:as}

We turn now our attention to the spectroscopy 
study of the $Sp(4)$ gauge theory 
with $n_f=3$ dynamical Dirac fermions on the 
antisymmetric representation. 
We measure the masses of the ground-state mesons using the operators listed in table~\ref{tab:MesonOps}.  
For the meson analogous to the $\rho$ in QCD, 
the physical states have overlap with two different operators ${\cal O}_V$
and ${\cal O}_T$, hence we also  implement the generalised eigenvalue problem~\cite{Blossier:2009kd},
to extract the mass of the excited state (analogous to the $\rho^{\prime}$ in QCD).
In our preliminary studies of the bare parameter 
space via the mass scan~\cite{Lee:2018ztv}, 
we found a first-order bulk phase transition 
at strong coupling and determined the lower bound 
in the lattice coupling for the weak coupling regime 
as $\beta \gsim 6.6$.

We start by studying  finite size effects. In Fig~\ref{fig:SP4AS3_finite_size}, we present the pseudoscalar mass in lattice units measured at six different lattice volumes with isotropic spatial extents $L/a=8,12,16,18,20,24$ 
and common temporal extent $T/a=54$. We find that the mass of pseudoscalar mesons converges
for  $m_{\text{ps}} L \gsim 8$, indicating that finite volume effects are  negligible for $m_{\rm{ps}}\gtrsim 8$. All the measurements reported in the following satisfy this condition. 
\begin{figure}
\centering
\includegraphics[scale=0.4]{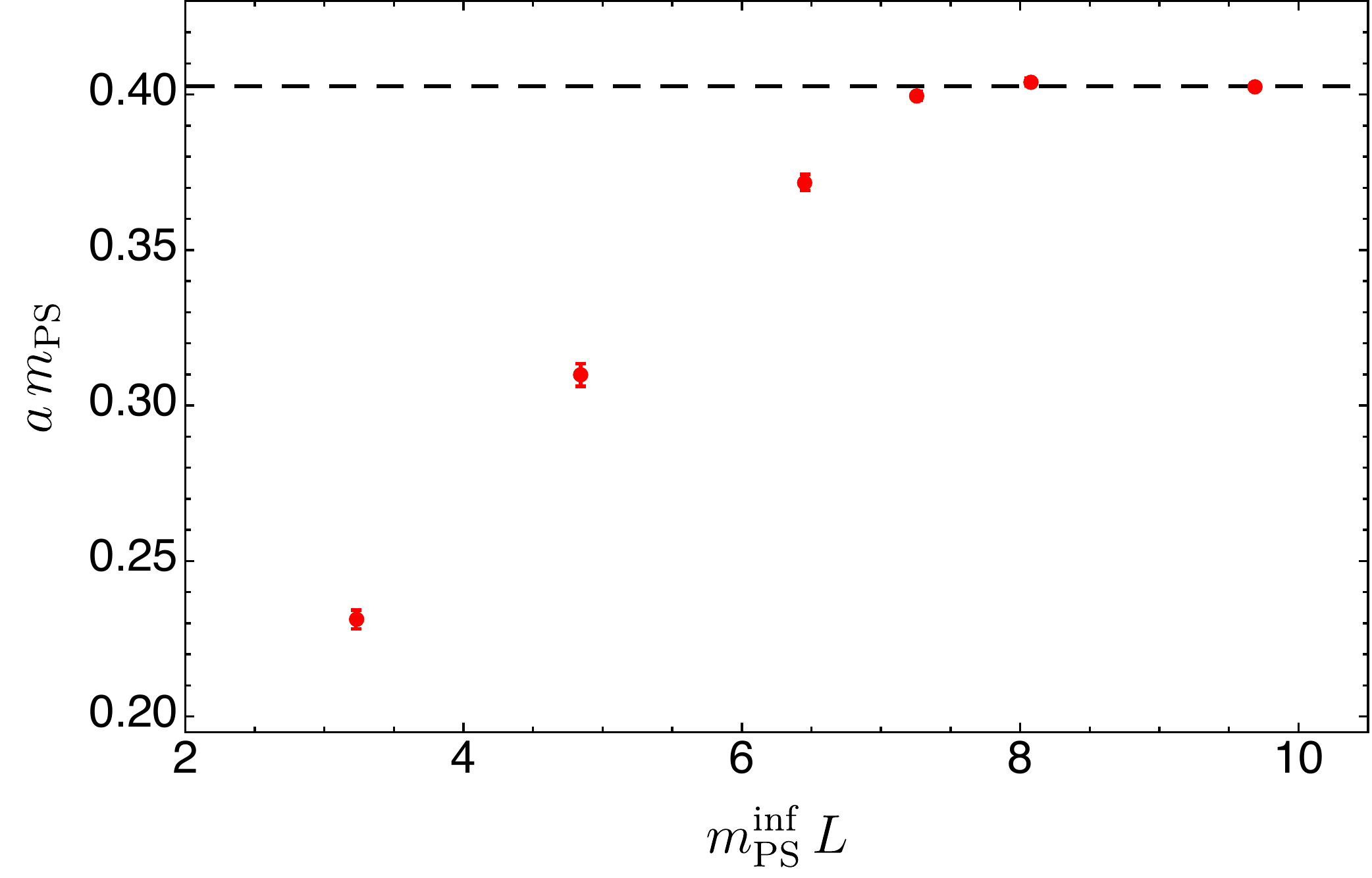}
\caption{Pseudoscalar meson mass  $am_{\text{PS}}$ (in lattice units) versus  $m_{\text{PS}} L$, where $L$ is the 
extension of a spatial direction, for varying $L/a$ and fixed $T/a=54$,
with $\beta = 6.8$ and $m_0 = -1.03$. The horizontal line is
  the measurement at the largest volume of $54\times 24^3$, taken to represent the infinite volume extrapolation.}
\label{fig:SP4AS3_finite_size}
\end{figure}

Finite volume effect results in a negative contribution
 to the pseudoscalar mass, which is the opposite to the typical lattice QCD results. 
 In fact, such a distinct  behavior can be understood in the context of chiral perturbation theory ($\chi$PT). 
Finite volume corrections are accounted for by loops  wrapping around each spatial extent of the lattice, 
and the resulting summations at finite volume
modify the  pseudoscalar mass.  At the next-to-leading order (NLO), 
one has
\beq
m_{\rm PS}^2= M^2 \left(
1 + a_M \frac{A(M)+A_{\rm FV}(M)}{F^2} + b_M(\mu) \frac{M^2}{F^2} + \mathcal{O}(M^4)
\right),
\label{eq:finite_nlo_mps}
\eeq
with 
\beq
A(M)=-\frac{M^2}{16 \pi^2} {\rm log}\frac{M^2}{\mu^2}
~\textrm{and}~
A_{\rm FV}(M) \overset{ML\gg 1}{\longrightarrow} -\frac{3}{4\pi^2}
\left(\frac{M\pi}{2L^3}\right)^{1/2}{\rm exp}[-ML],
\label{eq:chiral_logs}
\eeq
where $M$ and $F$ are the pseudoscalar mass and decay constant (at leading order)
 and $\mu$ is the renormalisation scale. 
The result $A_{\rm FV}(M)$ for the finite sum  has been obtained from a cubic 
box of size $L$ with periodic boundary condition (see the Appendix of Ref.~\cite{Arndt:2004bg}),
and is independent
 of the details of the theory. But, the coefficient $a_M$ reflects the symmetry
  breaking pattern~\cite{Bijnens:2009qm}:
\beqs
a_M = \begin{cases}
\frac{1}{2}-\frac{1}{2n_f},~&{\rm for}~SU(2N_f)\rightarrow SO(2N_f), \\
-\frac{1}{n_f},~&{\rm for}~SU(N_f)\times SU(N_f)\rightarrow SU(N_f), \\
-\frac{1}{2}-\frac{1}{n_f},~&{\rm for}~SU(2N_f) \rightarrow Sp(2N_f).
\end{cases}
\label{eq:coeff_a}
\eeqs
The theory with $n_f=3$ belongs to the first class,
and  has different sign compared to the other two cases.
The analytical results in Eqs.~(\ref{eq:finite_nlo_mps}), (\ref{eq:chiral_logs}) and (\ref{eq:coeff_a}) are hence
 consistent with the negative corrections displayed in~\Fig{SP4AS3_finite_size}.

\begin{figure}

\centering
     \begin{subfigure}[b]{0.49\textwidth}
         \centering
            \includegraphics[width=\textwidth]{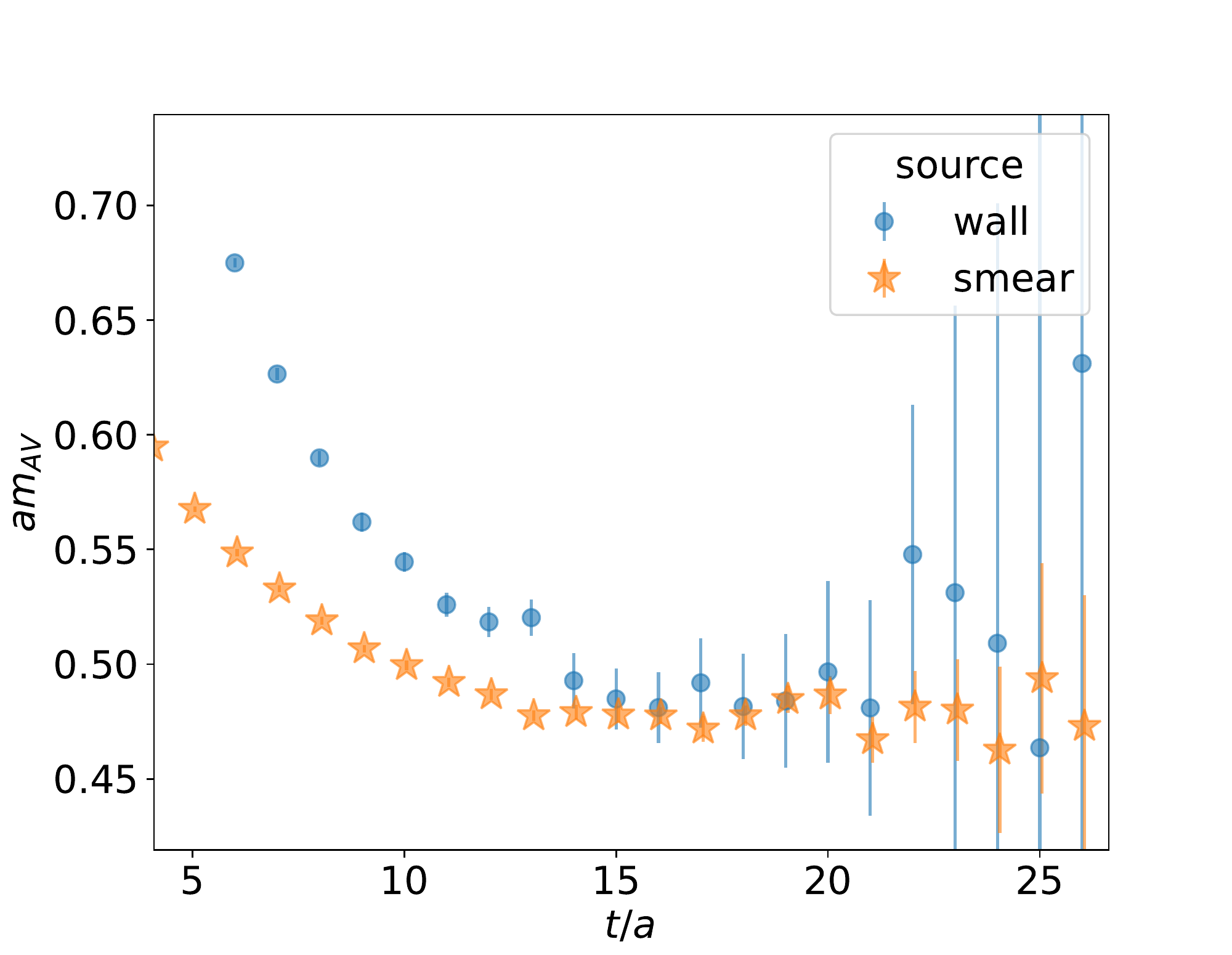}
            \caption{} \label{fig:SP4AS3_smear_v_wall}
     \end{subfigure}
     \hfill
     \begin{subfigure}[b]{0.49\textwidth}
         \centering
         \includegraphics[width=\textwidth]{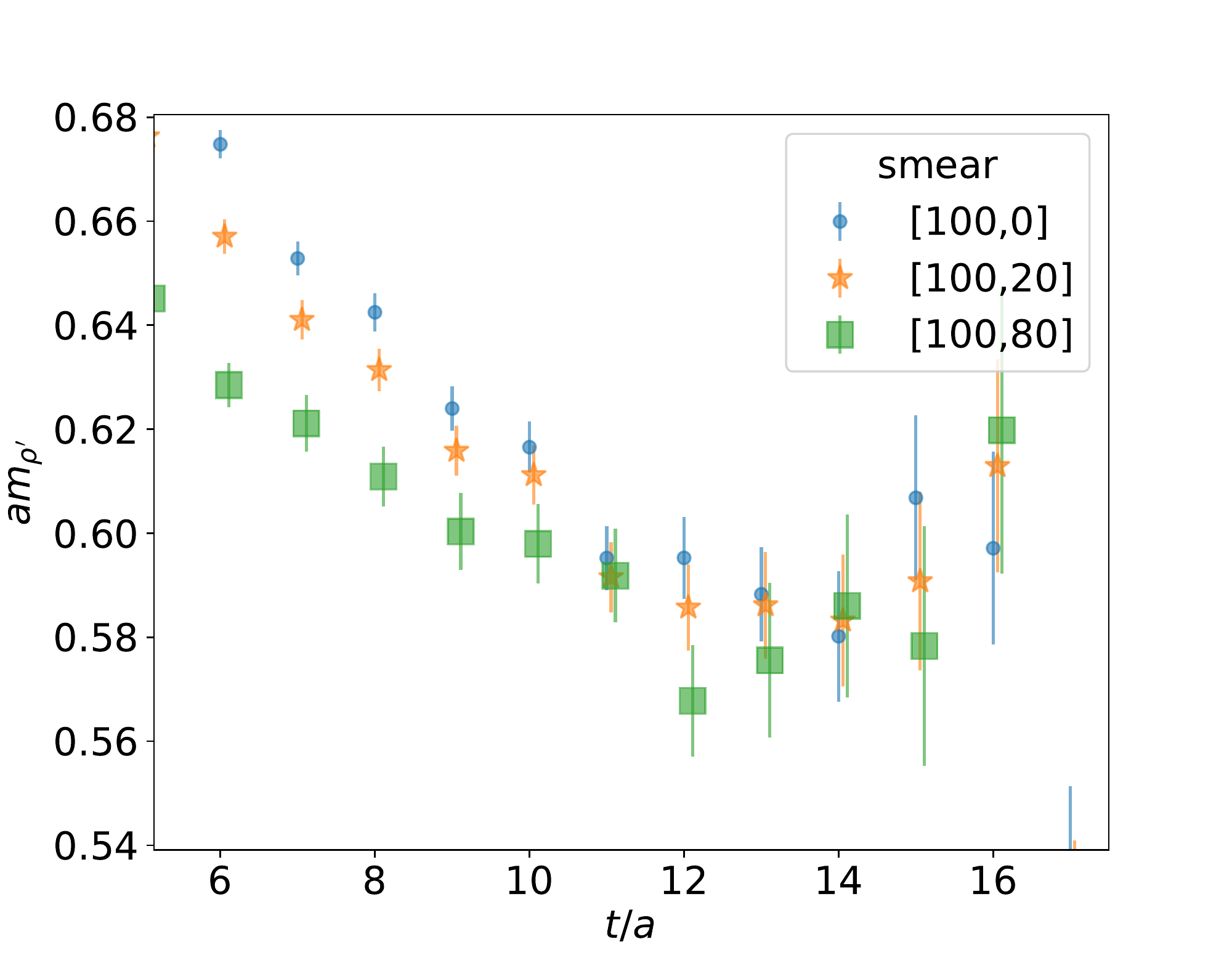}
         \caption{} \label{fig:rhoE2_smear}
       \end{subfigure}
    \caption{{\bf(a)} Comparison of axial-vector meson effective mass plot with
     $Z_2$-wall source and smeared source.  The number of iterations at the sink for
      axial vector meson is tuned to optimise signal.  {\bf(b)} Effective mass plot of excited $\rho$ meson 
     at different levels of sink smearing. 
 The smearing parameters are $\varepsilon=0.16$ and $N_{\text{W}}=100$ at the source, 
 while $N_{\text{W}}=0, 20, 80$ for the sink---values 
 in the square bracket in the legend.
   Both panels present calculations  performed on a $54\times 32^3$ lattice, with $\beta=6.8$ and $m_0 = -1.04$. }
\end{figure}

In our previous studies~\cite{Bennett:2017kga,Bennett:2019jzz,Bennett:2019cxd}, 
we used stochastic $Z_2$-wall sources to calculate correlation functions~\cite{Boyle:2008rh}. Unfortunately, as we can see in 
Fig~\ref{fig:SP4AS3_smear_v_wall}, the effective mass of the axial-vector mesons
 is noisy at large euclidean time. It is even harder to extract the mass of excited rho using the wall source. 
 To overcome this difficulty, we employ the well-known lattice technique dubbed Wuppertal smearing~\cite{Gusken:1989qx},
  to improve signals. The smearing function is defined as follows:
\begin{equation}\label{eq:Wuppertal_smearing}
q^{(n+1)}(x) = \frac{1}{1+6\varepsilon} \left [ q^{(n)}(x) + \varepsilon \sum_{\mu= \pm1}^{\pm 3} U_\mu(x)q^{(n)}(x+\hat{\mu}) \right]
\end{equation}
where $q^{(0)} (x) = \delta (x)$ and $q^{(n)}$, with $n=0\,,\cdots\,, N_{\text{W}}$
 are the smeared source resulting from an iterative process. 
The tunable  parameters $\varepsilon$ and $N_{\text{W}}$ are step size and number of iterations respectively. 
We supplement the source smearing by smoothening  
the gauge links with 
APE smearing~\cite{APE:1987ehd}:
\begin{equation}\label{eq:APE_smearing}
    U^{(n+1)}_\mu(x) = P \left\{ (1-\alpha)U^{(n)}_\mu(x) + \frac{\alpha}{6} S^{(n)}_\mu(x) \right \},
\end{equation}
with $n=0\,,\cdots\,,N_{\text{APE}}$ where we smeared link is 
\begin{equation*}
    S_\mu(x) = \sum_{\pm \nu \neq \mu} U_\nu(x)U_\mu(x+\hat{\nu})U^\dagger_\nu(x+\hat{\mu}).
\end{equation*}
This process removes short-distance fluctuations of the gauge fields. 
The effective mass is sensitive to the parameters $\varepsilon$ and $N_{\text{W}}$
of the Wuppertal smearing, which need to be optimised.  The effect of
APE smearing  on the effective mass is less dramatic, as long as the smeared 
spacial plaquette is close to $1$. We take $\alpha = 0.4$ and $N_{\text{APE}}=50$, 
for all ensembles in this study. 

Fig.~\ref{fig:SP4AS3_smear_v_wall} shows that the effective mass after smearing displays
the earlier appearance of a longer plateau,  and smaller errors. We arrived at  optimal smearing 
by tuning the number of iterations at the sink while fixing the number at the source. If we could not obtain 
the desired plateau, we would 
change the smearing level at the source. Fig.~\ref{fig:rhoE2_smear} illustrates the effect of different levels of 
sink smearing.  The adoption of a multi-source strategy further improves statistical precision.

 \begin{figure}
\centering
\includegraphics[scale=0.4]{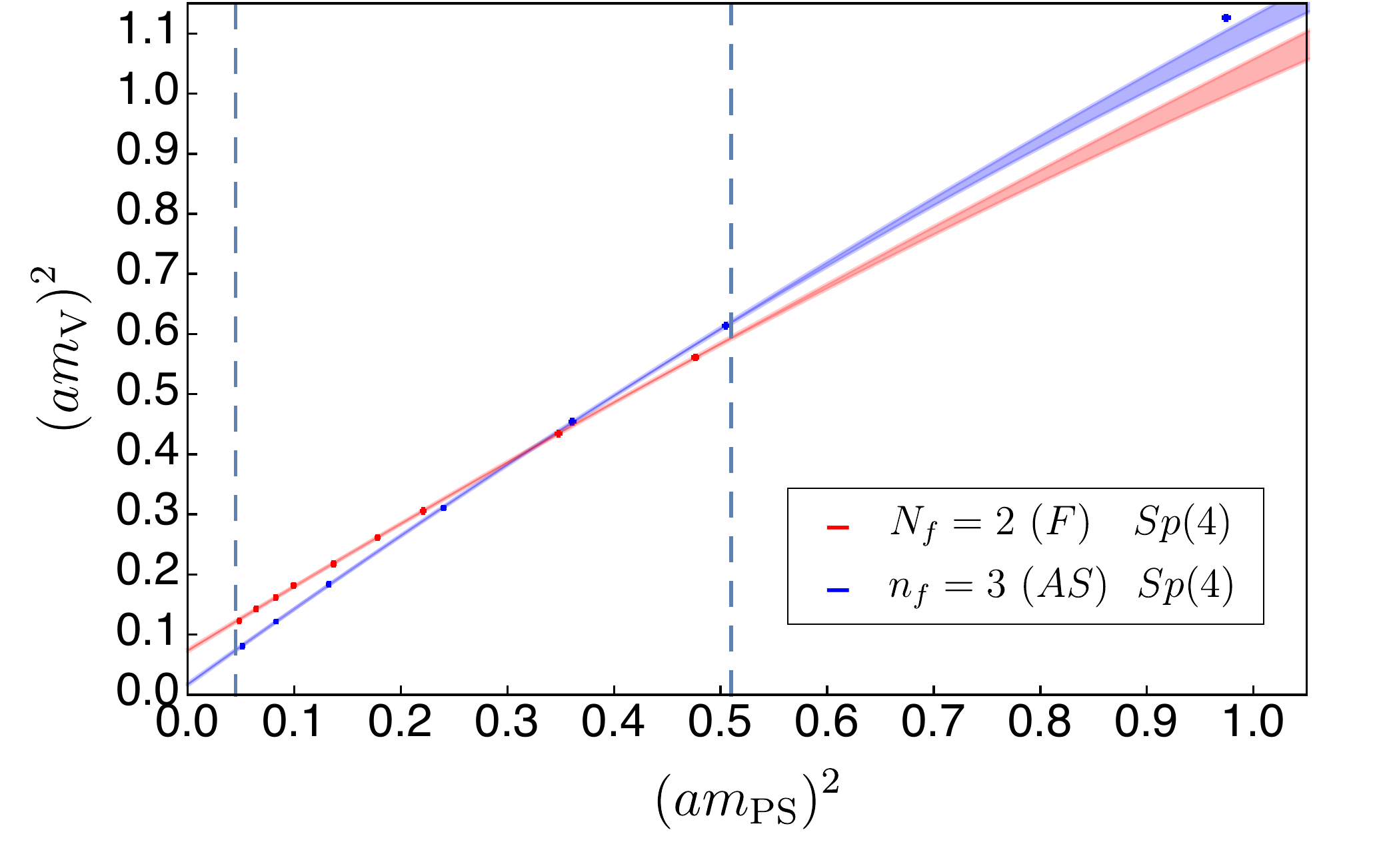}
\caption{Comparison of the massless extrapolations 
of the vector meson masses,
obtained with  $\beta=7.2$ with fundamental fermions,
and $\beta=6.7$ with antisymmetric fermions. The vertical dash lines demarcate the fit range.}
\label{fig:SP4AS3_m2v_vs_m2ps_F_vs_AS}
\end{figure}

\begin{figure}[t]
\centering
\includegraphics[scale=0.4]{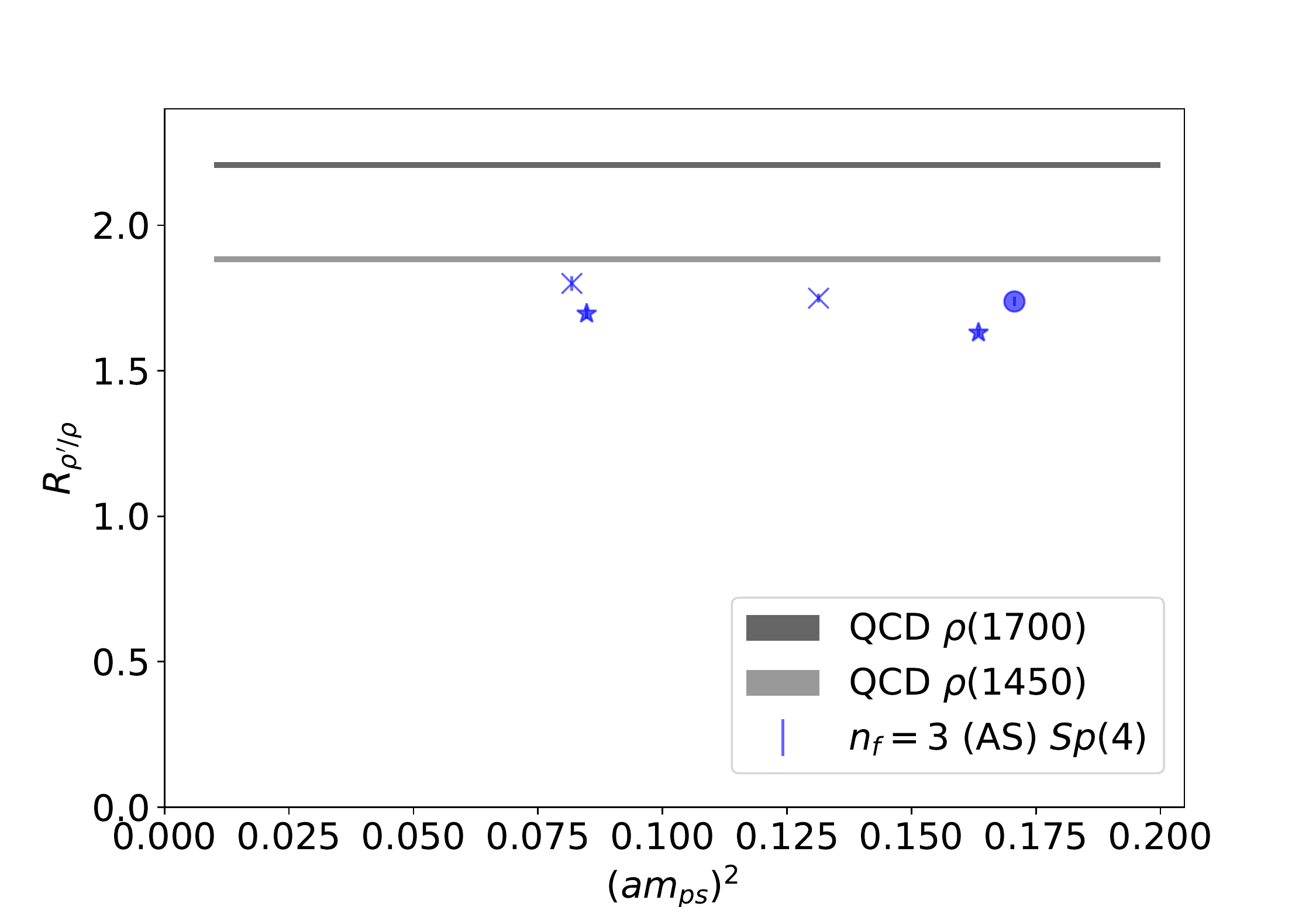}
\caption{The  mass ratio  $R_{\rho^{\prime}/\rho}$ of excited vectors states to the ground state 
for varying $(am_{\text{PS}}^2)$. The blue dots are our calculations in antisymmetric representation. The circle, x-shaped dot and star  refer to different choices of
$\beta=6.65, \, 6.7$, and $6.8$, respectively. The grey bands denote the experimental value in QCD~\cite{ParticleDataGroup:2020ssz}. }
\label{fig:SP4AS3_R_rho_Mps2}
\end{figure}

One noticeable finding of our study is that it is more difficult  to decrease the ratio $m_{\text{ps}}/m_{\text{v}}$
with antisymmetric matter compared to fundamental fermions. 
 We find $m_{\text{ps}}/m_{\text{v}} \simeq 0.8$ for parameter choices giving the lightest pseudoscalar meson mass. 
To generate ensembles with lighter pseudoscalar meson mass, we would require larger lattices,
 to suppress finite size effects, hence increasing the 
 demand for computation resources.  
Fig.~\ref{fig:SP4AS3_m2v_vs_m2ps_F_vs_AS} shows a comparison 
of the mass dependence of the vector meson mass squared
with our earlier work on the theory
 with fundamental matter~\cite{Bennett:2019jzz},
 at fixed lattice couplings.
 We find that the antisymmetric case approaches 
 the massless limit with a sharper slope and a tiny 
 chiral value of $(a m_V^\chi)^2$.
 This phenomenon might be an early indication that
  the theory with fermions in the antisymmetric representation is closer to
 the conformal window, and thus deserves further exploration, by extending the
 measurements to a larger set of ensembles.

Another interesting preliminary result pertains to the ratio of masses of excited vector mesons to the ground state.
 Fig.~\ref{fig:SP4AS3_R_rho_Mps2} shows that this ratio is about $1.7$ in the antisymmetric representation.  
 This is slightly lower than in QCD~\cite{ParticleDataGroup:2020ssz}.
Yet, direct comparison is not possible, as
we have not yet performed the continuum extrapolation.

\begin{figure}
     \centering
     \begin{subfigure}[b]{0.49\textwidth}
         \centering
         \includegraphics[width=\textwidth]{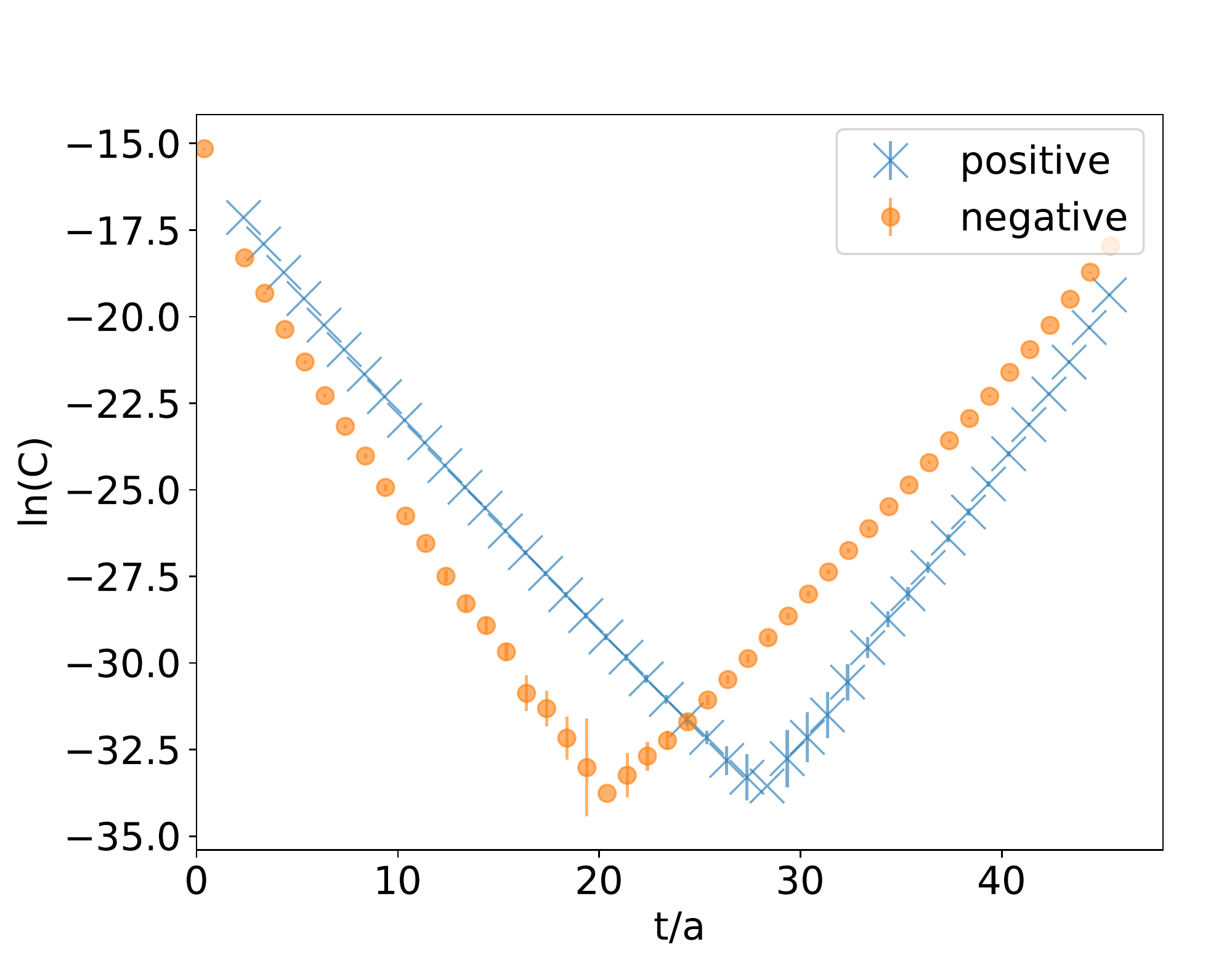}
     \end{subfigure}
     \hfill
     \begin{subfigure}[b]{0.49\textwidth}
         \centering
         \includegraphics[width=\textwidth]{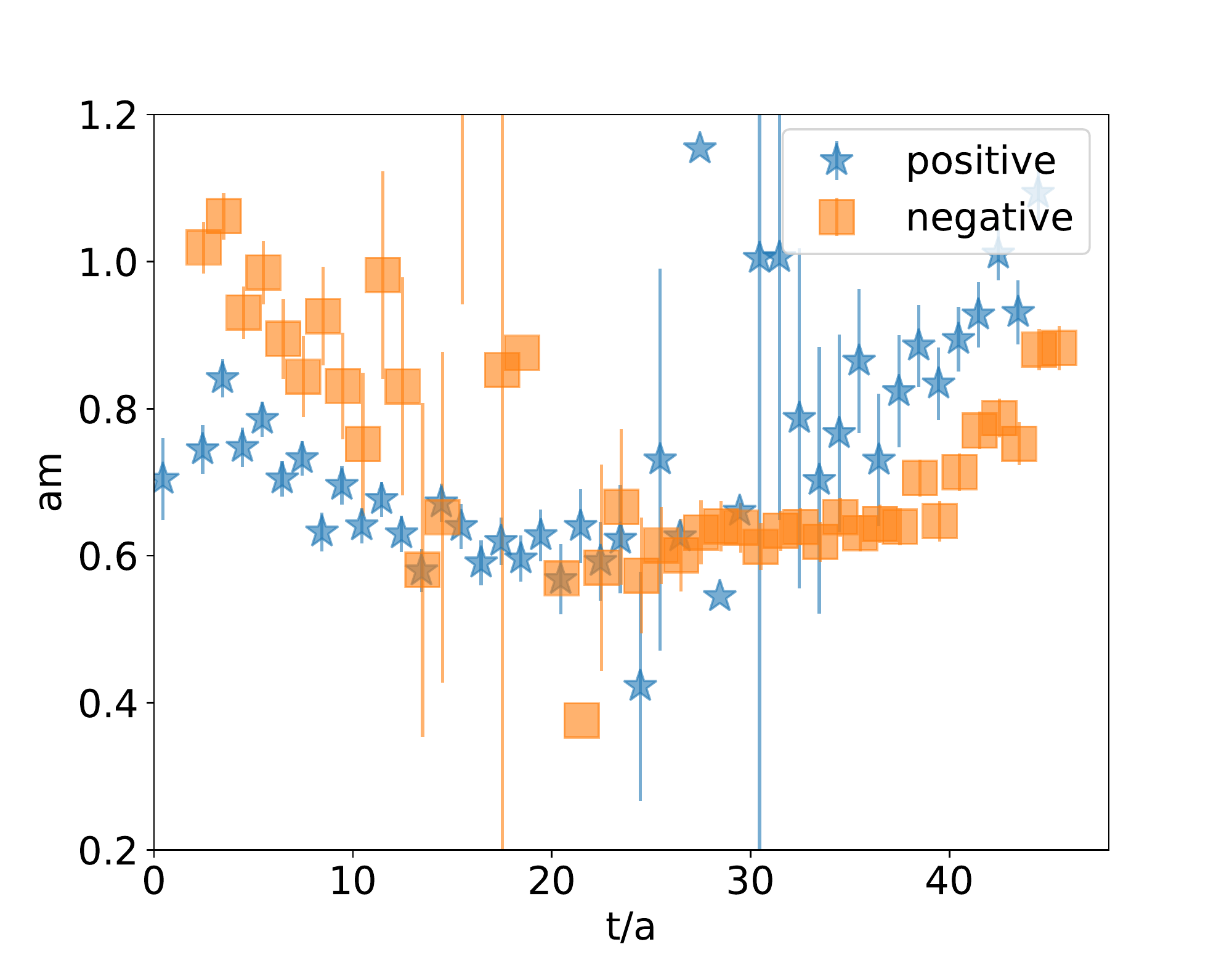}
       \end{subfigure}
    \caption{The log plot of the chimera baryon correlators (left panel) and their effective mass plot (right panel) with the parity projections.}
     \label{fig:SP4AS3_CB_parity}
\end{figure}

As a first step towards studying the spectrum of chimera baryons, we use these same ensembles, partially quenching the 
theory by treating the $N_f=2$ fundamental fermions as non-dynamical---we expect the $n_f=3$ antisymmetric fermions 
to have a stronger effect on the dynamics.
To enable a comparison, we tune the valence fundamental-fermion 
masses such that $m_{\text{PS}}/m_{\text{V}} \sim m_{\text{ps}}/m_{\text{v}}$---the former being 
the mass
ratio for the mesons made of  quenched fundamental fermions, the latter measured with dynamical antisymmetric ones. 
The effective mass of the chimera baryons defined  in Eq.~(\ref{Eq:cb_dq})
 is presented in Fig.~\ref{fig:SP4AS3_CB_parity}, after 
 parity projection with $P_{\pm}=(1 \pm \gamma_0 )$. 
 The correlation functions exhibit an asymmetric behaviour:
 the backward propagator with a given parity agrees
 with the forward  propagator with the opposite parity,
as is the case for nucleons in lattice QCD~\cite{Aarts:2015mma}.
 A comparison of the effective mass  of chimera baryons and mesons is displayed in Fig~\ref{fig:SP4AS3_CB_spec}.

\begin{figure}
\centering
\includegraphics[scale=0.5]{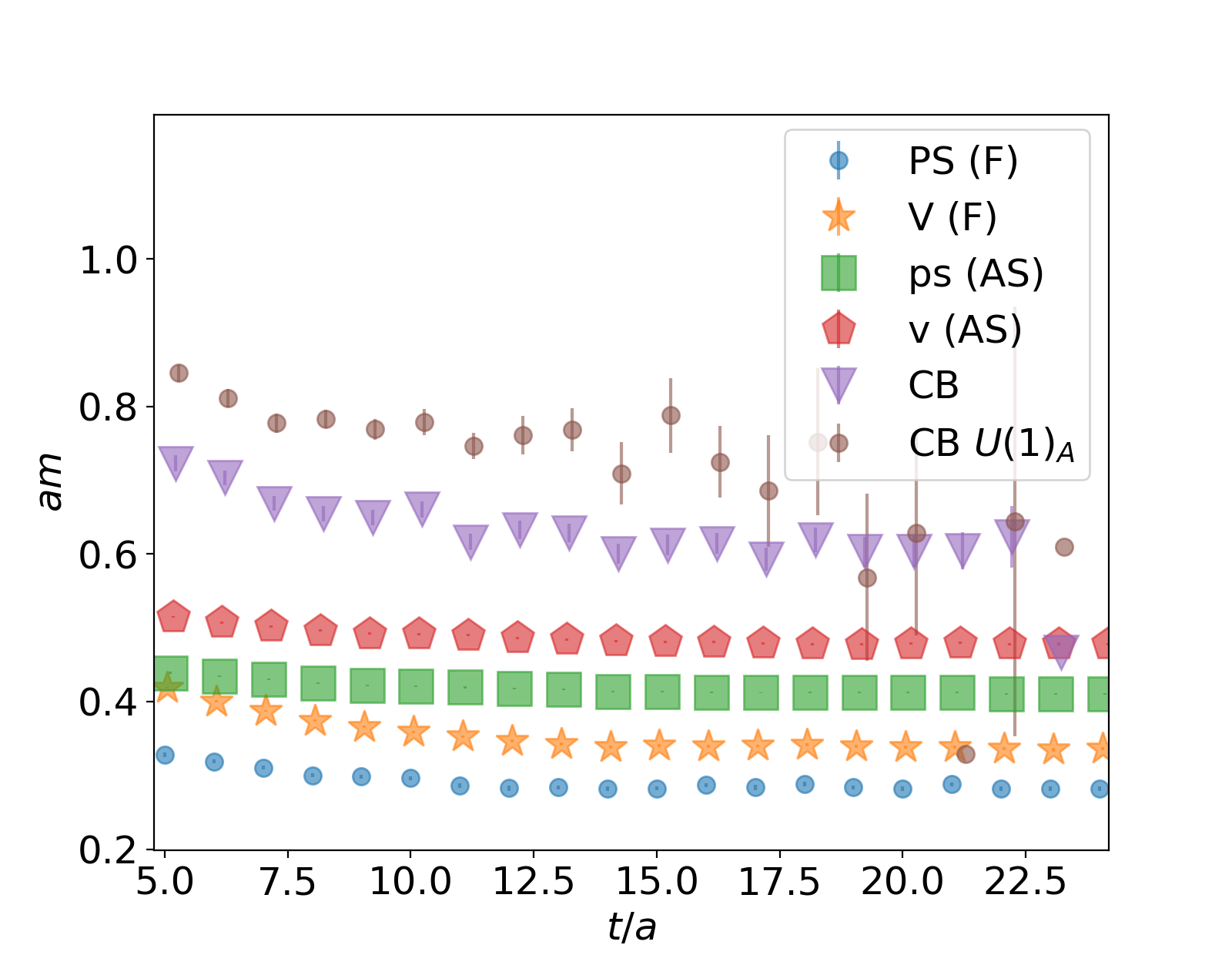}
\caption{Effective masses plot measured from $48\times24^3$ lattice with $\beta=6.65$ and $m_0=-1.07$ using dynamical antisymmetric fermions. The quenched fundamental bare mass is $m_{\text{f}}=-0.734$. The blue and orange are pseudoscalar and vector mesons with quenched fundamental fermions. The green and red are pseudoscalar and vector mesons with dynamical antisymmetric fermions. The purple and brown are chimera baryons and their $U(1)_A$ partner.}
\label{fig:SP4AS3_CB_spec}
\end{figure}

\subsection{Dynamical studies with $N_f=2$ fundamental and $n_f=3$ antisymmetric Dirac fermions}
\label{sec:multirep}

We modified the HiRep code~\cite{DelDebbio:2008zf} to perform dynamical calculations
with both fermions in the fundamental and antisymmetric representation, with the matter content 
require by the CHM in Ref.~\cite{Barnard:2013zea}, and we have been carrying on an
extensive programme of testing of the resulting numerical algorithms.
Besides the volume, the lattice theory depends on three couplings:  $\beta$ and
the masses of fermions in the fundamental and antisymmetric representations, 
denoted as $m_0^{\rm f}$ and $m_0^{\rm as}$, respectively. We hence explored
the parameter space  to identify the phase transitions.

\begin{figure}[t]
\begin{center}
\includegraphics[width=.55\textwidth]{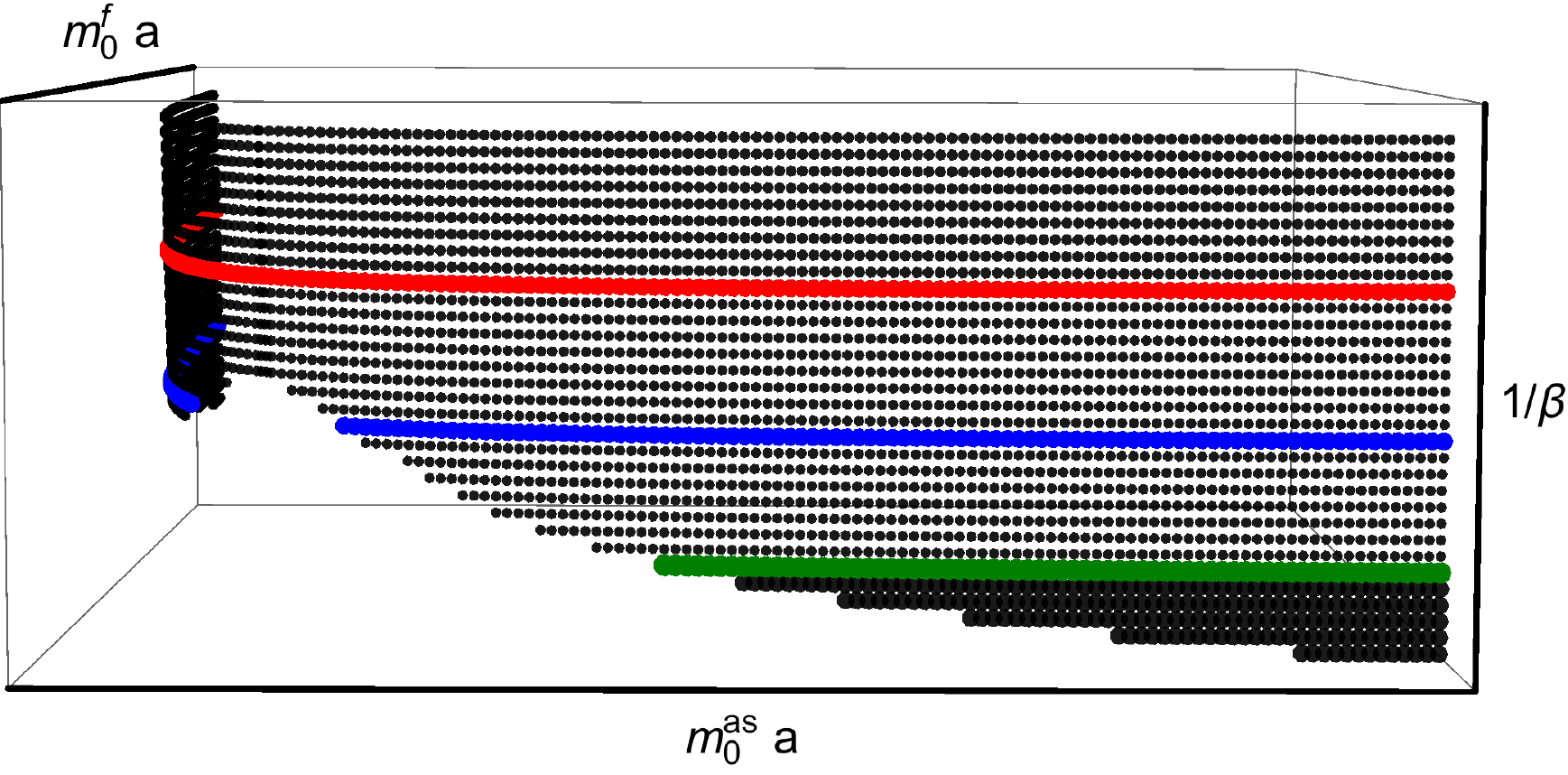}
\caption{%
\label{fig:phase_diagram}%
Cartoon  of the  phase space for the $Sp(4)$ gauge theory coupled to fermions
in both fundamental (F) and antisymmetric (AS) representations, as a functions of the 
lattice parameters $\beta$,  $m_0^{\rm f}$, and $m_0^{\rm as}$. A first order bulk
phase transition takes place on a surface with boundaries inside the three-dimensional space.
}
\end{center}
\end{figure}

Fig.~\ref{fig:phase_diagram} displays a three-dimensional
 cartoon of the phase space of the $Sp(4)$ theory with mixed representation.
The black surface denotes the first order phase transition.
The three (coloured) lines correspond to the first order lines at fixed representative choices of $\beta$.
For small $\beta$ (red line)  a phase transition occurs for all the possible  fermions masses.
With moderate $\beta$  (blue lines), the phase transition disappears when both fermions are light,
and the line of phase transitions has an end point.
The boundary of the first order surface is asymmetric,
and for even larger $\beta$ the phase transition depends only on the mass of the antisymmetric fermions.
We reported elsewhere  that the asymptotic values of the critical coupling in the quenched limit of either the fundamental
antisymmetric fermions are different,
with $\beta_{\rm cr}^{\rm f}\sim 6.7$ \cite{Bennett:2017kga} and $\beta_{\rm cr}^{\rm as}\sim 6.5$ \cite{Lee:2018ztv}.

\begin{figure}[t]
\begin{center}
\includegraphics[width=.48\textwidth]{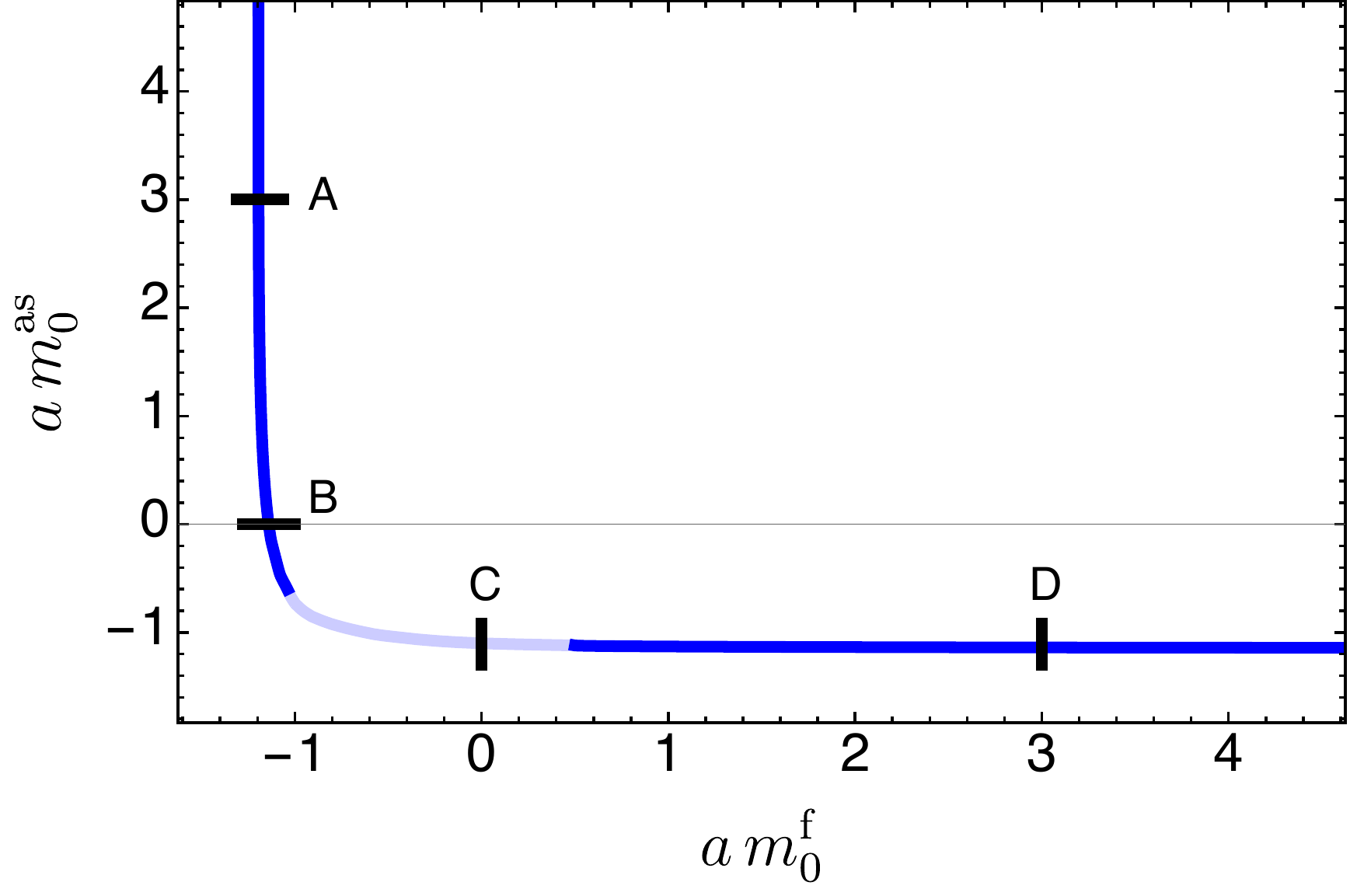}
\includegraphics[width=.48\textwidth]{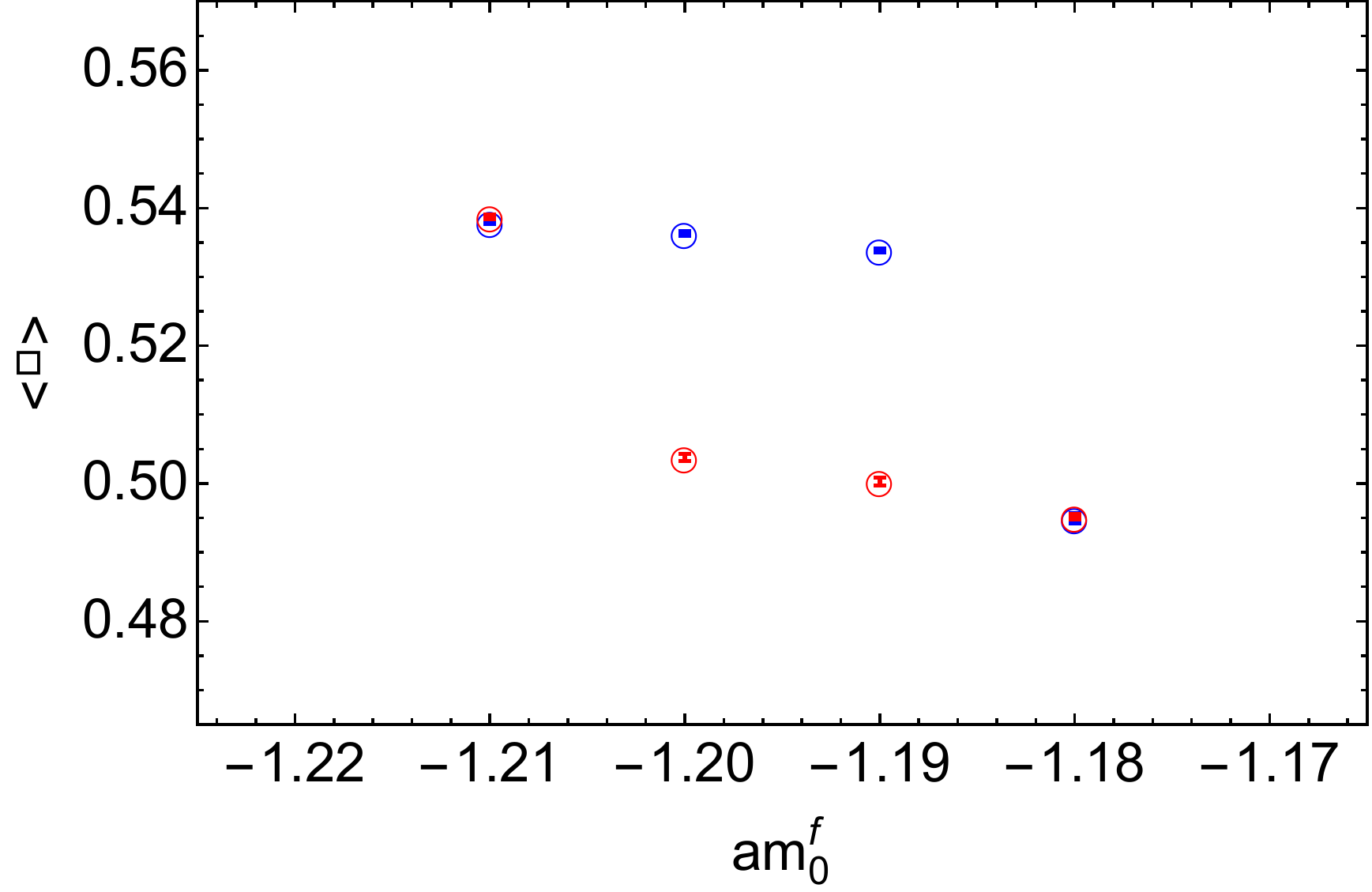}
\includegraphics[width=.48\textwidth]{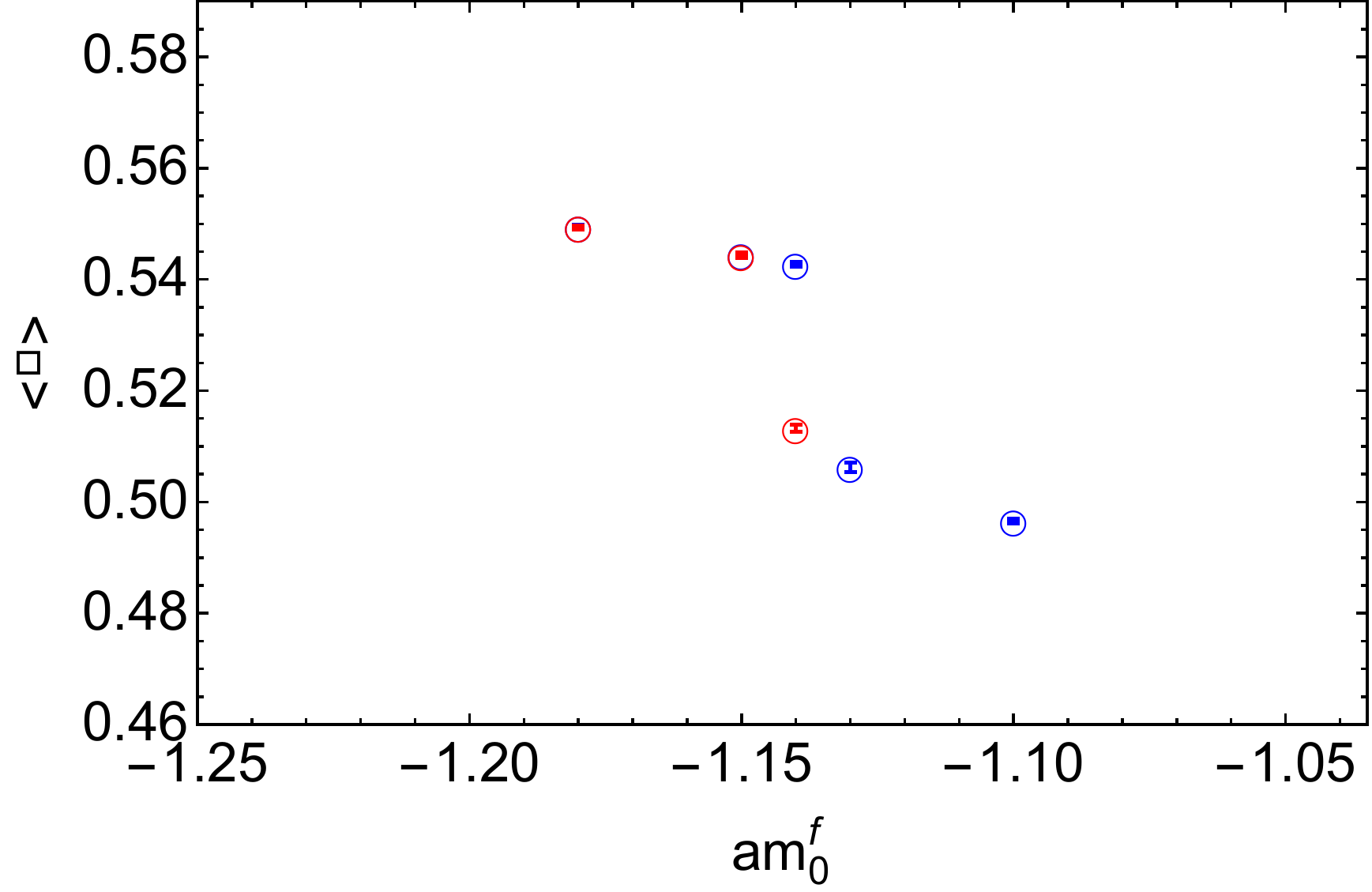}
\includegraphics[width=.48\textwidth]{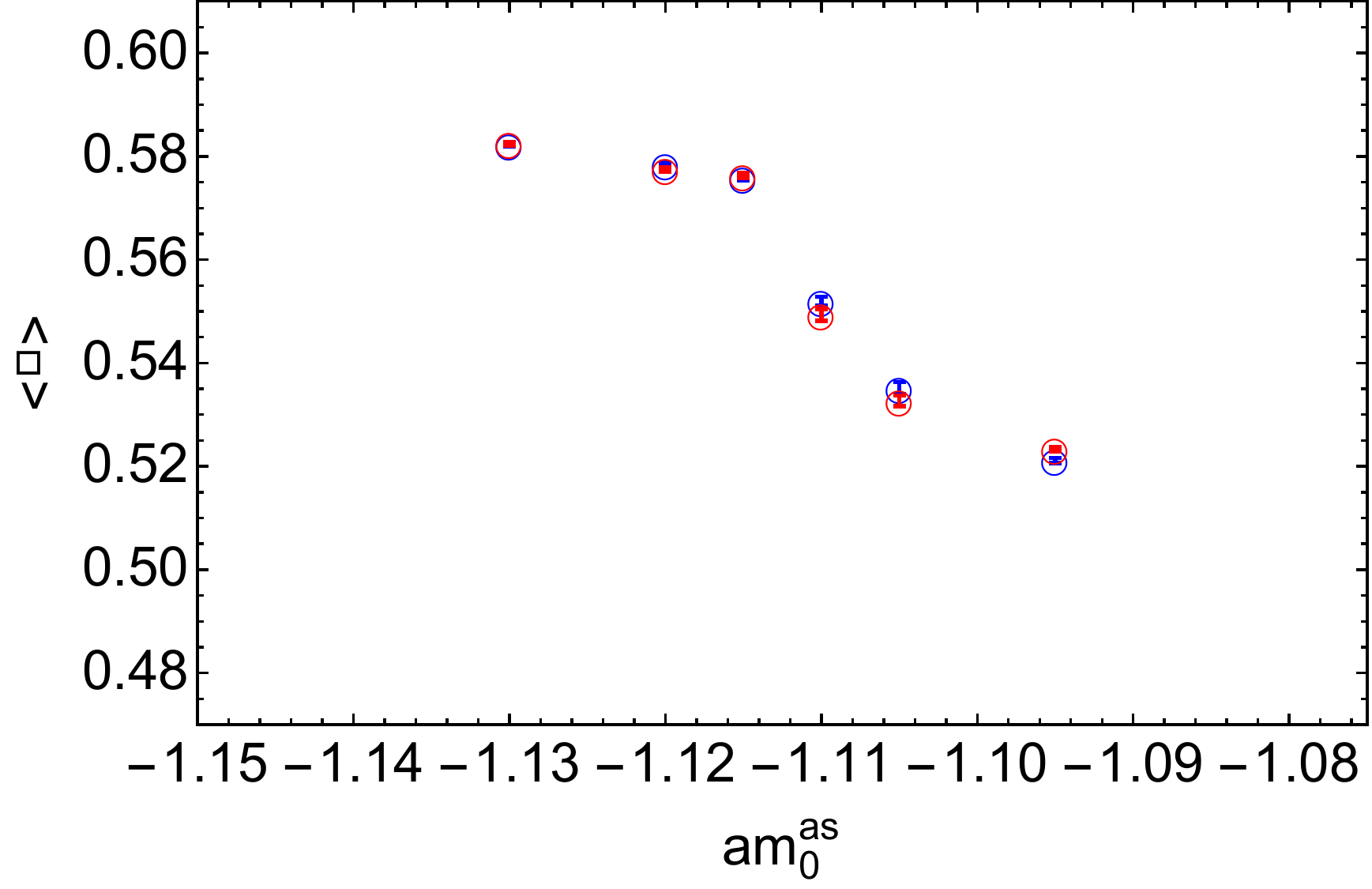}
\includegraphics[width=.48\textwidth]{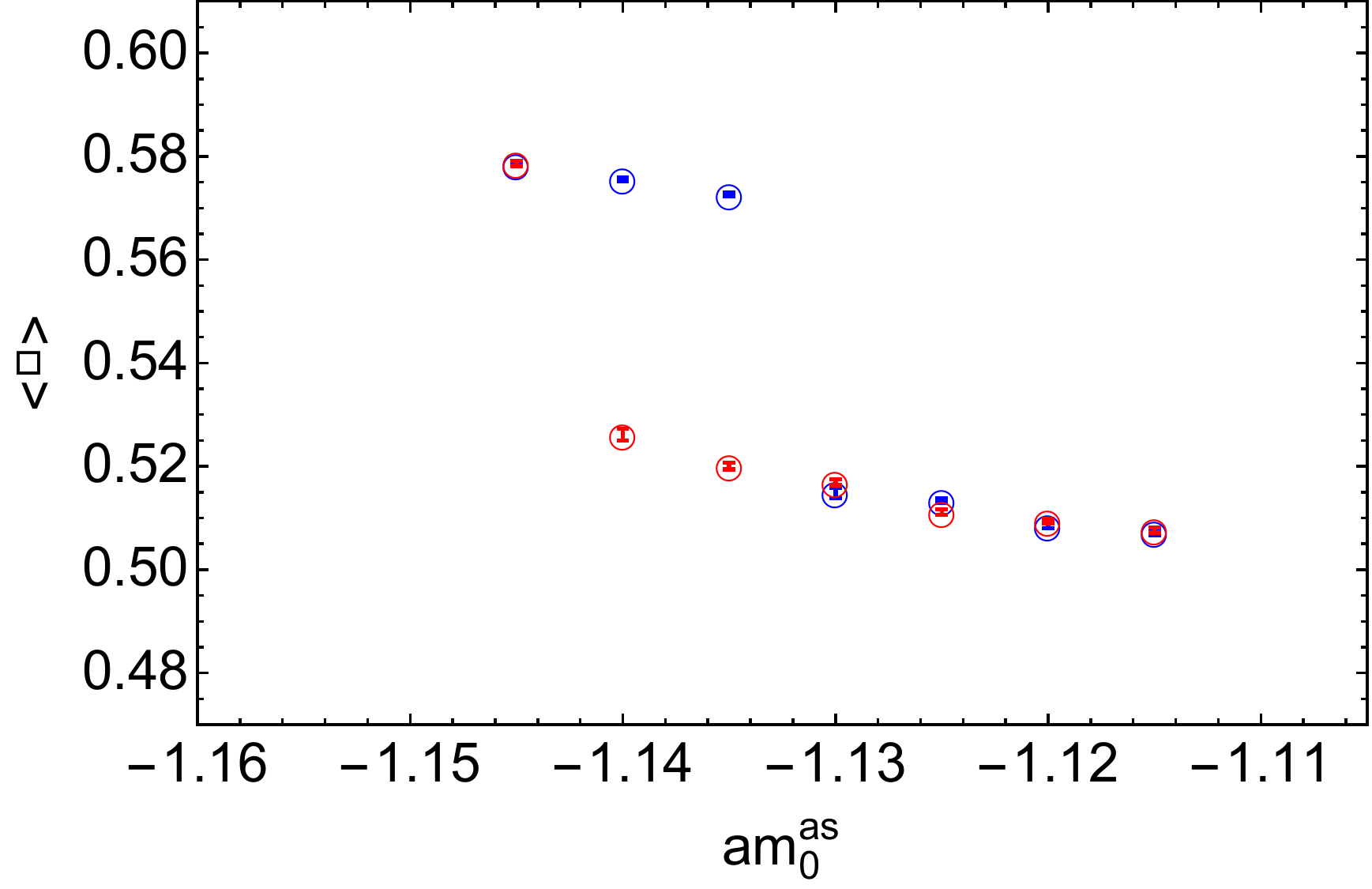}
\caption{%
\label{fig:phase_diagram_b6p4}%
The top-left panel displays a section at fixed $\beta=6.4$ of the phase structure of the two-representation $Sp(4)$ theory
from Fig.~\ref{fig:phase_diagram}.
The (blue) solid line 
denotes the first order phase transition, while the light blue line denotes a crossover.
From top-right to bottom, the other four panels show the average plaquette values, 
which are extracted for the calculations with initial configuration given by
unit (blue) or random (red) matrices, for parameter choices adjusted to
cross the phase boundaries at the points A, B, C and D in the top-left panel.
Numerical results are obtained from the lattice volume of $8^4$.
}
\end{center}
\end{figure}

In Fig.~\ref{fig:phase_diagram_b6p4}, we fix $\beta=6.4$, to match with the blue line in Fig.~\ref{fig:phase_diagram},
and show the value of the average  plaquette for four representative choices 
of phase-space trajectories that cross the blue lines at points A, B, C, D.
In three cases, we find strong evidence of  hysteresis, indicating the presence of a first-order phase transition,
while for C  the absence of  hysteresis suggests that the theory is in a cross-over region.
These results are consistent with earlier evidence that
the theories with fermions in either representation
show a first-order transition at $\beta=6.4$ \cite{Bennett:2017kga,Lee:2018ztv}.

\begin{figure}[t]
\begin{center}
\includegraphics[width=.47\textwidth]{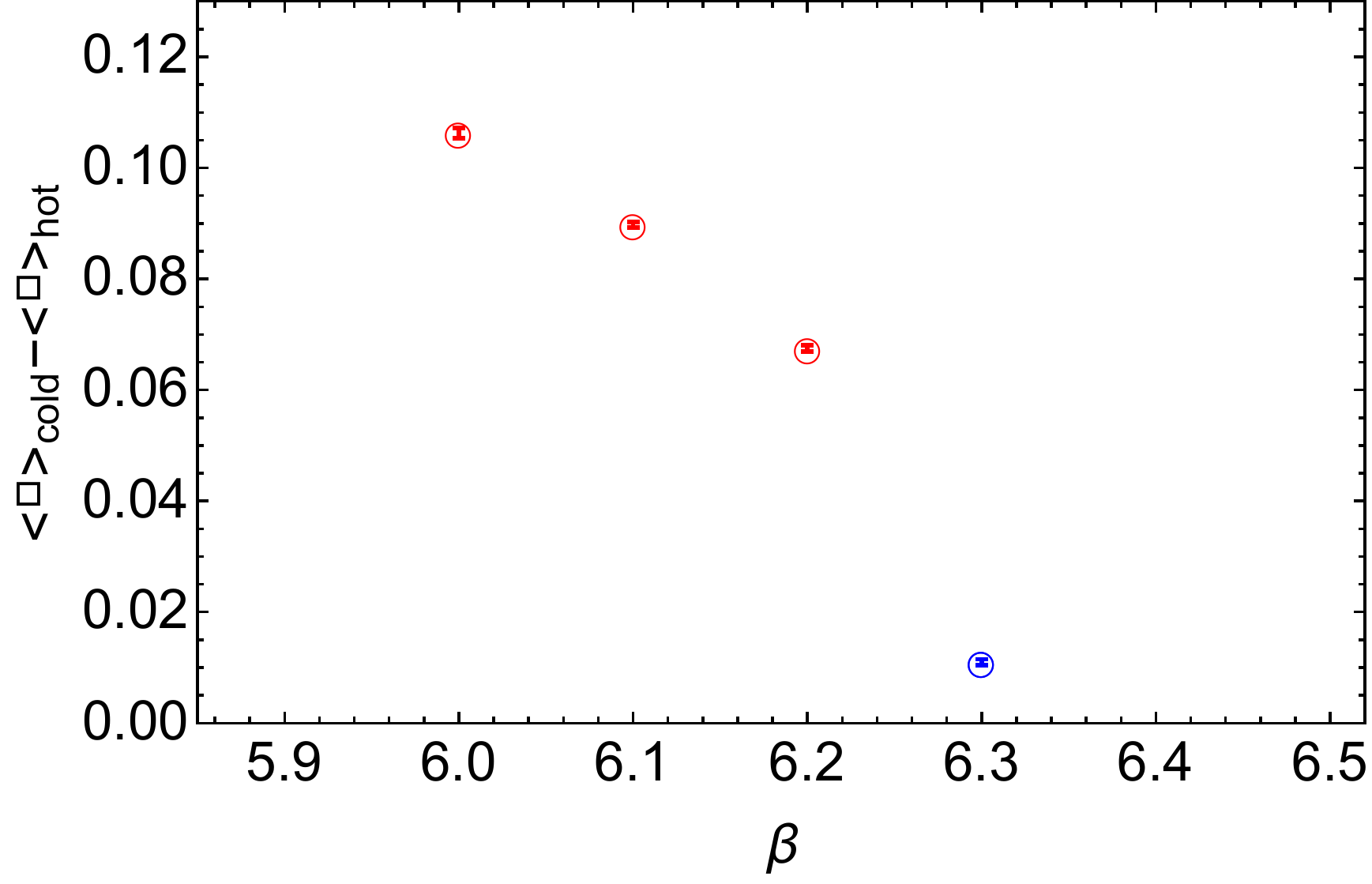}
\includegraphics[width=.49\textwidth]{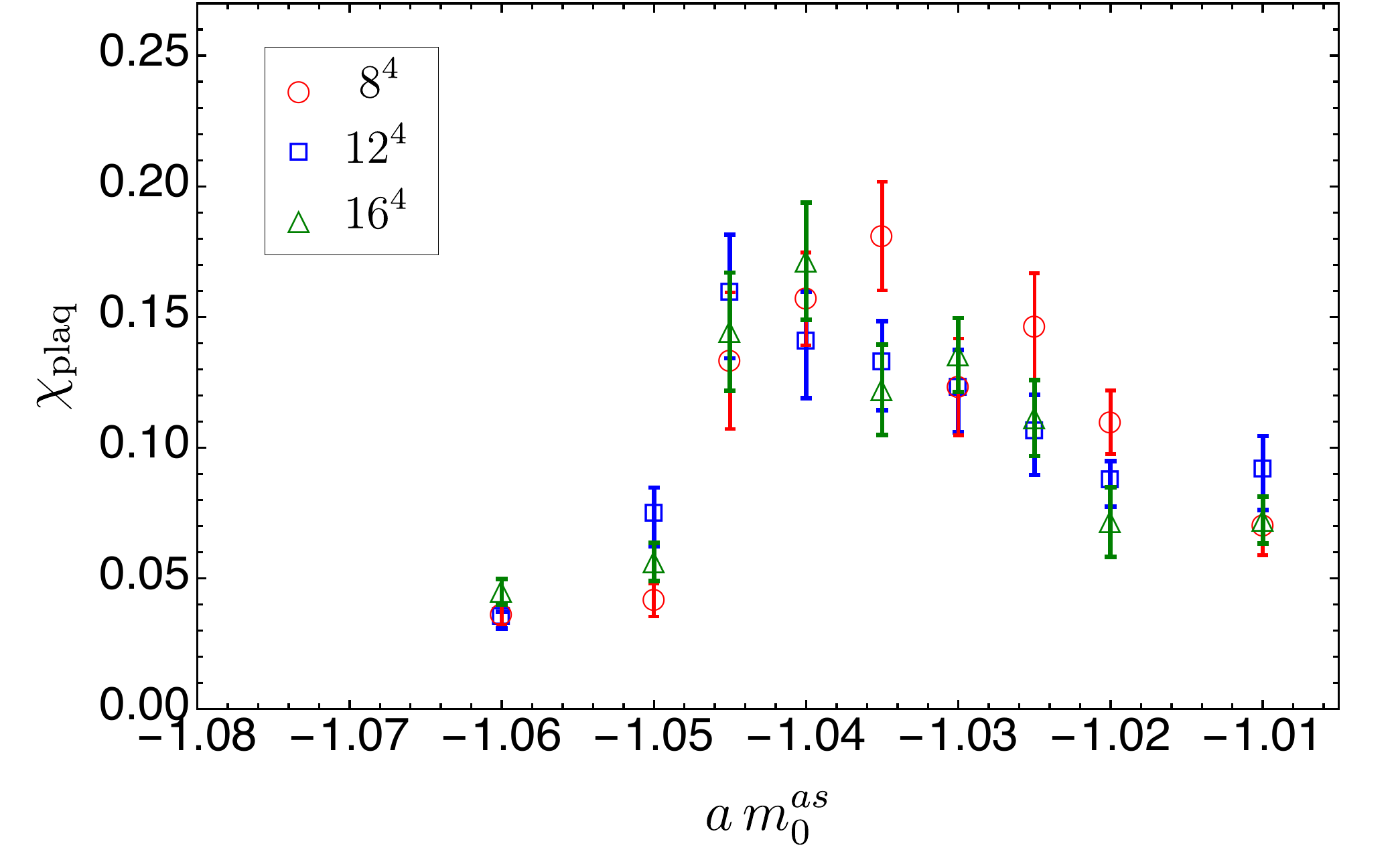}
\caption{%
\label{fig:end_point}%
Left panel: the difference of the average plaquette values
between cold (unit) and hot (random)
 initial configurations, at fixed bare mass  $a m_0^{\rm f}=-0.6$ for the fundamental fermion.
Red and blue symbols refer to  lattice volumes of $8^4$ and $12^4$, respectively.
Right panel:  plaquette susceptibilities at fixed coupling  $\beta=6.4$ and bare mass of $a m_0^{\rm f}=-0.6$,
for several different volumes.
}
\end{center}
\end{figure} 

Fig.~\ref{fig:end_point} illustrates our strategy to identify
a critical value of $\beta$, corresponding to the end of the first order surface,
at given values of the bare mass parameters.
We fix the mass $a m_0^{\rm f}=-0.6$, of the fundamental fermions
and compute the difference of the average plaquette values 
between cold and hot starts, for varying choices of  $\beta$.
The left panel of the figure shows that
 $\langle \Box \rangle_{\rm{cold}} - \langle \Box \rangle_{\rm{hot}}$ is nonzero for $\beta\lsim 6.3$,
but  is consistent with zero for $\beta=6.4$. 
In the right panel, we show that having fixed $\beta=6.4$, by computing the plaquette susceptibility $\chi_{\rm{plaq}}$
we find that its peak   is independent  of the volume,
consistently with a crossover.
We hence led to conclude that for $\beta \gsim 6.4$ the mixed-representation theory is
in the weak-coupling regime.

\begin{figure}[t]
\begin{center}
\includegraphics[width=.5\textwidth]{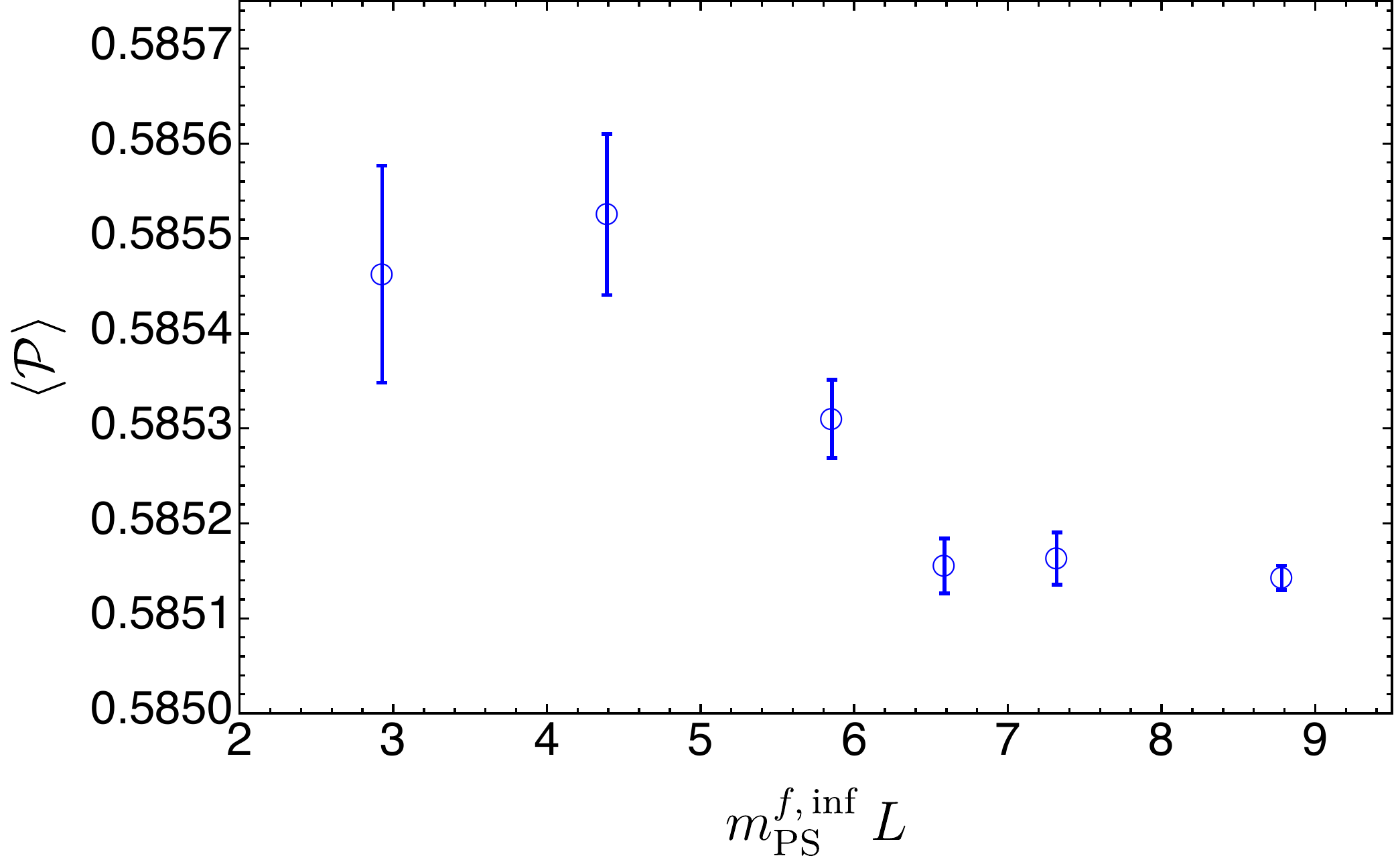}
\includegraphics[width=.47\textwidth]{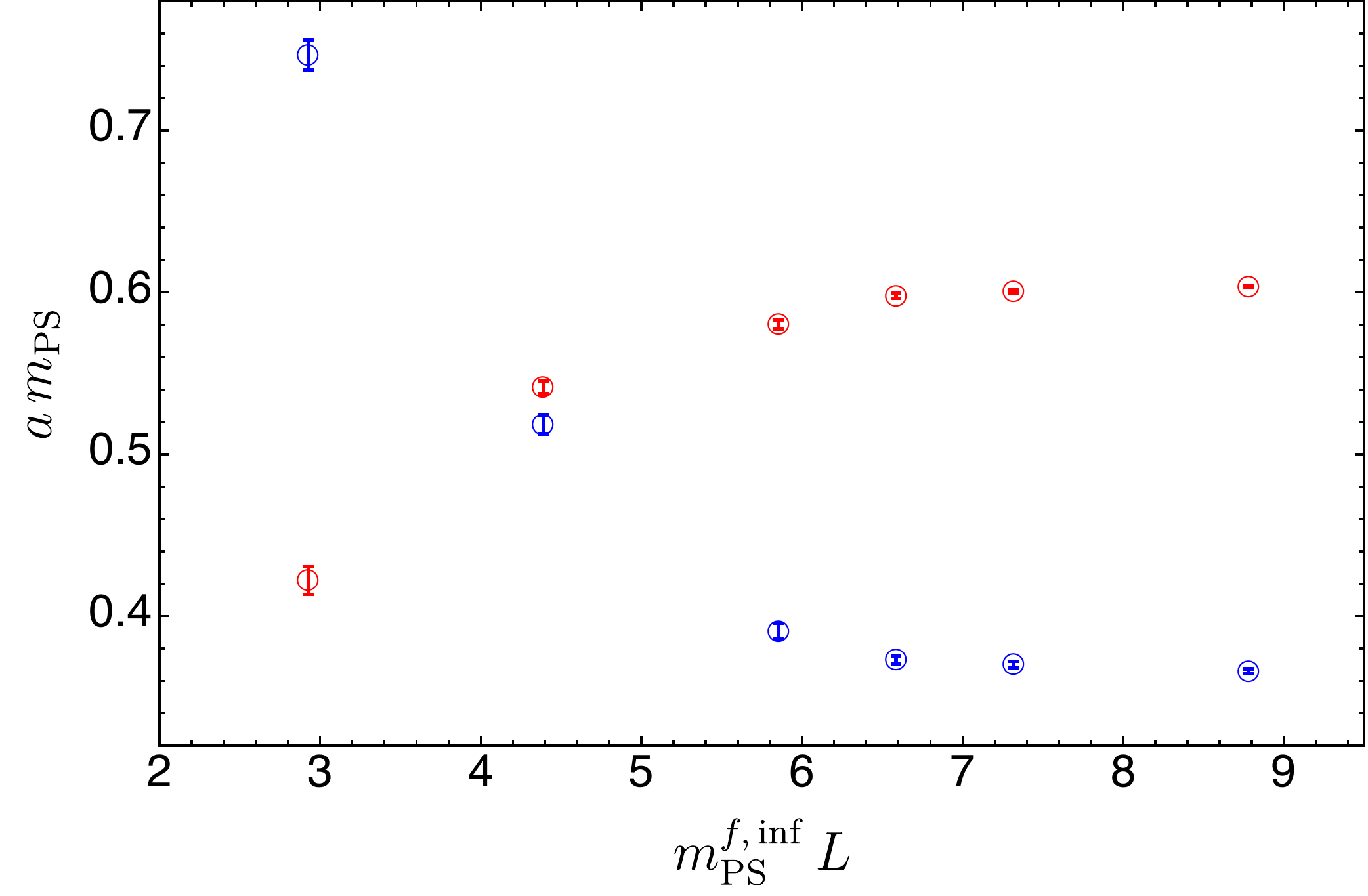}
\caption{%
\label{fig:FV}%
Finite volume effects displayed by the average plaquette (left panel)
and the mass of pseudoscalar mesons (right panel) in which the constituent fermions
are in the fundamental (blue) and antisymmetric (red) representations, for
 $\beta=6.5$, $a m_0^{\rm f}=-0.71$, $a m_f^{\rm as}=-1.01$.
}
\end{center}
\end{figure}

We next turn our attention to finite size effects. 
We start by fixing $\beta=6.5$, $a m_0^{\rm f}=-0.71$, and $a m_0^{\rm as}=-1.01$, and we vary the volume.
In the left panel of Fig.~\ref{fig:FV}, we show the average plaquette 
measured on  lattice volumes varying from $36\times 8^3$ to $48\times 24^3$, 
in terms of $m_{\rm PS}^{\rm inf}L$,  the mass of pseudoscalar meson with
 constituents given by the fermions in the fundamental representation---the lightest state in the spectrum--in
 units of the lattice size. 
 $am_{\rm PS}^{\rm inf}$ denotes the mass measured at the largest available volume. 
 Finite volume correction to the plaquette are no larger than a per mile effect, and
are negligible  for $m_{\rm PS}^{\rm inf} L \gsim 6.5$. 

In the right panel of the figure, we present the masses of the pseudoscalar mesons composed of 
constituent fermions in the fundamental and antisymmetric representations,  measured in different volumes. 
Finite volume corrections are at the percent level and comparable to the statistical errors when 
$m_{\rm PS}^{\rm inf} L \gsim 7$. Preliminary analyses not reported here confirm
 that  finite size corrections to the masses of chimera baryons and other mesons in other channels, 
 and to the pseudocalar decay constant, are negligible
under the same condition to the lattice volume. 
As observed earlier, finite volume corrections to the pseudoscalar mass have opposite signs depending on the constituent fermions, consistently with the predictions of  chiral perturbation theory at finite volume.

\begin{figure}[t]
\begin{center}
\includegraphics[width=.48\textwidth]{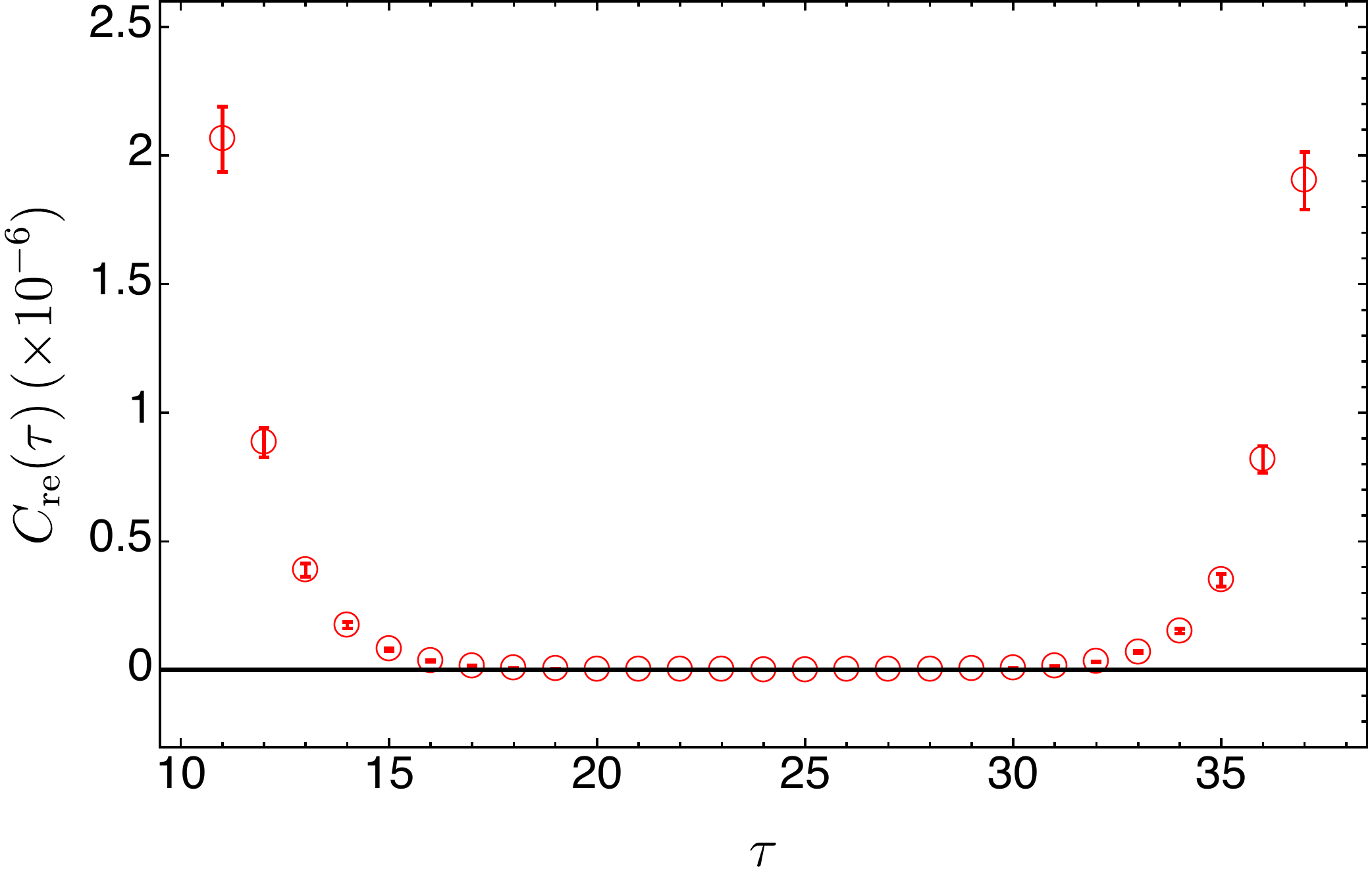}
\includegraphics[width=.48\textwidth]{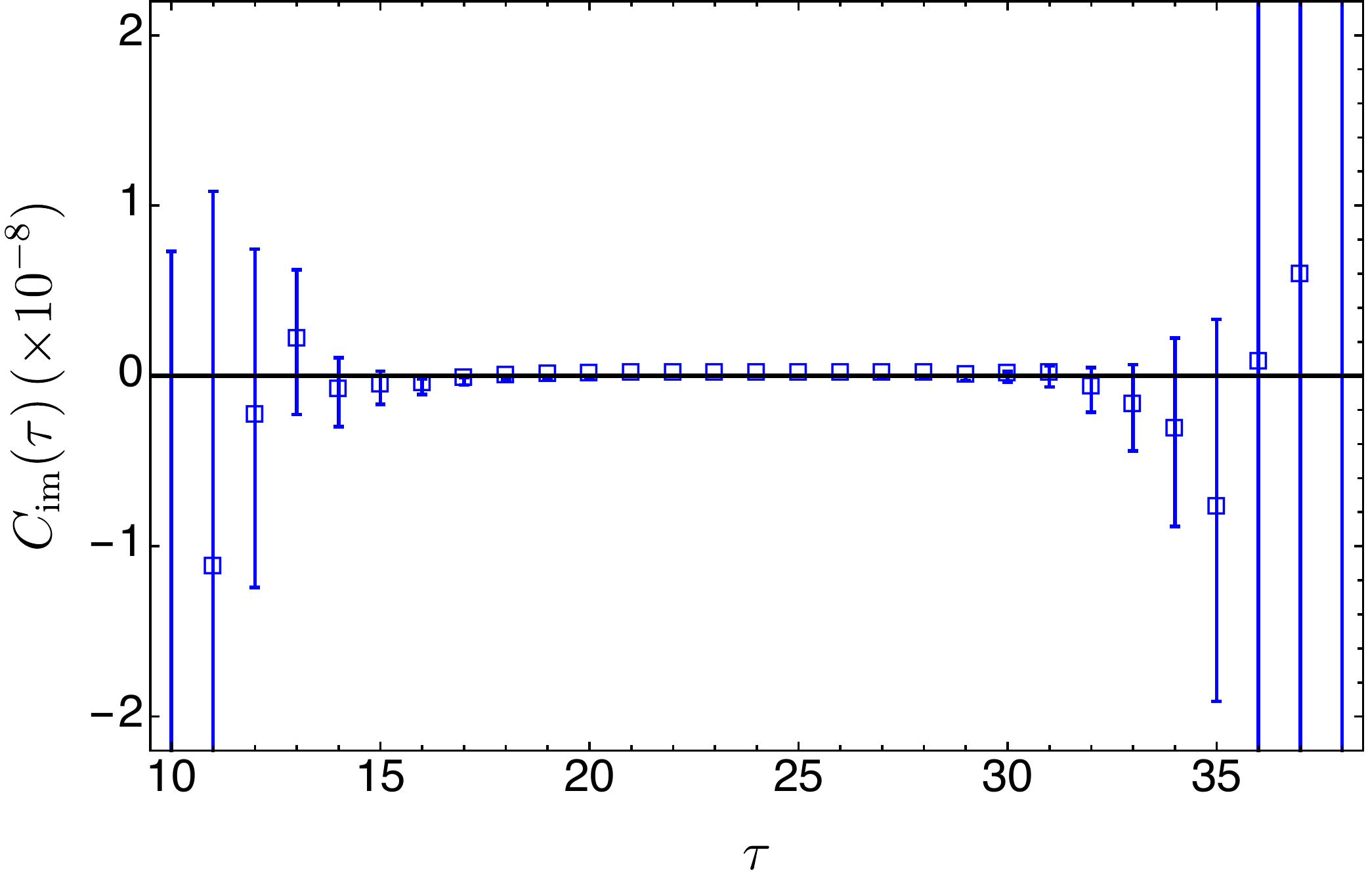}
\includegraphics[width=.48\textwidth]{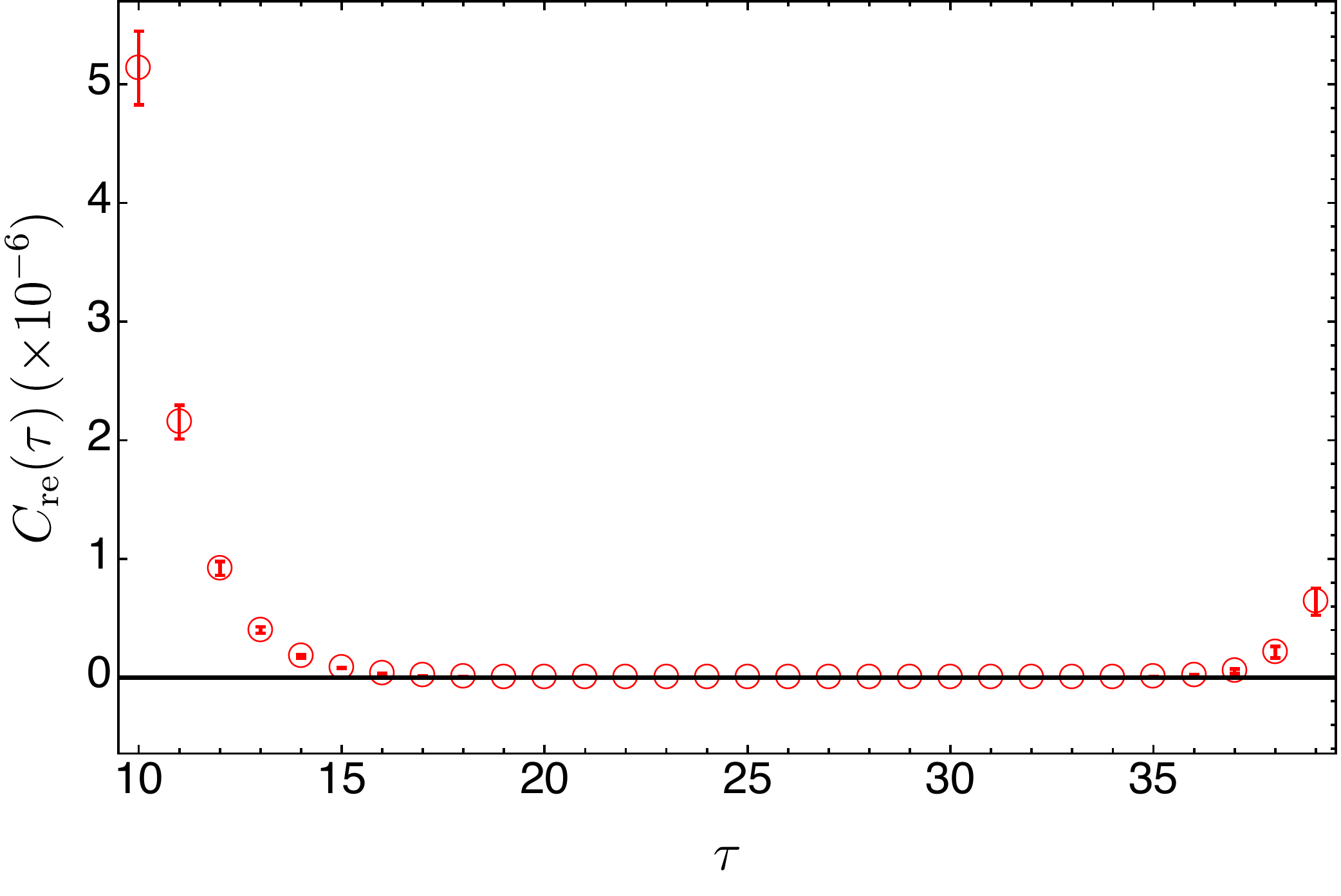}
\includegraphics[width=.48\textwidth]{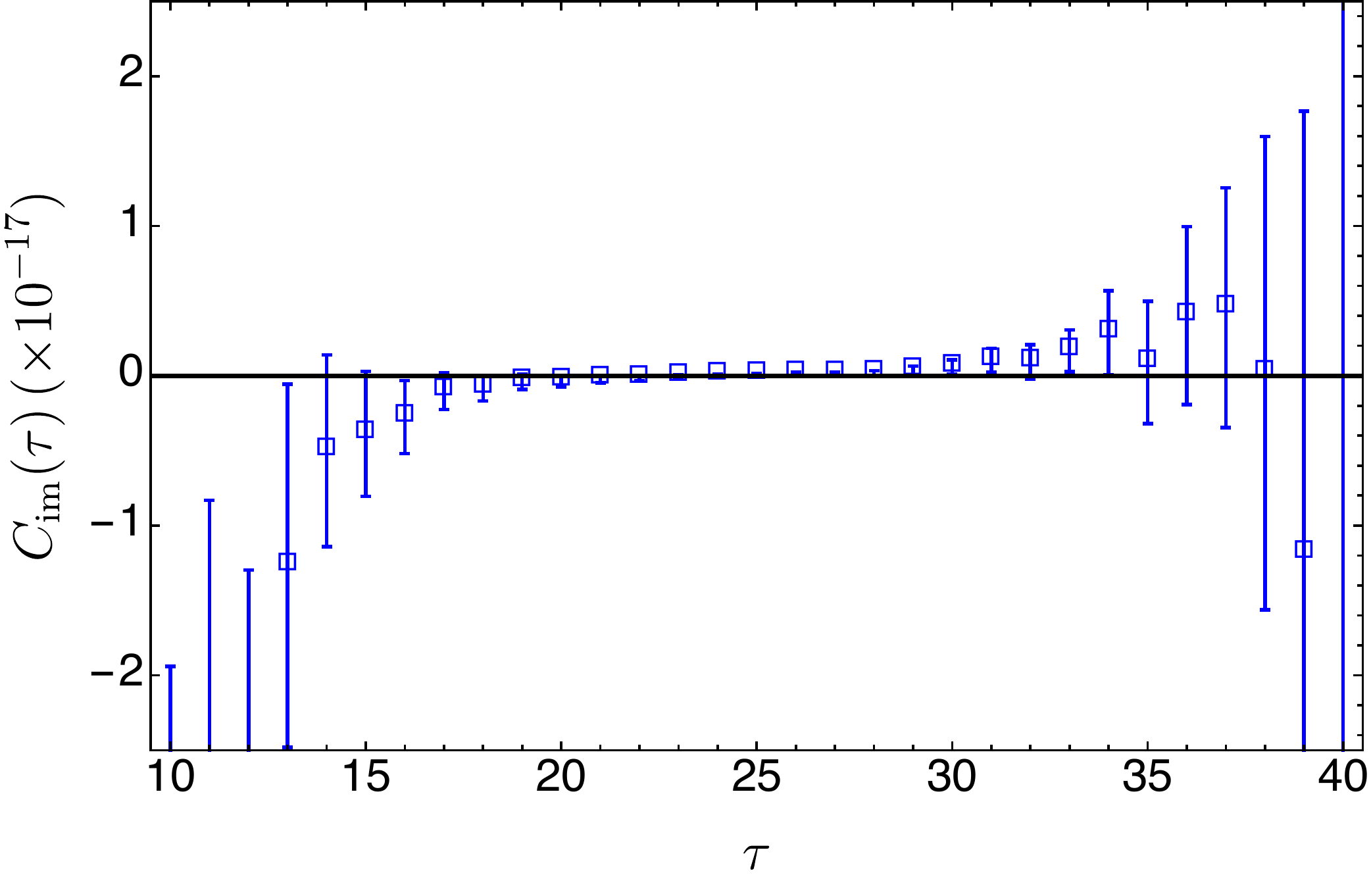}
\caption{%
\label{fig:corr_chimera}%
Real and imaginary parts of the two-point Euclidean correlation function of chimera baryons in the 
dynamical mixed-representation theory, before (upper panels)
and after (bottom panels) parity projection. 
The lattice parameters are $\beta=6.5$, $a m_0^{\rm f}=-0.71$, $a m_f^{\rm as}=-1.01$, 
and the lattice volume is $48\times 24^3$.
}
\end{center}
\end{figure}

We performed the first calculations of the spectrum of chimera baryons in the $Sp(4)$ theory 
with mixed-representation dynamical matter fields ($N_f=2$ and $n_f=3$), and 
carried out several non-trivial tests. The upper panels of Fig.~\ref{fig:corr_chimera}
display the real and imaginary parts of the correlation function.
While the former shows a clear signal of exponential decay,  
and a beautiful symmetry between forward and backward propagation,
 statistical fluctuations in the latter are larger than the signal.
The lower panels of Fig.~\ref{fig:corr_chimera} are obtained by computing 
 the real and imaginary parts of the correlators after acting with the parity projector onto the baryon interpolating 
 operators. In this case, the real part  $\mathcal{C}_{\rm re}(\tau)$ is asymmetric in respect to
 Euclidean time: the forward and backward propagators correspond to the even and odd parity states, respectively. 
 The imaginary part is again just statistical noise, but after parity 
 projection the fluctuations of the imaginary part 
are now about ten
 orders of magnitude smaller than the error on the real part.

\begin{figure}[t]
\begin{center}
\includegraphics[width=.48\textwidth]{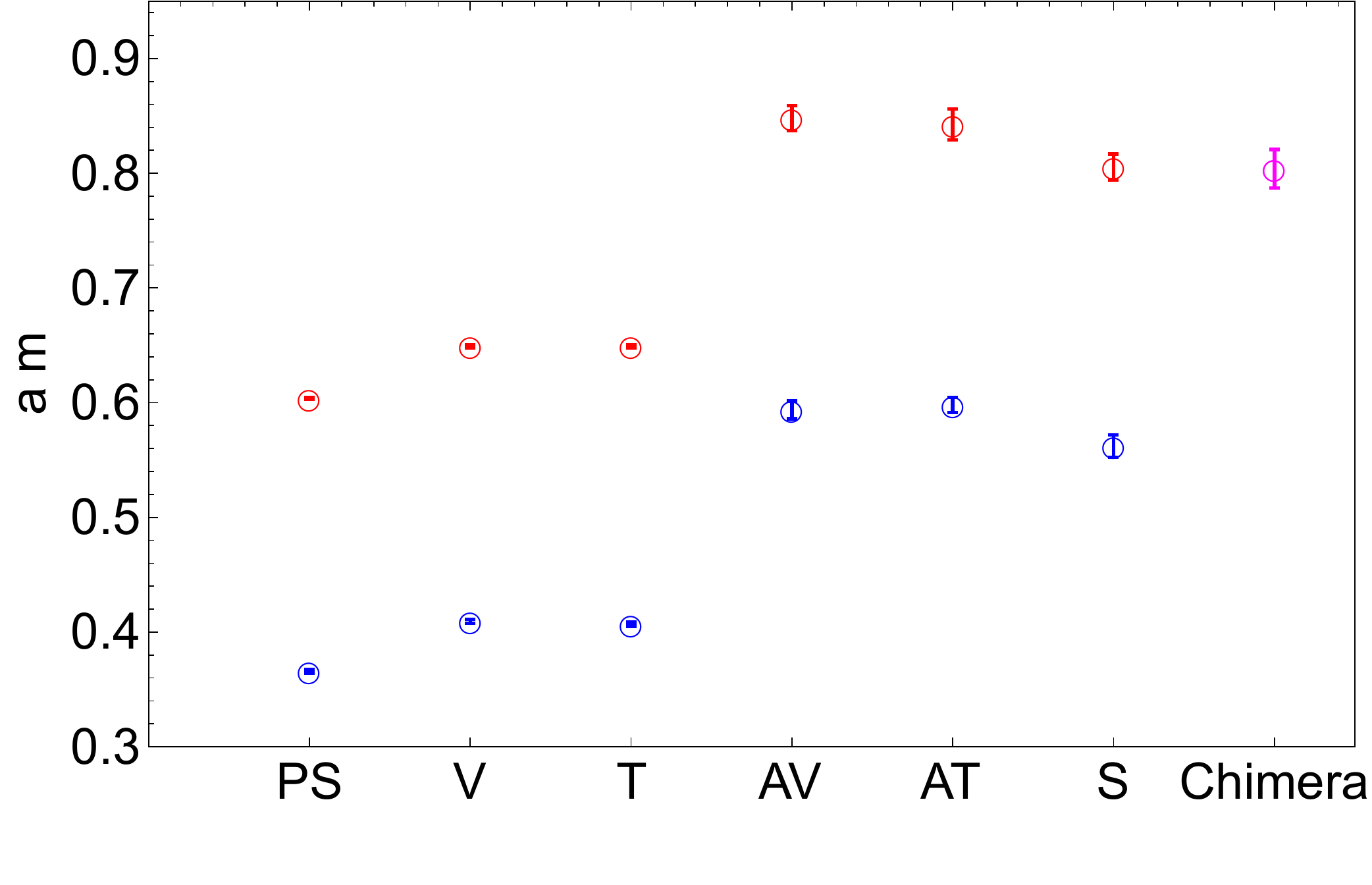}
\caption{%
\label{fig:full_spectrum}%
Illustrative example of mass spectrum of  composite states in the 
mixed-representation dynamical $Sp(4)$ theory.
The lattice parameters are $\beta=6.5$, $a m_0^{\rm as}=-1.01$, $a m_0^{\rm f}=-0.71$,
and the lattice volume is $48\times 24^3$.
Blue and red symbols denote flavoured mesons composed of the fundamental and antisymmetric representation constituents, respectively,
in different channels---pseudoscalar (PS), vector (V), tensor (T), axial vector (AV), axial tensor (AT), and scalar (S).
The magenta symbol denotes the chimera baryon with $J^P=\frac{1}{2}^+$, with interpolating operator
consisting of one antisymmetric and
two fundamental  fermions.
}
\end{center}
\end{figure}

We conclude with an illustrative example of the spectrum
obtained for fixed lattice parameters  in Fig.~\ref{fig:full_spectrum}.
We present the masses of flavoured spin$-0$ and spin$-1$ mesons with the constituents either in the fundamental 
or in the antisymmetric representation, as well as
the mass of the parity-even chimera baryon. 
With these choices of lattice parameters, the mesons composed of fundamental-representation
 constituents are much lighter than those of antisymmetric-representation ones,
yet the  mass difference is approximately independent of the meson. 
The lightest chimera baryon has mass similar to the scalar meson composed of 
 antisymmetric-representation fermions.

\section{Outlook}
\label{sec:conclusion}

We summarised preliminary results for an extensive programme 
of explorations on the lattice of 
gauge theories with $Sp(2N)$ group.
We will follow up this Proceedings contribution
with a series of publications reporting more extensive studies,
and definite physics results,
yet these preliminary measurements illustrate our processes 
and our current stage of progress in the programme.

In the case of the pure Yang-Mills theories, 
we will collect a larger number of ensembles,
 perform the continuum limit extrapolations of 
 the $Sp(2N)$ topological susceptibility for $N=1,\,2,\,3,\,4$,
extrapolate to $N\rightarrow +\infty$, and compare to available 
literature on the $SU(N_c)$ theories, possibly
by using the string tension $\sigma$ as a 
physical comparison scale.

We will complete the extensive programme of computing the 
spectrum of mesons
in the fundamental, as well as 2-index symmetric and antisymmetric representations, 
in the quenched approximation,
by extending it to the $N=4$ case,
performing combined continuum and chiral extrapolations, and combining the 
results for  $N=1,\,2,\,3,\,4$ towards approaching the large-$N$ limit.

For dynamical fermions in the antisymmetric representation of $Sp(4)$, 
we saw that the approach to the chiral limit is
slow, and hence we will study a larger number of ensembles, 
moving closer both towards lower mass, and 
towards the continuum. We will study the extrapolations to the 
continuum limit for the spectrum of mesons,
and include also some of the excited states in the spectrum. A
 systematic study of the spectrum of partially quenched chimera baryons composed of 
one fermion in the 2-index antisymmetric and two (quenched)
in the fundamental representation is under way.

For the fully dynamical $Sp(4)$ theory with mixed-representation fermions, for which 
we chose the field content to match the CHM in Ref.~\cite{Barnard:2013zea},
our next step requires completing 
 an extensive study of the general properties of the lattice theory,
reaching beyond the scan of parameter space presented here. 
We will then be in a position to start performing systematic
studies of the spectrum of the theory 
that are relevant both to composite Higgs and partial top compositeness.

The completion and combination of the aforementioned
measurements has potential
transformative effect on the current understanding of
$Sp(2N)$ gauge theories, providing unprecedented level 
of quantitative information about them. 
Besides the specific application to the  CHM 
 we investigate, this information
can be used as a starting point for the characterisation of other CHMs,
or models
of dark matter with strong-coupling origin,
and might have  applications in other fields of study.

\acknowledgments

The work of E.~B. has been funded by the Supercomputing Wales project, which is part-funded by the European Regional Development Fund (ERDF) via Welsh
Government and by the UKRI Science and Technologies Facilities Council (STFC) Research Software Engineering Fellowship EP/V052489/1. J.~H. is supported by the STFC Consolidated
Grant No. ST/P00055X/1, by the College of Science,
Swansea University, and by the Grant No. STFC-DTG
ST/R505158/1. The work of D.~K.~H. was supported by
Basic Science Research Program through the National
Research Foundation of Korea (NRF) funded by the
Ministry of Education (NRF-2017R1D1A1B06033701).
The work of J.~W.~L. is supported in part by the National
Research Foundation of Korea funded by the Ministry of
Science and ICT (NRF-2018R1C1B3001379) and in part
by Korea Research Fellowship program funded by the
Ministry of Science, ICT and Future Planning through
the National Research Foundation of Korea
(2016H1D3A1909283). The work of C.~J.~D.~L. is supported
by the Taiwanese MoST Grant No. 105-2628-M-009-003-
MY4. The work of B.~L. and M.~P. has been supported in part
by the STFC Consolidated Grants No. ST/P00055X/1 and No. ST/T000813/1. B.~L. and M.~P. received funding from
the European Research Council (ERC) under the European
Union’s Horizon 2020 research and innovation program
under Grant Agreement No. 813942. The work of B.~L. is
further supported in part by the Royal Society Wolfson
Research Merit Award No. WM170010 and by the
Leverhulme Trust Research Fellowship No. RF-2020-4619. The work of D.~V. is supported in part by the INFN HPCHTC project and in part by the Simons Foundation under the
program “Targeted Grants to Institutes” awarded to the
Hamilton Mathematics Institute. D. V. thanks C. Bonati,
M. D’Elia, and L. Gallina for useful discussions. Numerical
simulations have been performed on the Swansea SUNBIRD
cluster (part of the Supercomputing Wales project) and AccelerateAI A100 GPU system,
on the local HPC clusters in Pusan National
University (PNU) and in National Chiao-Tung University
(NCTU), and on the Cambridge Service for Data Driven
Discovery (CSD3). The Swansea SUNBIRD system and AccelerateAI are part funded
by the European Regional Development Fund (ERDF) via
Welsh Government. CSD3 is operated in part by the
University of Cambridge Research Computing on behalf
of the STFC DiRAC HPC Facility (www.dirac.ac.uk). The
DiRAC component of CSD3 was funded by BEIS capital
funding via STFC capital Grants No. ST/P002307/1 and
No. ST/R002452/1 and STFC operations Grant No. ST/
R00689X/1. DiRAC is part of the National e-Infrastructure.

\end{document}